\documentclass[traditabstract]{aa} 
\pdfoutput=1
\usepackage[varg]{txfonts}
\usepackage{graphicx,rotating,amssymb,amsmath}
\usepackage{booktabs}
\usepackage{subfig}
\usepackage{float,capt-of}
\usepackage{natbib}
\usepackage{multirow}
\usepackage[english]{babel}
\usepackage{morefloats}
\usepackage{color}

\newcommand\cii{[\ion{C}{ii}]~}
\newcommand\kms{km~s$^{-1}$}
      
\newcommand\oi{[\ion{O}{i}]~}
\newcommand\nii{[\ion{N}{ii}]~}
\begin{document}
   \title{Very extended cold gas, star formation and outflows in the halo \\ of a bright QSO at $z>6$}

   \author{C. Cicone
          \inst{1, 2} 
          \and
           R. Maiolino  \inst{1, 2}
	\and
    	   S. Gallerani \inst{3}
	 \and
	 R. Neri \inst{4}
	 \and 
	 A. Ferrara \inst{3}
	 \and 
	 E. Sturm \inst{5}
	 \and
	 F. Fiore \inst{6}
	 \and 
	 E. Piconcelli \inst{6}
	 \and
	 C. Feruglio \inst{4}
	}
      \institute{Cavendish Laboratory, University of Cambridge 19 J. J. Thomson Avenue, Cambridge CB3 0HE, UK\\ 
      	\email{c.cicone@mrao.cam.ac.uk}
	  \and Kavli Institute for Cosmology, University of Cambridge, Madingley Road, Cambridge CB3 0HA, UK\\ 
	  \and Scuola Normale Superiore, Piazza dei Cavalieri 7, 56126 Pisa, Italy \\ 
	  \and Institute de Radioastronomie Millimetrique (IRAM), 300 Rue de la Piscine, F-38406 St. Martin d'Heres, Grenoble, France\\ 
	  \and Max Planck Institute f\"{u}r Extraterrestrische Physik (MPE), Giessenbachstra\ss e 1, 85748, Garching, Germany\\ 
	   \and Osservatorio Astronomico di Roma (INAF), via Frascati 33, I-00040 Monteporzio Catone, Italy \\ 
	      }

   \date{Received xx xx xx / Accepted: xx xx xx }

\abstract{Past observations of QSO host galaxies at $z >6$ have found cold gas and star 
formation on compact scales of a few kiloparsecs.
We present new high sensitivity IRAM PdBI follow-up observations
of the \cii 158$\mu$m emission line and FIR continuum in the host galaxy of SDSS~J1148+5251, a luminous QSO at redshift 6.4189. 
We find that a large fraction of the gas traced by \cii is at high velocities, up to $\sim$1400 \kms relative to the systemic velocity,
confirming the presence of a major outflow indicated by previous observations.
The outflow has a complex morphology
and reaches a maximum projected radius 
of $\simeq$30 kpc. The extreme spatial extent of the outflow allows us, for the first time
in an external galaxy, to estimate mass-loss rate,
kinetic power and momentum rate of the outflow as a function of the projected distance 
from the nucleus and the dynamical time-scale. These trends reveal multiple outflow events during the past 100 Myr,
although the bulk of the mass, energy and momentum appear to have been released more recently, within the past $\sim$20~Myr.
Surprisingly, we discover that also the quiescent gas at systemic velocity is extremely extended. 
More specifically, we find that, while 30\% of the \cii within $v\in$(-200, 200) \kms traces a compact component that is not resolved
by our observations, 70\% of the \cii emission in this velocity range is extended, 
with a projected FWHM size of 17.4$\pm$1.4 kpc.
We detect FIR continuum emission associated with both the compact and the extended \cii components, although 
the extended FIR emission has a FWHM of
11$\pm$3 kpc, thus smaller than the extended \cii source. 
Overall, our results indicate that the cold gas traced by \cii is distributed up to $r\sim$30 kpc in the host galaxy of SDSS~J1148+5251.
A large fraction of extended \cii is likely associated with star formation occurring on large scales, but the \cii source extends
well beyond the FIR continuum, and additional multi-wavelengths 
observations are needed in order to clarify the origin of this most extended \cii.}

 
   \maketitle

%

\section{Introduction}

Recent millimetre observations have revealed that the host galaxies of high redshift luminous QSOs 
are quite compact, with extremely intense star formation confined within a few kiloparsecs in their nuclear regions, fed
by large amounts of dense gas (e.g. see \citealt{Carilli+Walter13} for a recent review). 
Moreover, high resolution interferometric observations of these high redshift
sources have suggested the presence of regularly rotating and relatively undisturbed disks \citep{Carniani+13,Wang+13},
although in some cases they appear to be highly turbulent \citep{DeBreuck+14}.
However, recent theoretical studies have concluded that, in order to reconcile the extreme properties (BH mass, dust and gas mass)
of these high redshift quasars with early star formation occurring with a standard IMF, the dynamical masses of these objects
should be a factor of 3-30 higher than the maximum values allowed by observations (\citealt{Valiante+11, Valiante+14, Calura+14}).
This suggests that past millimetre observations may have missed a significant fraction of baryons in high redshift QSO host galaxies. 

Only in a few high-redshift QSOs and radio galaxies, observations have revealed the presence 
of some diffuse gas at a distance of several kiloparsecs from the nucleus,
possibly associated with gas-rich companions and suggestive of galaxy merger processes
(e.g. \citealt{Ivison+08,Ivison+12,Gallerani+12,Emonts+14}). Within the massive
haloes of high-z radio galaxies, cold gas emission can be detected up to tens of kiloparsecs 
offset from the galactic nucleus, often aligned with the extended radio jets \citep{Nesvadba+09, Emonts+11,Emonts+14}.

The collisionally-excited fine structure transition of C$^{+}$ ${\rm ^{2}P_{3/2}\rightarrow ^{2}P_{1/2}}$ at rest-frame wavelength
of 157.741~$\mu$m (hereafter \cii 158$\mu$m) is conveniently redshifted, at $z>4$, into the sub-millimetre/millimetre atmospheric windows.
The \cii 158$\mu$m is one of the brightest lines in the spectrum of star forming galaxies, as it is the dominant coolant of the ISM
over a wide range of physical conditions. The ionisation potential of carbon is 11.26 eV, hence C$^+$, HI and even H$_2$ usually coexist in 
the same medium, and indeed neutral hydrogen is the main collisional partner of C$^+$. 
Although \cii emission is commonly associated with star formation (e.g. \citealt{Sargsyan+14, DeLooze+14}), the large variety of physical conditions in which
C$^+$ can be produced and excited makes its interpretation not straightforward. This has been clearly shown by recent 
detailed studies of the spatial distribution of various gas tracers in the Milky Way and other few nearby galaxies,
including Herschel observations of [\ion{C}{ii}]. 
In particular, in our Galaxy,
only $\sim$50\% of the \cii emission is
associated with dense photon-dominated regions (PDRs) exposed to the FUV radiation field of young and massive stars 
\citep{Pineda+13}.
A large fraction of \cii ($\sim$30\%) in the Milky Way is linked to ``CO-dark'' molecular hydrogen, which dominates the gas mass of
diffuse molecular clouds, but it can also be an important component of some unshielded dense molecular clouds with large envelopes where
the formation of CO is inhibited and most of the gas-phase carbon is found in C$^0$ or C$^+$ atoms 
\citep{Wolfire+10,Langer+10,Langer+14,Velusamy+10,Pineda+13}.
Finally, \cite{Pineda+13} have shown that $\sim$20\% of \cii is linked to diffuse atomic clouds (in particular to the cold neutral medium, CNM), and a low fraction ($\sim$4\%) is associated with warm ionised medium (WIM), although
the WIM contribution to \cii emission may be significantly larger according to \cite{Velusamy+Langer14}.
Further complicating its interpretation, \cii can be enhanced by shocks \citep{Appleton+13}.
Intriguingly, \cii seems to be also a good tracer of outflows and, especially, of molecular gas in outflow
\citep{Fischer+10, Contursi+13,Kreckel+14}.

The first high redshift source in which \cii emission was ever detected \citep{Maiolino+05}, SDSS~J1148+5251 at redshift 6.4189, is one of the
most powerful quasars in the Universe, with an intrinsic bolometric luminosity of $L_{\rm AGN} = 5.63 \times 10^{13}~L_{\odot}$ \citep{Schneider+14}. 
This is the perfect laboratory to study the formation of the first massive galaxies and the
coevolution of galaxies and black holes at the end of the reionization epoch. SDSS~J1148+5251 was discovered
through SDSS photometry \citep{Fan+03}, and since then it has
been the subject of extensive multi-wavelength studies, aimed at constraining the extreme properties of its host galaxy 
and of its cosmological environment (e.g. \citealt{Bertoldi+03,Walter+03,Walter+04,Willott+03,Riechers+09,Gallerani+10,Decarli+12,Carilli+Walter13}).
\cite{Walter+09} observed the \cii emission line in SDSS~J1148+5251 using the most extended configuration of the IRAM Plateau de Bure Interferometer (PdBI), and 
derived a size of 1.5 kpc, which, combined with the extreme FIR luminosity of this source, suggests the presence of a kpc-scale
``hyper-starburst'' in its nucleus, fed by large amounts of molecular gas.
Moreover, SDSS~J1148+5251 was also the first high redshift galaxy in which a powerful quasar-driven
outflow was discovered \citep{Maiolino+12}.
The outflow was revealed by the presence of extremely broad wings of the \cii emission line, tracing gas at high velocities
($v$ up to 1200 \kms), which appeared marginally resolved at a 2 arcsec resolution, with a measured (projected) diameter of $\sim16$~kpc.

In this work we present new follow-up observations of the \cii emission line and FIR continuum in SDSS~J1148+5251, obtained
with the C and D configurations of the IRAM PdBI (resulting in a synthesized beam size of 1.3\arcsec$\times$1.2\arcsec 
for the \cii observations and
of 1.1\arcsec$\times$1.0\arcsec for the FIR continuum observations), 
which enable us to investigate the properties of this source on large scales,
with a sensitivity much higher than previous observations.
Throughout the paper, we adopt a standard $\Lambda$CDM cosmological model with $H_0$ = 67.3 \kms Mpc$^{-1}$, 
$\Omega_{\Lambda}$ = 0.685, $\Omega_{\rm M}$ = 0.315 \citep{Planck_cosmology+13}.
With the adopted cosmology, at the redshift of SDSS~J1148+5251, the physical scale is
5.623 kpc~arcsec$^{-1}$, and the age of the Universe is 0.85~Gyr.

\begin{figure*}[tb]
	\centering
    \includegraphics[clip=true, trim=3.5cm 4.7cm .2cm 7.2cm,angle=90,width=.65\columnwidth]{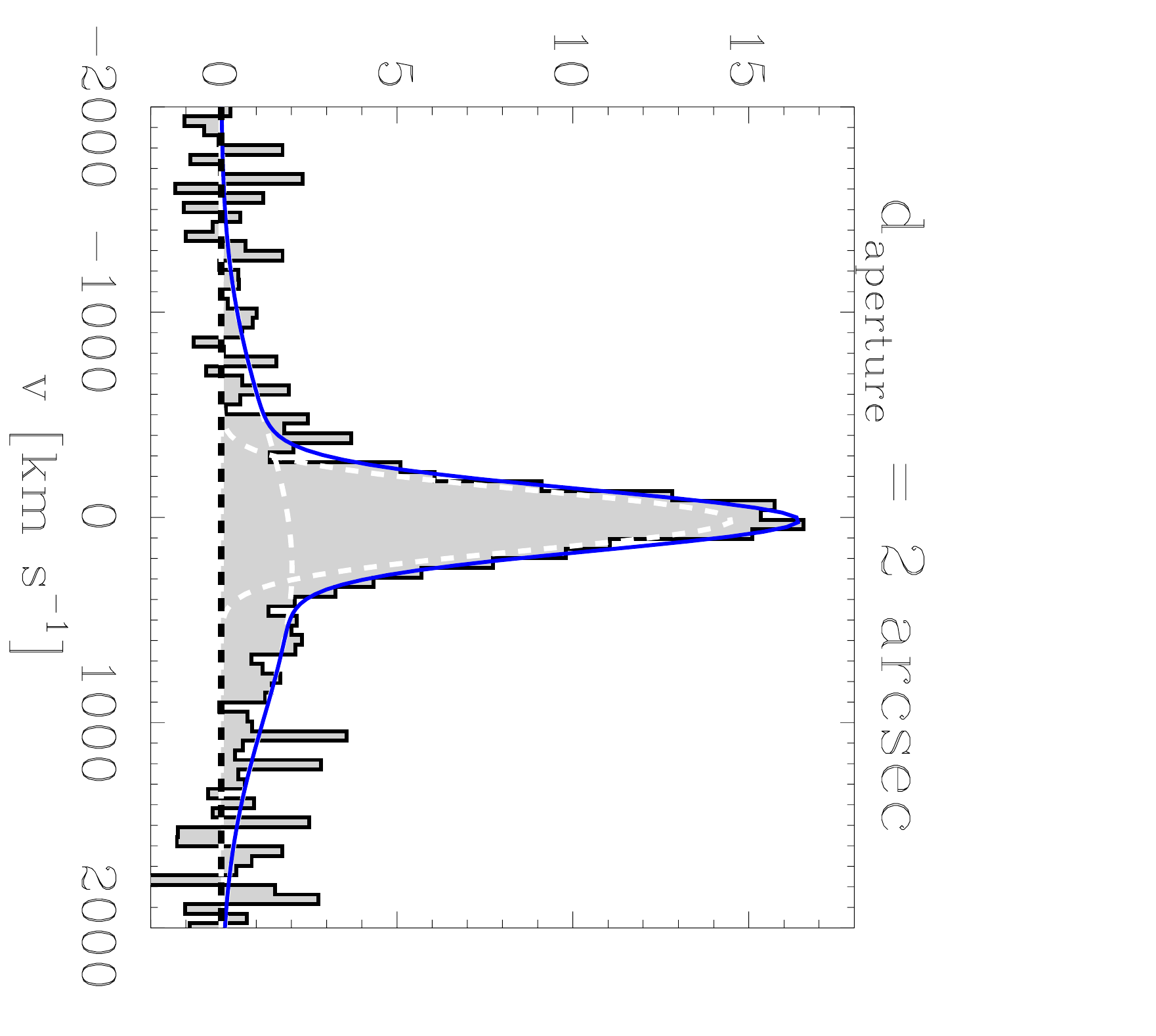}\quad
    \includegraphics[clip=true, trim=3.5cm 4.7cm .2cm 7.2cm,angle=90,width=.65\columnwidth]{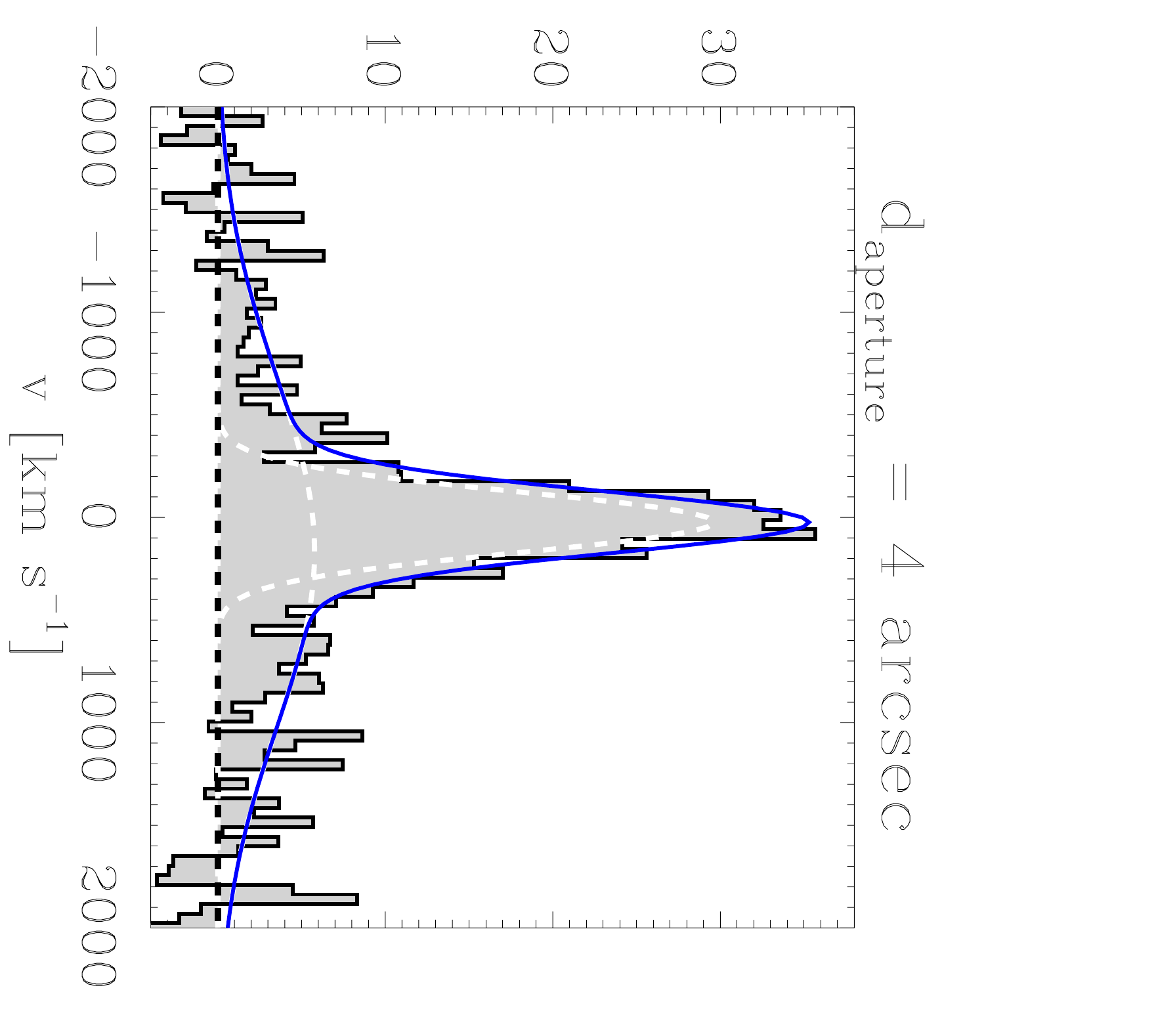}\quad
    \includegraphics[clip=true, trim=3.5cm 4.7cm .2cm 7.2cm,angle=90,width=.65\columnwidth]{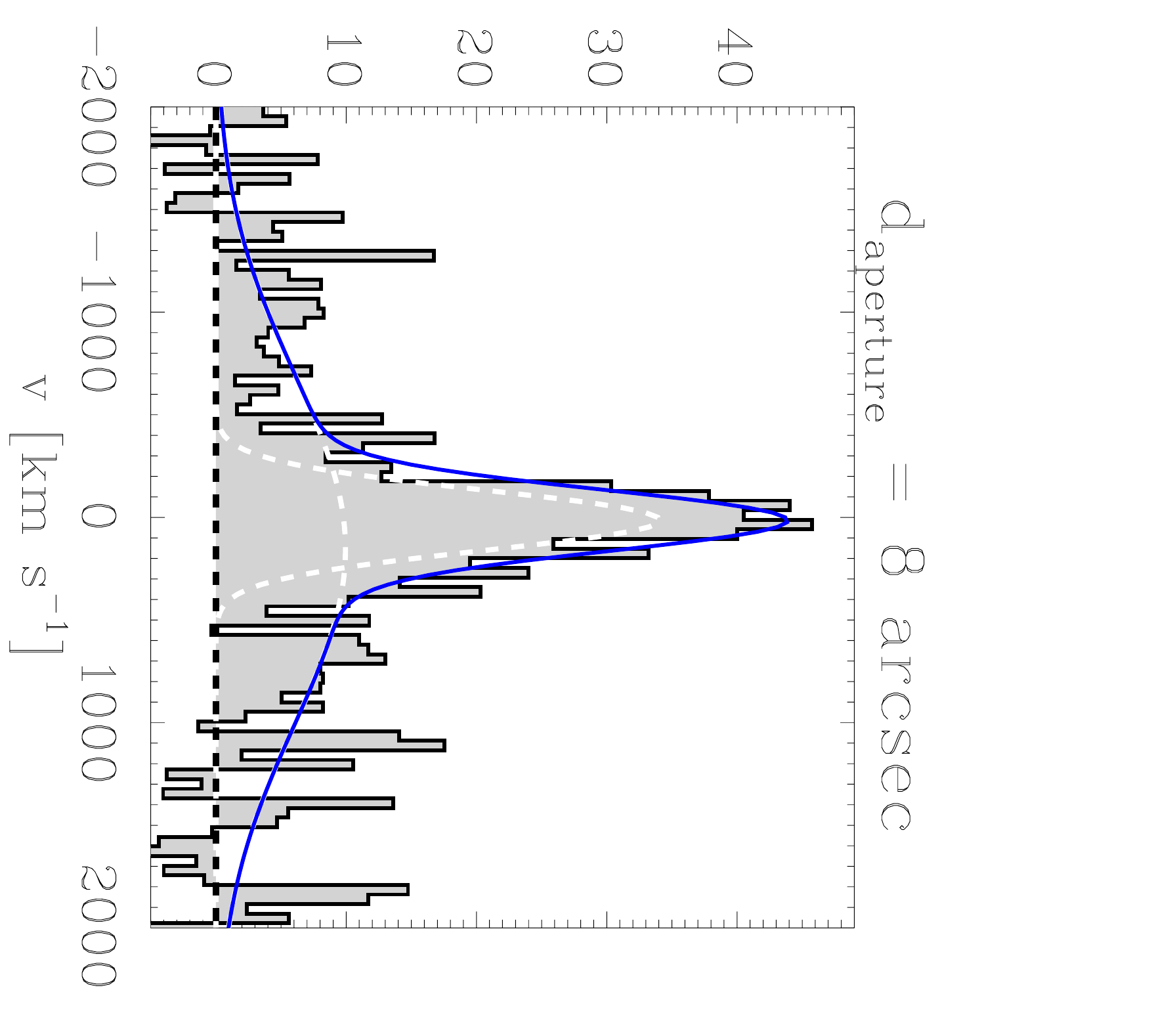}\quad
     \caption{IRAM PdBI continuum-subtracted spectrum of the \cii 158$\mu$m emission line of SDSS~J1148+5251. 
     		The three spectra shown have been extracted using different apertures, 
		namely circular apertures with diameters of 2 arcsec ({\it left panel}), 4 arcsec ({\it central panel})
		and 8 arcsec ({\it right panel}).
		Note the different scales on the y-axes in the three panels. 
		For display purposes, the spectra have been re-binned by a factor of 2, corresponding to channels of 46.8 \kms. 
		The Gaussian fits to the line profiles (using a narrow and a broad Gaussian 
		component to fit respectively the narrow core, tracing quiescent gas, and the broad wings, tracing the outflow) are performed on the original 
		non-binned spectra and the fits results are reported in Table~\ref{table:gauss_fit}. }
   \label{fig:spectra}
\end{figure*}

\begin{table*}
 \begin{minipage}{175mm}
  \caption{Results of the Gaussian fits to the \cii158$\mu$m emission line spectrum of SDSS J1148+5251:}
  \label{table:gauss_fit}
\begin{tabular}{@{}lccccccccc@{}}
\hline
\hline
 Aperture$^{\dag}$     		&        \multicolumn{4}{c}{Narrow component} &  \multicolumn{4}{c}{Broad component} &  \multicolumn{1}{c}{Total line}  \\
     		& $v$     & $\sigma_v$ &  $S_{\rm peak}$ & $I_{\nu}$ & $v$    &     $\sigma_v$  & $S_{\rm peak}$ & $I_{\nu}$ & $I_{\nu}$ \\
		& [\kms] 	& [\kms] 	   &  [mJy] 		& [Jy \kms] & [\kms] & [\kms] 	   &  [mJy] 		& [Jy \kms] & [Jy \kms] \\
\hline
\\
2 arcsec 	& 13 $\pm$ 9 &  146 $\pm$ 11 & 14.5 $\pm$ 0.9 & 5.3 $\pm$ 0.5 & 250 $\pm$ 160 & 730 $\pm$ 170 & 2.0 $\pm$ 0.6 & 3.7 $\pm$ 1.4 &  9.0 $\pm$ 1.5 \\
4 arcsec   & 24 $\pm$ 13 & 148 $\pm$ 16 &  30 $\pm$ 3 & 11.0 $\pm$ 1.5 & 170 $\pm$ 170 & 900 $\pm$ 200 & 5.8 $\pm$ 1.6 & 12 $\pm$ 5 & 23 $\pm$ 5 \\
8 arcsec & 14 $\pm$ 19 & 150 $\pm$ 20 &  34 $\pm$ 4 &  13 $\pm$ 3 &  200 $\pm$ 300 & 900 $\pm$ 300 &  10 $\pm$ 3 & 21 $\pm$ 9 &  34 $\pm$ 9  \\
 \hline
\end{tabular}
\end{minipage}

\begin{flushleft}
\small
\textbf{Notes:} 
$^\dag$  Diameter of the circular aperture used to extract the spectrum, as in Fig.~\ref{fig:spectra}. 
\end{flushleft}
\end{table*}

\section{Observations}\label{sec:observations}

We have observed with the IRAM PdBI the \cii 158$\mu$m line in the host galaxy of the luminous quasar 
SDSS~J1148+5251. The pointing and phase centre of our observations corresponds to the SDSS optical position
of the quasar (RA=11:48:16.64, Dec=52:51:50.30).
At the redshift of the source (z=6.4189, \cite{Maiolino+05}), the \cii 158$\mu$m transition is redshifted
into the 1mm band, at an observed frequency of 256.175 GHz.
The observations were obtained with the C configuration of the IRAM PdBI (with 6 antennas covering baselines from 16.6m to 176.0m) between 
November 2012 and April 2013 and were combined with the previous 2011 observations in D configuration 
\citep{Maiolino+12}, for a total on-source-time of 30~hrs. We reach a sensitivity of 0.46 mJy beam$^{-1}$ in
channels of 100~\kms, which is unprecedented for IRAM PdBI observations of high redshift quasars at this frequency
(\cite{Wang+13} reach a similar line sensitivity using ALMA).
The resulting synthesized beam-size using natural weighting is 1.3\arcsec$\times$1.2\arcsec,
and the PdBI primary beam, giving the field of view, at these frequencies is 19.7\arcsec$\times$19.7\arcsec.
Data reduction and analysis were performed by using the CLIC and MAPPING softwares within the GILDAS package. 
The flux calibration accuracy in the 1mm band is about 20\% \citep{PdBI_Cookbook}. 
We used the wide-band (WideX) correlator offered by the PdBI, which provides a spectral resolution 
of 20~MHz (corresponding to 23.4~\kms) over its full bandwidth of 3.6~GHz. 
The 3.6~GHz spectral window was centred at 256.172~GHz (i.e. on the \cii 158$\mu$m line) 
for all tracks but two, for which the centre was instead set at 256.672~GHz, hence providing 
an additional 0.5~GHz-wide window on the continuum blueward of the line.

In this work we also make use of ancillary C+D configuration IRAM PdBI observations of the FIR continuum of
SDSS~J1148+5251 at 262~GHz (rest frame wavelength $\lambda = 154.2~\mu$m) and 259.4~GHz (rest frame 
wavelength $\lambda = 155.8~\mu$m), carried out between May 2010 and December 2012. 
By merging these two datasets at 262~GHz and 259.4~GHz, we obtain a high sensitivity
(1$\sigma$ rms noise of 0.082 mJy beam$^{-1}$) map of the FIR continuum of SDSS~J1148+5251
at an average rest-frame wavelength $\lambda = 155~\mu$m (average observed frequency of 261~GHz). 
The resulting synthesized beam-size using natural weighting is 1.1\arcsec$\times$1.0\arcsec.
The total on source time for these ancillary FIR observations is 28 hrs.
These deep FIR continuum observations are part of a parallel
IRAM PdBI programme. The dataset at 262~GHz has been
presented by \citet{Gallerani+14}, who report the serendipitous detection of the
CO(17-16) and OH$^+$ emission lines in SDSS~J1148+5251.

\section{Results}

\begin{figure*}[tb]
\centering
\hspace{-2cm}
    \includegraphics[angle=270,width=1.5\columnwidth]{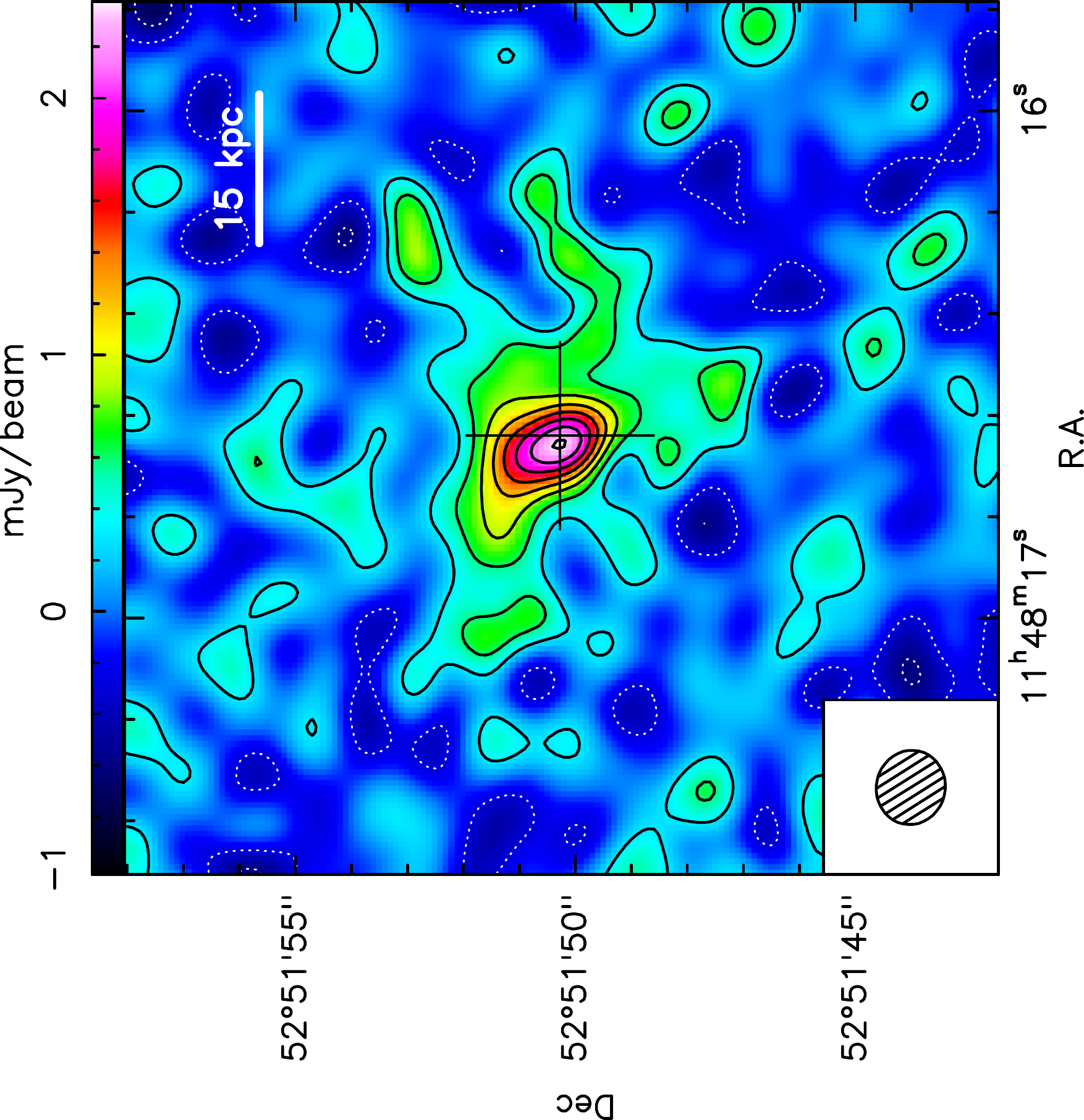}\\
     \caption{IRAM PdBI continuum-subtracted map of the total \cii158$\mu$m emission of SDSS J1148+5251,
     		integrated within $v\in(-1400, 1200)$ \kms. Negative and positive contours are in steps of 3$\sigma$ (1$\sigma$
     		rms noise is 0.26 Jy beam$^{-1}$ km s$^{-1}$).
     		The synthesized beam (1.3\arcsec$\times$1.2\arcsec) is shown in the bottom left corner of the map. The cross
		indicates the pointing and phase centre, corresponding to the optical position of the quasar. }
   \label{fig:total_map}
\end{figure*}

\begin{figure*}[tb]
	\centering
    \includegraphics[clip=true,angle=90,width=1.7\columnwidth]{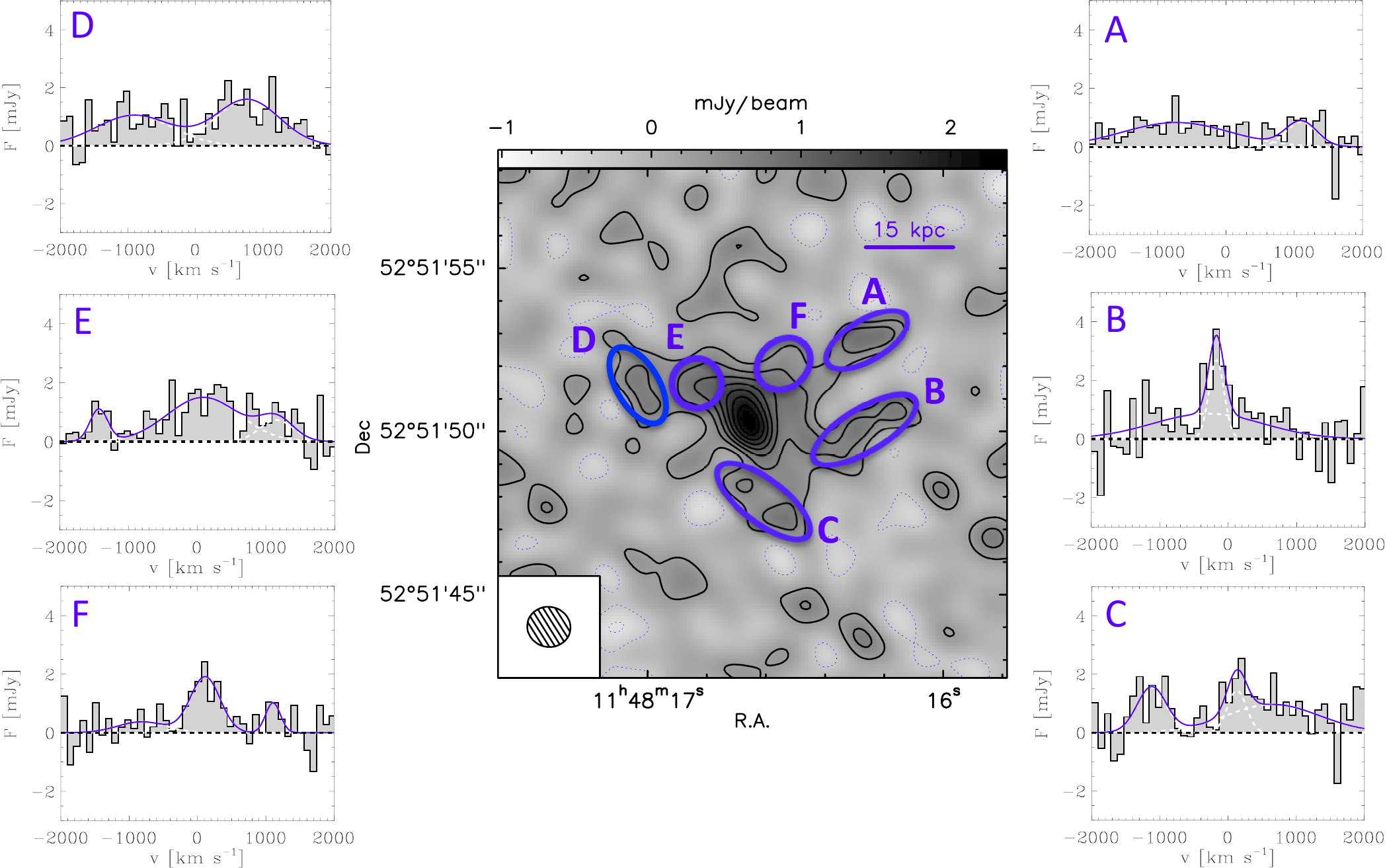}\\
     \caption{IRAM PdBI continuum-subtracted map of the total \cii 158$\mu$m emission of SDSS J1148+5251,
     		integrated within $v\in(-1400, 1200)$ \kms, along with spectra extracted from different positions on the source.
     		Negative and positive contours are in steps of 3$\sigma$ (1$\sigma$
     		rms noise is 0.26 Jy beam$^{-1}$ km s$^{-1}$). For display purposes, the spectra have been re-binned by a factor of 4, 
		corresponding to channels of 93.6 \kms. The Gaussian fits to the line profiles are performed on the original 
		non-binned spectra and the fits results are presented in Table \ref{table:pos_fit}.
		}
   \label{fig:position_spectra}
\end{figure*}

\subsection{The FIR continuum at 256~GHz}\label{sec:cont_256}

For the purpose of continuum subtraction, 
the rest-frame 157.7$\mu$m FIR continuum (observed frequency of 256 GHz)
is estimated by collapsing the line-free channels
of our wide-band IRAM PdBI observations, corresponding to the velocity
ranges of $v\in(-2700, -2400)$ km~s$^{-1}$, $v\in(-1900, -1500)$ km~s$^{-1}$
and $v\in(1500, 2100)$ km~s$^{-1}$. 
The channels between $v$ = -2400~km~s$^{-1}$ and $v$ = -1900~km~s$^{-1}$
are excluded from the continuum integration because
of the presence of a putative feature, detected at the $4.2\sigma$
level in the maps, whose position is shifted by 1.5~arcsec to the south 
with respect to the quasar optical position and whose origin is not clear from our data
(it could be be either part of the extended outflow discussed in Section~\ref{sec:outflow} 
or associated with another galaxy).

The {\it uv} plot of the continuum at 256~GHz can be fitted by a point
(unresolved) source\footnote{We note that, although the 
FIR continuum of SDSS~J1148+5251 is not resolved in the 256 GHz dataset, 
the FIR continuum emission appears well resolved in the deeper ancillary dataset at 261 GHz
($\lambda_{\rm rest} = 155~\mu m$, see also explanation in Section 2), as shown in Section~\ref{sec:FIR_cont}.}
centered at RA=11:48:16.62 and Dec=52:51:50.22, with a flux density of 3.3 mJy. 
We compare our new continuum flux density estimate with the MAMBO-II 1.2mm bolometric
observations of \cite{Bertoldi+03}, who report a flux density of ($5.0\pm0.6$)~mJy.
Because of the broad bandwidth of MAMBO-II, the flux expected
at our PdBI frequency of 256 GHz from the MAMBO-II observations of \cite{Bertoldi+03} must be modelled by
taking into account both the MAMBO-II transmission curve, which determines the spectral response of the bolometer
at different frequencies, and the FIR SED of SDSS~J1148+5251, which, at FIR wavelengths,
can be modelled by using a modified blackbody curve with emissivity $\beta = 1.6$ \citep{Leipski+14}.
For simplicity, we perform the calculation under the Rayleigh-Jeans approximation and by assuming 
$\tau_{\nu} \ll 1$ so that the flux density scales as $S_{\nu} \propto \nu^{(\beta + 2)}$, 
where the optical depth is parametrized as a power law in frequency,
$\tau_{\nu}=(\nu/\nu_0)^{\beta}$, following \cite{Draine06} \footnote{$\lambda_0=c/\nu_0$ is the wavelength where
the optical depth is unity, expected to be $\lambda_0\simeq100~\mu$m.}.
We obtain an expected flux density at 256 GHz of ($4.8\pm 0.6$)~mJy, to be compared with
our measured value of ($3.3\pm0.7$)~mJy, where the error includes a 20\% flux calibration uncertainty.
We note that there is a small inconsistency between our new
157.7$\mu$m continuum estimate and the flux density expected at these wavelengths from the
bolometric observations of \cite{Bertoldi+03}. This discrepancy was not evidenced 
in \cite{Maiolino+12} because of the larger error of their continuum measure, which was based on
the D-configuration data only.
This small inconsistency with \cite{Bertoldi+03} is likely due to
the presence of additional sources close to SDSS~J1148+5251 contributing to the 
10.7\arcsec MAMBO-II beam. 
In particular, one source $\sim$10\arcsec north-west 
of SDSS~J1148+5251 was clearly detected by Herschel \citep{Leipski+10,Leipski+13}, and it
is also marginally detected by our PdBI observations (Figure~\ref{fig:appendix_newsource}).
In Appendix~\ref{sec:additional_source} we further test this hypothesis and we
show that, when modelling the FIR SED of SDSS~J1148+5251 
by considering only the Herschel-PACS continuum measurements of \cite{Leipski+13}, 
corrected for the contaminating north-western source, and the PdBI measurements by 
\cite{Riechers+09} and \cite{Gallerani+14} at other frequencies, the flux
density expected at the MAMBO-II frequencies (by taking into account the full transmission curve) 
would be of 3.8 mJy, which in turn would predict
a flux density at 256 GHz of (3.6 $\pm$ 0.4) mJy, consistent with our PdBI measurement. 

\subsection{The total \cii 158$\mu$m emission}\label{sec:total}

Following the standard procedure, we subtract 
the continuum (estimated in Section~\ref{sec:cont_256}) directly from the $uv$ data, and we extract from
the continuum-subtracted and cleaned data cube the spectrum of the \cii 158$\mu$m emission
line of SDSS~J1148+5251. In Figure~\ref{fig:spectra} we show the resulting continuum-subtracted
spectra extracted using 
three circular apertures with diameters of 2~arcsec, 
4~arcsec, and 8~arcsec. Note that the three plots in Fig.~\ref{fig:spectra} have different scales
on the y-axis.
Each spectrum in Fig.~\ref{fig:spectra} is fit by using two Gaussians,
e.g., a narrow Gaussian to fit the central core of the line, tracing quiescent gas, and a broad
Gaussian to fit the broad wings, tracing the outflow discovered by \cite{Maiolino+12}. 
The fit with two Gaussians is a simplistic model for the 
\cii 158$\mu$m emission profile of SDSS~J1148+5251, however it constitutes a good starting point for our study.
The fit results are reported in Table~\ref{table:gauss_fit}. 
The widths of the two Gaussians used to fit the \cii 158$\mu$m spectral profile
show no significant variations between spectra extracted from different apertures\footnote{We note
that all the six parameters of the fit are left as free parameters in the spectral fitting procedure. In particular, 
we did not constrain the fits performed on the spectra extracted from different apertures to have same
widths, central velocities and/or amplitudes.},
and they are consistent with \cite{Maiolino+12}.

Figure~\ref{fig:spectra} and Table~\ref{table:gauss_fit} clearly show that the contribution from an extended gas
component to the total \cii 158$\mu$m flux 
in SDSS~J1148+5251 is significant. Indeed, the total integrated flux estimated by using the spectrum extracted from 
the 8~arcsec diameter aperture is significantly larger than the flux retrieved using the 
4~arcsec aperture, and is almost {\it four} times as large as the flux measured from the 2 arcsec aperture.
Such flux increment is observed in both the narrow and the broad Gaussians that are fit to the emission line profiles,
suggesting that both components trace emission extended on {\it extremely} large scales 
($d\gtrsim 4$~arcsec on sky, corresponding to $d\gtrsim 22$~kpc).
However, we note that the largest flux variation with aperture size is observed in the broad
wings of the \cii 158$\mu$m line, confirming that the 
massive outflow is {\it very} extended.

The complex morphology and extreme spatial extent of the \cii 158$\mu$m emission
around this $z\sim6.4$ QSO are clearly revealed by the map in Figure~\ref{fig:total_map},
obtained by collapsing the velocity channels spanning the full 
velocity extent of the emission
line profile, i.e. from $v=-1400$ \kms to $v=1200$ \kms. The map shows a filamentary \cii 158$\mu$m source, with
structures extending in the radial direction up to the extraordinary (for a QSO host galaxy at $z=6.4$) (projected) distances of 
$r\sim$30~kpc from the
QSO position (indicated by the cross).

The fits performed on the spectra extracted
from different apertures (Figure~\ref{fig:spectra} and Table~\ref{table:gauss_fit})
suggest that a large fraction of the \cii 158$\mu$m emission at a distance $r>2$~arcsec from the QSO
(i.e. the emission that is not retrieved by using the 4 arcsec diameter aperture, 
but that is recovered with the largest 8 arcsec diameter aperture) is associated with the broad wings of the 
\cii line, tracing a powerful outflow \citep{Maiolino+12}. 
We further investigate this in Fig.~\ref{fig:position_spectra},
where we show the 
same map as Fig.~\ref{fig:total_map}, in which
we have marked the position of six regions (indicated with A-F) that
are at a distance of $r\gtrsim2$~arcsec from the emission peak. These regions 
represent the brightest features within the extended \cii 158$\mu$m source.
The spectra extracted from regions A-F (by using the elliptical apertures
indicated in the map) are also shown in Fig.~\ref{fig:position_spectra}. The spectral fitting, performed
by using multiple Gaussians (the fit results are reported in Table~\ref{table:pos_fit} in Appendix~\ref{sec:position_spectra}),
shows that features A-F are characterised, on average, 
by high velocity dispersions and velocity offsets, suggesting an outflow origin.

However, there is also a large fraction of extended \cii 
associated with ``narrow'' emission at systemic velocity (e.g. features B, C, F in Fig.~\ref{fig:position_spectra}), indicating the presence of extended dynamically ``quiescent'' (i.e. non outflowing) \cii-emitting gas
outside of the central ``hyper-starburst'' detected by \cite{Walter+09}. This extremely interesting - and unexpected - 
new prospect is investigated in Section~\ref{sec:core}.
Interestingly, the narrow ($\sigma_v = 110\pm 50$~\kms)
feature in spectrum B with 
peak flux density of $S_{\rm peak} = 2.7 \pm 1.0$~mJy
is blue-shifted at a velocity of $v = -170 \pm 40$~\kms
 (Table~\ref{table:pos_fit}). The shift in velocity, 
together with the low velocity dispersion
and large distance from the nucleus, suggests that this feature may have a merger/inflow origin. 
This hypothesis is however difficult to test with the current data. 

\subsection{Comparison with previous observations}\label{sec:comparison}

\begin{table}
 \begin{minipage}{80mm}
  \caption{Comparison with previous \cii observations of SDSS J1148+5251:}
  \label{table:comparison}
\begin{tabular}{@{}lccc@{}}
\hline
\hline
Ref.     			&  $\sigma_v$$^\dag$ 	& $S_{\rm peak}$  	&   $I_{\nu}$   	\\
				& [\kms] 	         	& [mJy] 		       	& [Jy \kms]  	 	\\
\hline
\cite{Maiolino+05} 	&   150 $\pm$ 20 	 & 12 			&   4.1 $\pm$ 0.5 	\\
\cite{Walter+09}  	&    122 $\pm$ 12	&  12.7 $\pm$ 1.1	&  3.9 $\pm$ 0.3	\\
\cite{Maiolino+12}$^\ddag$     &   150 $\pm$ 20 & 23 $\pm$ 2  &  14 $\pm$ 3		\\ 
This work $^\ddag$	& 150 $\pm$ 20  &	 44 $\pm$ 5 			&	34 $\pm$ 9		\\
\hline
\end{tabular}
\end{minipage}

\begin{flushleft}
\small
\textbf{Notes:} $^\dag$ The values of $\sigma_v$ reported for \cite{Maiolino+12} and for the present work refer to the {\it narrow} component 
to allow a comparison with earlier studies that did not detect the broad component. $^\ddag$ Values obtained from Gaussian fitting of
the spectra extracted from the largest apertures used in these works.
\end{flushleft}
\end{table}

We listed in Table~\ref{table:comparison} all previous and current \cii observations of SDSS~J1148+5251.
We note that the {\it total} \cii 158$\mu$m velocity-integrated flux of $34\pm9$~Jy~\kms that we estimate
from our new IRAM PdBI observations is significantly larger than {\it any} previous flux measurement of \cii
obtained for this source. 
Such large discrepancy deserves further clarification, as these new observations
significantly change the scenario presented by previous \cii 158$\mu$m observations of 
SDSS~J1148+5251 and of other high redshift quasars.

The first but only apparent flux inconsistency that we evidence is with the D-configuration data presented
in \cite{Maiolino+12}. The total flux derived by using the spectrum extracted from a 6~arcsec-diameter aperture 
shown in \cite{Maiolino+12} is over a factor of two lower than our new estimate (Table~\ref{table:comparison}). 
This is in part due to the use of a smaller aperture by \cite{Maiolino+12}, which likely missed a fraction
of the most extended emission.
However, we have also re-analysed 
the \cite{Maiolino+12} dataset (which is also included in the final dataset used in this work), and we have found that the bulk of the
flux loss was due to cleaning problems associated with the data cubes, with lower sensitivity, from which
the spectra were extracted. Indeed, the discrepancy between the two datasets
is much reduced (down to a $\lesssim 20$\% flux discrepancy, likely associated with 
flux calibration issues in the 1mm band, \citealt{PdBI_Cookbook}) when estimating the fluxes from the {\it uv} data.
 
The discrepancy with \cite{Walter+09}, who derived a flux of only $3.9\pm0.3$~Jy~\kms, 
can instead be almost entirely explained by the complex and extended nature of the source
(Figure~\ref{fig:total_map}).
These authors indeed observed SDSS~J1148+5251 with the most extended configuration
of the IRAM PdBI, therefore lacking short baselines and completely resolving out
the extended \cii 158$\mu$m emission. As a result, they only detected the 
\cii 158$\mu$m flux associated with the central ``hyper-starburst'', which is
concentrated within a very compact region of size $\sim1.5$~kpc.
Moreover, due to their narrow bandwidth of $\Delta\nu =1$~GHz, the broad wings associated with
the outflow could not be identified by \cite{Walter+09}, and were subtracted from the line 
as continuum emission. This implies that also the flux of the narrow component was underestimated 
because the continuum level was overestimated, thereby causing an additional significant flux loss with respect to
our new observations.

The origin of the discrepancy with the single-dish (IRAM 30m) flux measure by \cite{Maiolino+05} is more
uncertain. Similarly to \cite{Walter+09}, \cite{Maiolino+05} had a bandwidth of
only 1~GHz, and so they also subtracted the broad wings from the baselines
as continuum emission. However, even taking into account the flux loss due to the
wings subtraction, there is still an inconsistency of more than a factor of 2 with respect
to our estimate (Table~\ref{table:comparison}). As a consequence, we suspect that the early \cii 158$\mu$m detection in
SDSS~J1148+5251 by \cite{Maiolino+05} was affected by some calibration issues.
In particular, pointing uncertainties may have played an important role. 
 At that time (2004), the lack of bright sources close to SDSS~J1148+5251 suitable for pointing calibration, had forced the use of pointing
calibration sources very far away from the target, which may have affected significantly the final pointing accuracy.
As a consequence, the pointing uncertainty of the IRAM 30m observations was probably comparable to or
possibly even larger than the nominal upper limit of 3\arcsec quoted by the observatory in 2004. Since the FWHM of the IRAM 30m beam-size
at 256 GHz is of only 9.6\arcsec, a pointing offset of 3\arcsec-4\arcsec
could have caused a significant flux loss.

\subsection{The resolved [CII] outflow}\label{sec:outflow}

\begin{figure*}[tb]
	\centering
    \includegraphics[angle=270,width=.75\columnwidth]{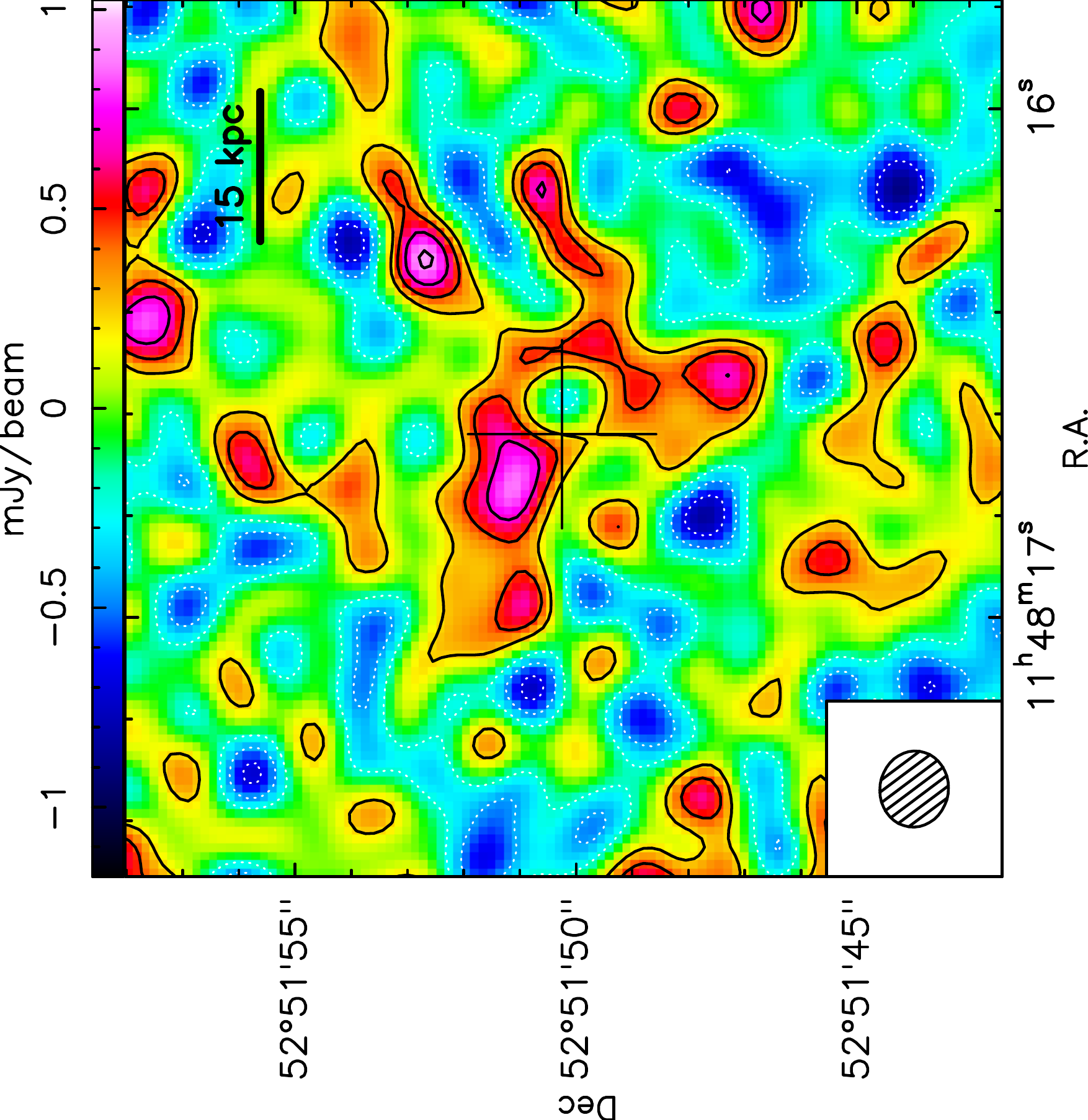}\quad
    \includegraphics[angle=270,width=.75\columnwidth]{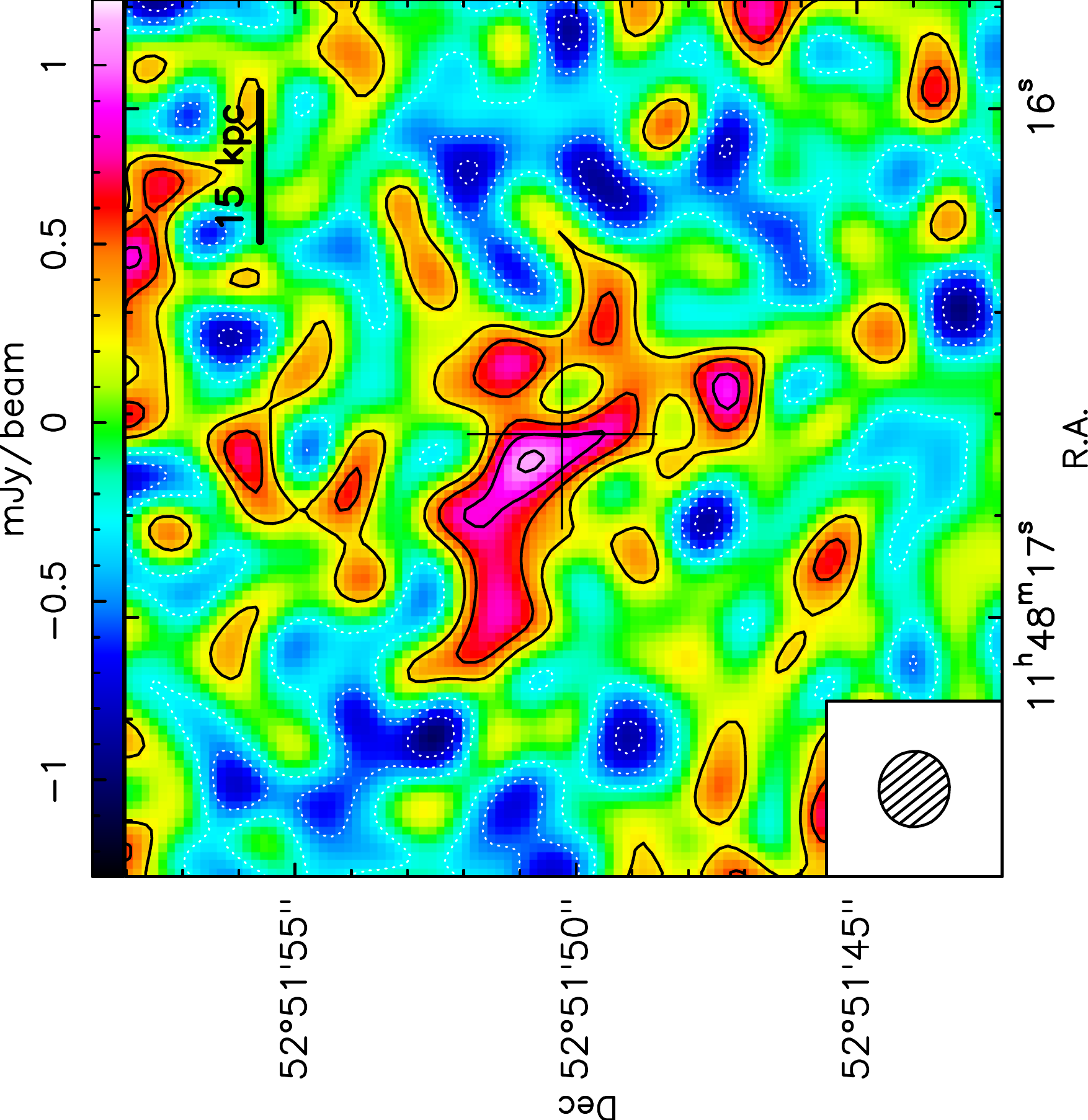}\\
    \includegraphics[angle=270,width=.75\columnwidth]{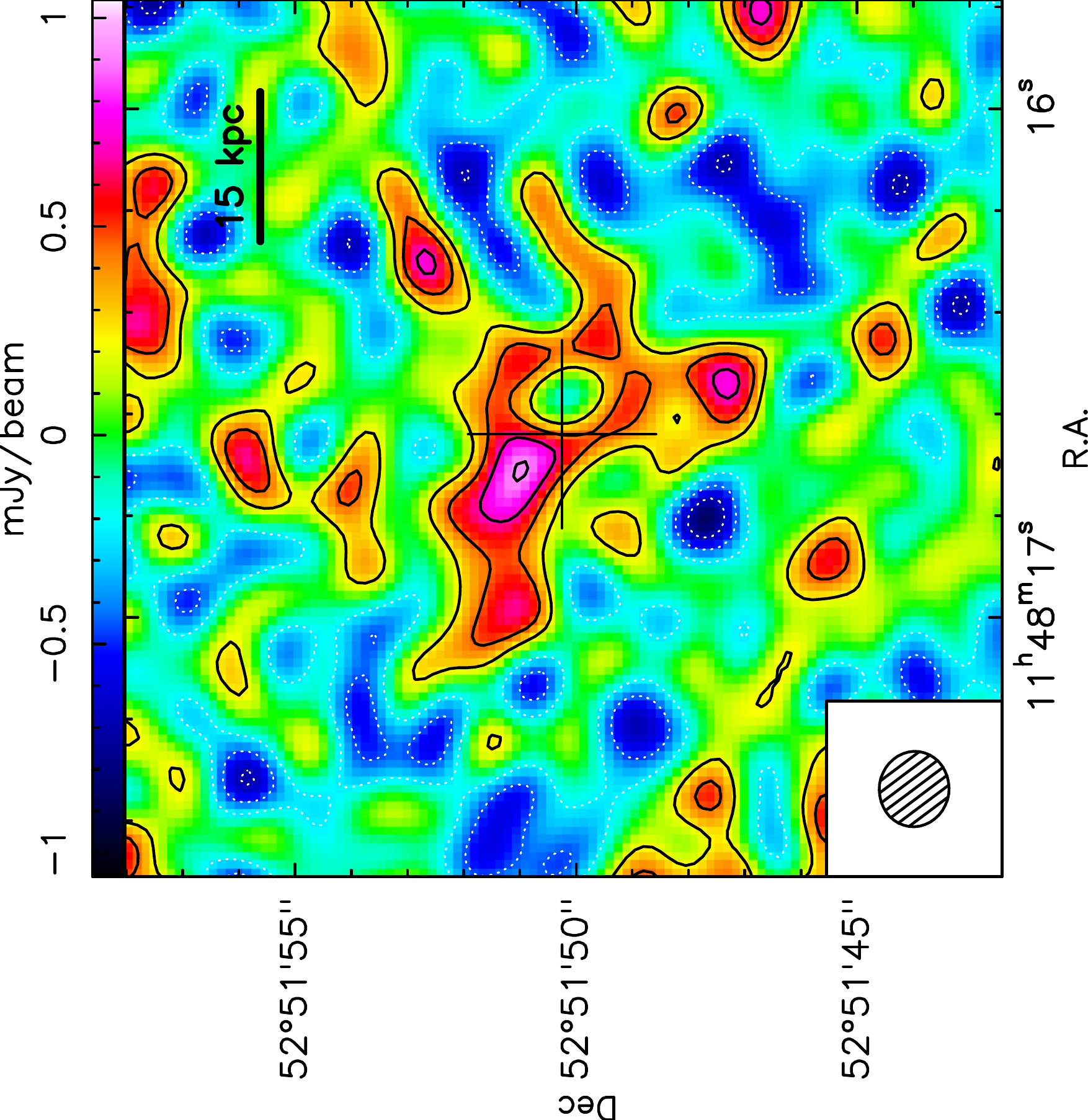}\quad
    \includegraphics[angle=270,width=.78\columnwidth,clip=true,trim=-2cm -1.6cm 0cm 0cm]{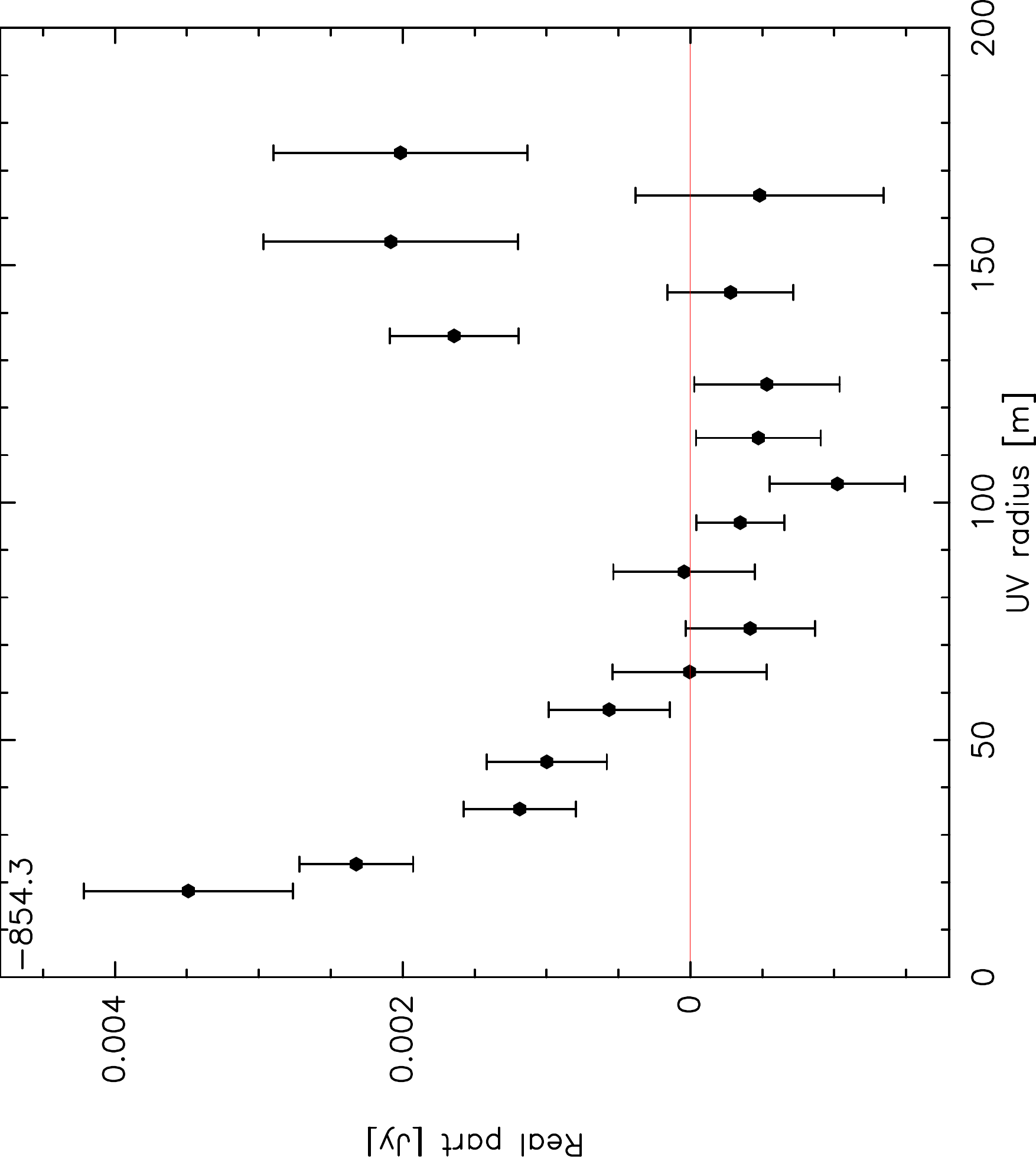}\\
     \caption{{\it Top panels:} IRAM PdBI continuum-subtracted maps of the blue ({\it left}) and red ({\it right}) wings
     of the \cii 158$\mu$m line of SDSS J1148+5251, integrated within $v\in(-1400, -300)$~\kms and
     $v\in(400, 1200)$~\kms, respectively. Negative and positive contours are in steps of 1.5$\sigma$ (1$\sigma$ rms is 
     0.17 Jy beam$^{-1}$ km s$^{-1}$ and 0.14 Jy beam$^{-1}$ km s$^{-1}$ for the blue and red wing, respectively).
     {\it Bottom panels:} Map of the blue and red wings merged together ({\it left}) and corresponding $uv$ plot ({\it right}). 
     In the map of the merged wings, negative and positive contours are in steps of 2$\sigma$ (1$\sigma$ rms is 0.21 Jy beam$^{-1}$ km s$^{-1}$).
     The $uv$ plot shows the real part of visibilities as a function of the $uv$ distance, binned in baseline
     steps of 10m. In all maps, the cross
		indicates the pointing and phase centre, corresponding to the optical position of the quasar.}
   \label{fig:wings_map}
\end{figure*}

The \cii 158$\mu$m spectrum of SDSS~J1148+5251 
exhibits extremely broad (FWHM = $2100\pm500$~\kms)
wings, superimposed on a narrower core with FWHM = $350\pm50$~\kms (Figure~\ref{fig:spectra}).
The broad wings of the \cii 158$\mu$m line, discovered by \cite{Maiolino+12}, are believed to trace the
most distant and powerful quasar-driven outflow ever detected. The outflow is also extremely extended.
This could already be inferred from the low spatial resolution (synthesised beam of 2.2\arcsec$\times$1.8\arcsec) 
observations presented in \cite{Maiolino+12}, which
marginally resolved the emission associated with the broad wings, providing a rough estimate for the
outflow size (FWHM) of 2.9 arcsec. 

Our new, higher resolution IRAM PdBI observations fully resolve the \cii 158$\mu$m outflow of 
SDSS~J1148+5251. In the top panels of Figure~\ref{fig:wings_map} we show the continuum-subtracted maps of the 
blue and red wings of the \cii line, integrated within $v\in(-1400, -300)$ \kms and $v\in(400, 1200)$ \kms,
respectively. We also report in the bottom-left panel of 
Fig.~\ref{fig:wings_map} the map of the wings merged together, similarly
to \cite{Maiolino+12}. 
As already suggested by the analysis of the spectra extracted from different apertures (Figure~\ref{fig:spectra}
and Table~\ref{table:gauss_fit}), the maps unambiguously show 
that the wings are extremely extended, and in particular that they are even more extended than what initially estimated by
\cite{Maiolino+12}. 
The bottom-right panel of Fig.~\ref{fig:wings_map} shows the real part of the visibilities of the broad
wings as a function of the $uv$ baseline radius (also known as ``$uv$ plot''). The steeply declining 
trend clearly confirms that the broad wings are extended and resolved
by our observations at high significance, and that they are not an artefact of the cleaning process.
We note that the detection of the broad and spatially extended \cii component, tracing the outflow, is confirmed
also by splitting the dataset into independent parts.

The fluctuations observed at large $uv$ radii in the $uv$ plot of the broad wings (bottom-right panel of Fig.~\ref{fig:wings_map})
indicate that the morphology of the outflow is complex, non-gaussian,
consistently with the complexity of the structures revealed by the maps. 
For example, both a disk-like and a ring-like brightness distributions would produce similar
positive and negative fluctuations as a function of $uv$ radius, although we found that neither
of these models provides a good fit to our data.

The maps of the blue and red wings of the \cii 158$\mu$m emission 
(upper panels of Fig.~\ref{fig:wings_map})
indicate that the blueshifted and the redshifted sides
of the outflow are generally co-spatial. Indeed, the two maps 
trace the same butterfly-shaped structure, although
some blobs appear brighter in the blue component than in the red one and viceversa. 
We note that the spatial correspondence between blueshifted and redshifted wings rules out
the hypothesis that companion galaxies with different systemic velocities are at the origin of the bulk of the broad 
\cii 158$\mu$m emission, leaving a massive and extended outflow as the only viable explanation.

\begin{figure}[tb]
	\centering
    \includegraphics[width=.8\columnwidth,trim=3.5cm 5cm 3cm 5cm,angle=90]{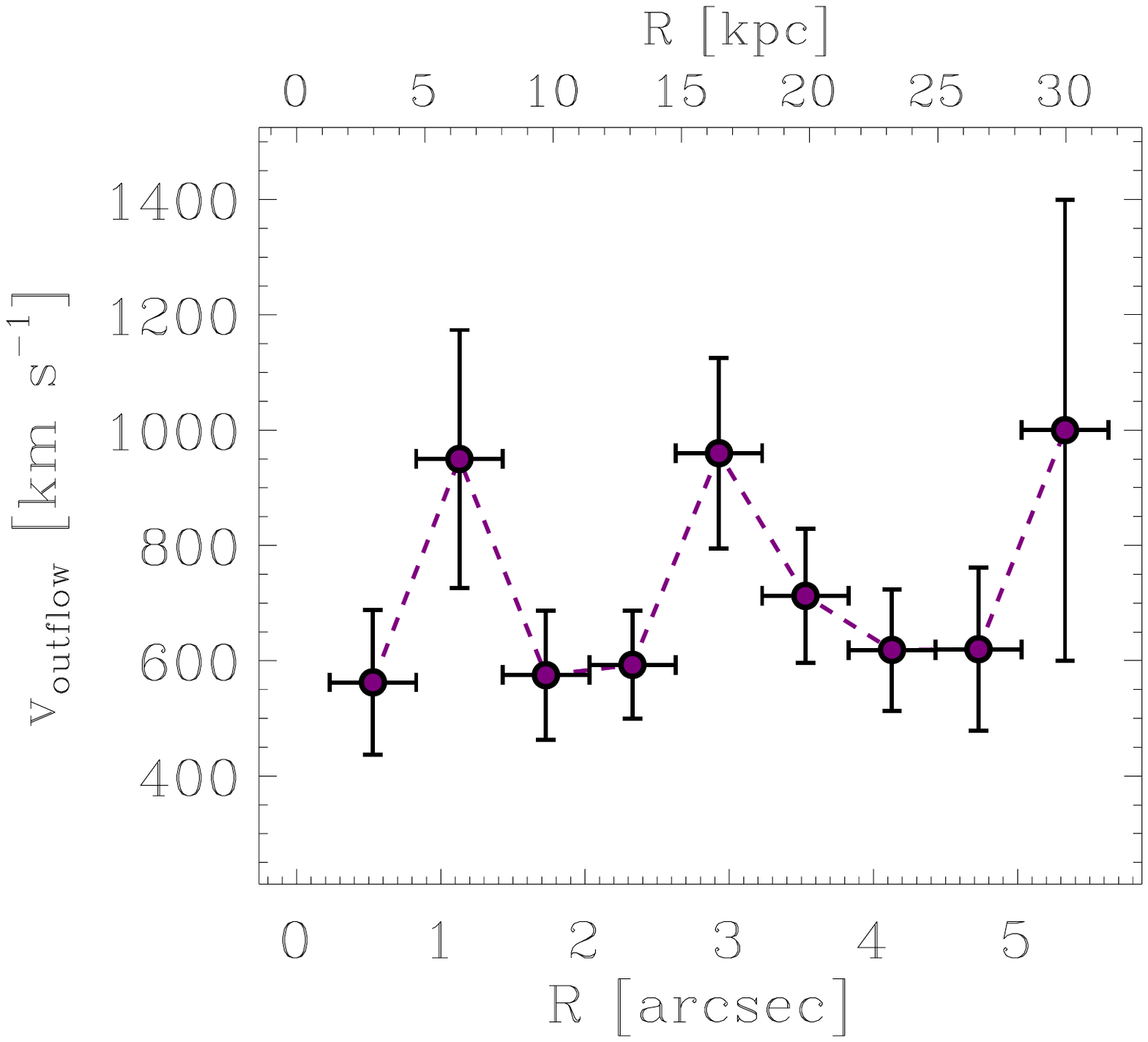}\\
     \caption{Outflow velocity as a function of the radial distance from the QSO position.
     	Both $R$ and $v$ are projected quantities.
    	To produce this plot, the outflowing clumps have been divided into bins of radial distance from the QSO ($\Delta R = 0.6$~arcsec, i.e. $\simeq 3$~kpc).
	The plot shows the average velocity of the clumps within each bin ($\langle v \rangle = \sum_{i=1}^{i=N}v_i$, where $N$ is the number
	of clumps in each bin). The error bars in the $x$ direction indicate the bin size. The error on $\langle v \rangle$ is given by 
	$\sigma_{\langle v \rangle} = (\sum_{i=1}^{i=N} \sigma_{v_i}^2)^{1/2}/N$, where
	$\sigma_{v_i}$ is the uncertainty on the average velocity $v_i$ of a clump in outflow, estimated as explained in Appendix~\ref{sec:outflow_channel_maps}.}
   \label{fig:outflow_vel}
\end{figure}

\begin{figure}[htb]
	\centering
    \includegraphics[clip=true,trim=3cm 1cm 3cm 1cm,angle=90,width=\columnwidth]{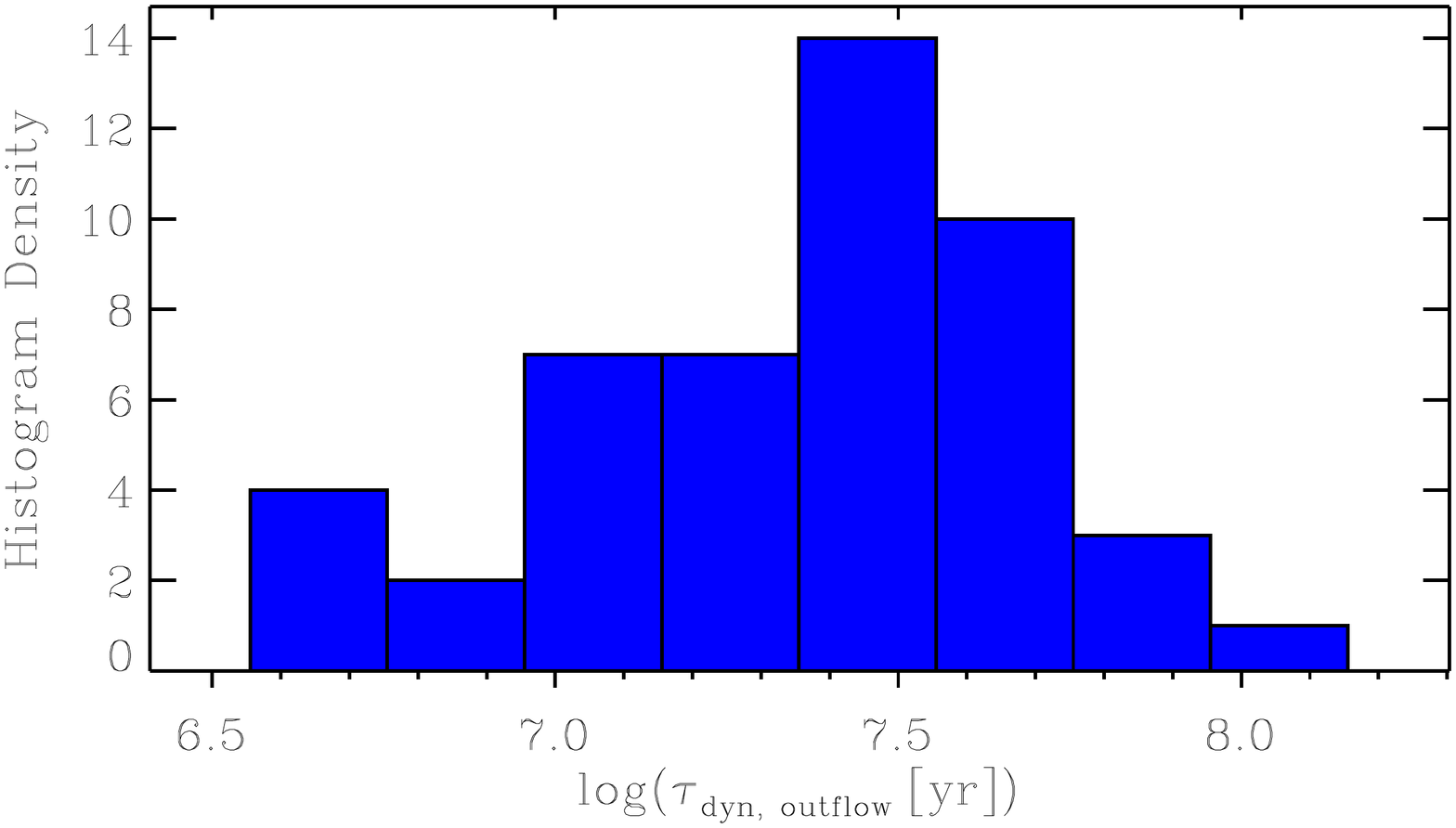}\\
       \caption{Distribution of dynamical time-scales (${\tau_{\rm dyn} = R/v}$, where $R$ is the projected distance from the central QSO
       		and $v$ is the outflow velocity) within the outflow, obtained for the individual outflowing
       		 clumps as described in Section~\ref{sec:outflow} and in Appendix~\ref{sec:outflow_channel_maps}.}
   \label{fig:t_dyn}
\end{figure}

\subsubsection{Outflow velocity and dynamical time-scale}\label{sec:outflow_dyn}

In \cite{Maiolino+12} the outflow was only marginally resolved, and the outflow mass-loss rate 
of ${\dot{M}_{\rm out} \gtrsim 3500~M_{\odot}~{\rm yr}^{-1}}$ was
estimated by assuming a spherical and/or multi-conical geometry. In the model adopted by \cite{Maiolino+12}, the outflow
rate was given by ${\dot{M}_{\rm out} = 3vM_{\rm out}/R}$, where $v$ is the outflow velocity, 
$M_{\rm out}$ is the mass of atomic gas in outflow (estimated from the \cii 158$\mu$m flux in the broad wings), and 
$R$ is the radius of the broad wings emission (obtained by fitting the $uv$ plot of the broad wings with a circular Gaussian model). 
However, our new, higher resolution IRAM PdBI observations allow us to 
fully resolve the outflow, hence enabling a much more detailed characterisation 
of the outflow properties, which was not possible with the previous dataset. 
The complex morphology revealed by the new maps of the outflow (Figure~\ref{fig:wings_map}) is
clearly {\it not consistent} with the simple spherical or multi-conical geometry assumed by \cite{Maiolino+12}.
Any attempt of modelling the \cii outflow in SDSS~J1148+5251 based on the current data 
would require a different, more complex geometry, and the uncertainties on such model would be very large. For this reason, in this paper, we 
decide to abstain from a simplistic modelling of the outflow geometry and instead we adopt
a different, ``resolved'' approach to estimate the outflow properties.

Our approach is described in detail in Appendix~\ref{sec:outflow_channel_maps}, and is based on 
measuring the dynamical time-scales of individual ``blobs'' of \cii emission within the outflow \footnote{We note that
in this work we use also the term ``clumps'' to refer to the blobs of \cii emission detected in outflow. The reader should be aware
that these clumps or blobs generally do not correspond to individual clouds, as the minimum physical size probed 
by our observations (given by the synthesized beam) is $\sim$6-7 kpc.}.
More specifically, we produce channel maps centred at different velocities, spanning the entire velocity range
corresponding to the red and blue wings of the \cii 158$\mu$m line, tracing the outflow in the host galaxy of SDSS~J1148+5251.
The outflow channel maps are shown in Figure~\ref{fig:outflow_channel_maps}. In these maps, we identify 48 
``clumps'' belonging to the outflow, as explained in Appendix~\ref{sec:outflow_channel_maps}. For 
each clump in outflow we estimate its average (projected) velocity ($v$), its (projected) distance from the QSO ($R$), and its \cii flux, which
is used to estimate the mass of (atomic) gas in outflow as explained in the next section (Section~\ref{sec:outflow_integrated}).
The dynamical time-scale of a clump in outflow
is therefore simply given by ${\tau_{\rm dyn} = R/v}$. We note that, by applying this definition, we are implicitly assuming that 
the velocity $v$ of the clumps
is constant throughout the outflow.

The relationship between the velocity and the distance from the QSO of the clumps in outflow
is shown in Figure~\ref{fig:outflow_vel}. 
The average velocity of the outflowing blobs does not depend on $R$, 
suggesting that the outflow is not significantly accelerating and/or decelerating. This also supports the definition
of dynamical time-scales given above.
The distribution of ${\tau_{\rm dyn}}$ within the outflow is reported in Fig.~\ref{fig:t_dyn}; the histogram shows
a slightly skewed left distribution, with median of
 $\tau_{\rm dyn}^{\rm median} = 10^{7.4}$~yr, but spanning a broad range from $\tau_{\rm dyn}^{\rm min} = 10^{6.6}$~yr to 
 $\tau_{\rm dyn}^{\rm max} = 10^{8.0}$~yr. This indicates that the outflow has been in place for $\sim$100~Myr, which is
a non-negligible fraction of the age of the Universe at this redshift (i.e. 850~Myr with the adopted cosmology).

One should note that our measurement of $\tau_{\rm dyn}$ neglects projection effects, more specifically, the fact
that the {\it observed} distance $R$ of the clumps from the centre is the projected distance, and that the {\it observed}
velocity is actually only the component along our line of sight. As a consequence, the dynamical time-scales should be
corrected by a factor $|\tan{\phi}|^{-1}$, i.e. $\tau_{\rm dyn}({\rm real}) = \tau_{\rm dyn}({\rm obs})/|\tan{\phi}|$, 
where $\phi$ is the angle between our line of sight and the outflow direction of each individual
cloud. Obviously $\phi$ is not known. 
However, in the absence of any additional information on the outflow geometry, 
the most simplistic assumption we can make 
is that of a uniform distribution of the outflowing clouds' velocities as a function of $\phi$. In this case, the correction
factor averaged over the entire solid angle, $\left\langle \frac{1}{|\tan{\phi}|}\right\rangle = \frac{1}{4\pi}\int _{4\pi}|\tan{\phi}|^{-1}~d\Omega$,
is equal to unity. Therefore, although the dynamical time-scales may need to be
corrected by a significant factor for the individual clumps, the global average trends may not be affected significantly. 

\begin{figure*}[tb]
	\centering
    \includegraphics[width=.58\columnwidth,trim=3.5cm 5cm 3cm 5.5cm,angle=90]{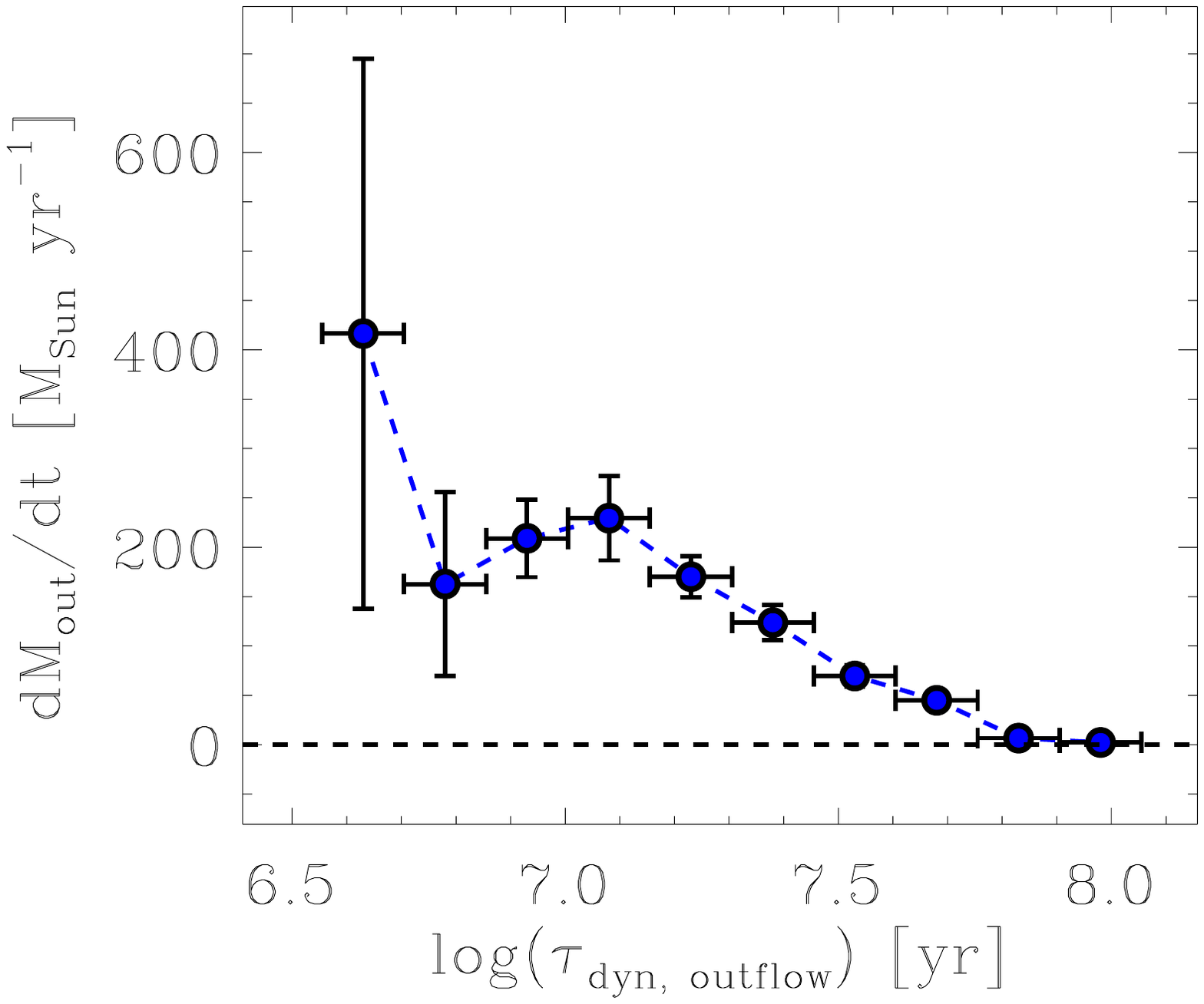}\quad
    \includegraphics[width=.58\columnwidth,trim=3.5cm 5cm 3cm 6cm,angle=90]{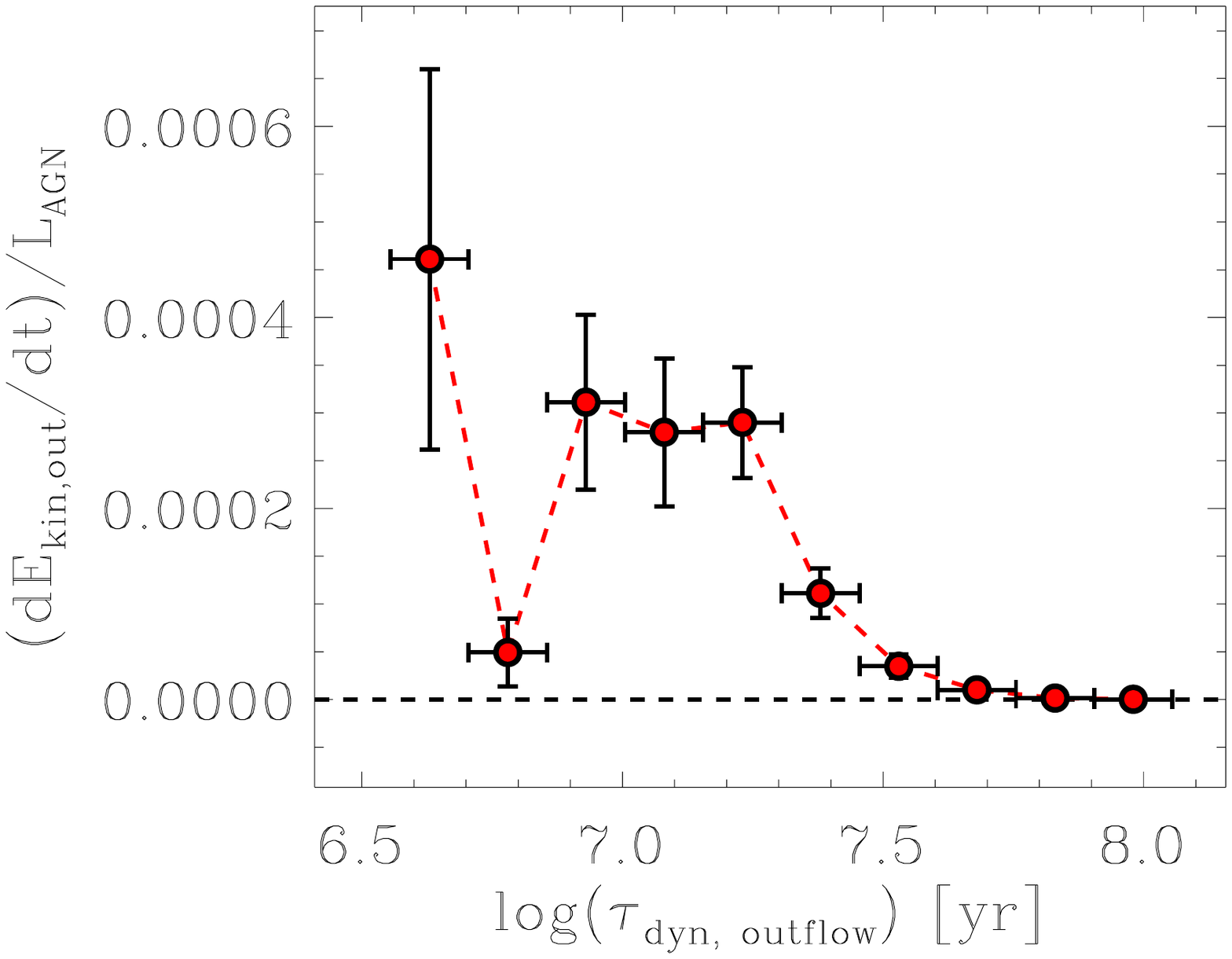}\quad
    \includegraphics[width=.58\columnwidth,trim=3.5cm 5cm 3cm 6.5cm,angle=90]{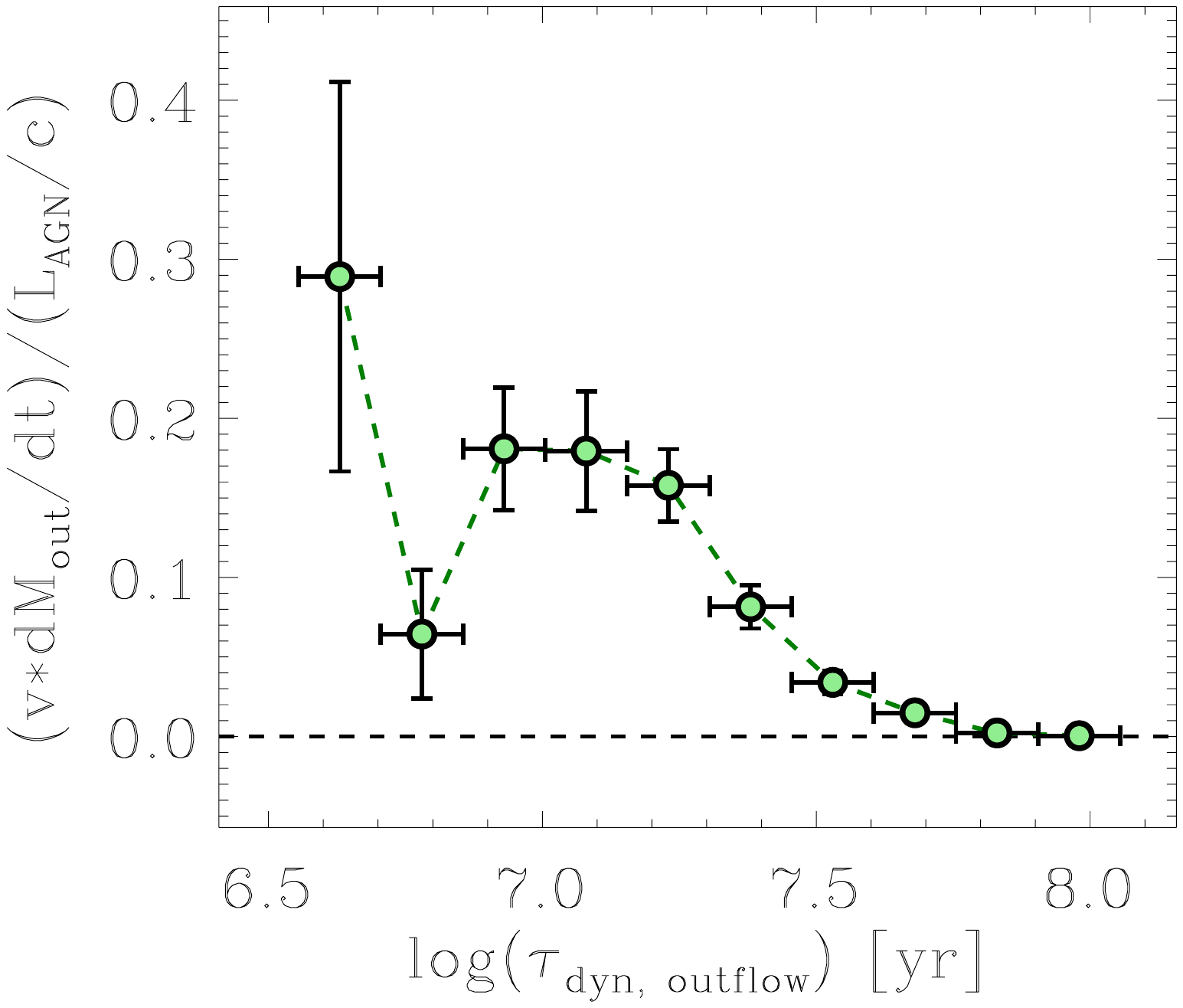}\\
     \caption{Outflow mass-loss rate $\dot{M}_{\rm out}$ ({\it left}), kinetic power ratio $(1/2\dot{M}_{\rm out}v^2)/L_{\rm AGN}$ ({\it centre}) 
     and momentum rate ratio $(\dot{M}_{\rm out}v)/(L_{\rm AGN}/c)$ ({\it right}) as a function of the dynamical time-scale of the clumps in outflow. 
     These plots have been obtained by dividing the outflowing clumps (see Section~\ref{sec:outflow}) in bins
     of $\Delta {\rm log}(\tau_{\rm dyn})$ = 0.15. The plots show, in correspondence of the average $\tau_{\rm dyn}$ of each bin, the total $\dot{M}_{\rm out}$,
     $(1/2\dot{M}_{\rm out}v^2)/L_{\rm AGN}$ and $(\dot{M}_{\rm out}v)/(L_{\rm AGN}/c)$ obtained by adding up the contribution of all the outflowing
     clumps within the bin. The error bars in the $x$ direction indicate the bin size, i.e. $\Delta x = \pm 0.075$. 
     The error bars in the $y$ direction are calculated using the statistical error propagation (i.e. by propagating the errors on $v$, $R$ and on the \cii fluxes 
     measured for each clump in outflow, which are estimated in Appendix~\ref{sec:outflow_channel_maps}). The errors do not take into account the uncertainty on the
     conversion from the \cii luminosity to atomic gas mass and the uncertainty on the AGN luminosity estimate \citep{Schneider+14}.}
   \label{fig:outflow_dyntime}
\end{figure*}

 \begin{figure*}[tb]
	\centering
    \includegraphics[width=.58\columnwidth,trim=3.5cm 5cm 3cm 5.5cm,angle=90]{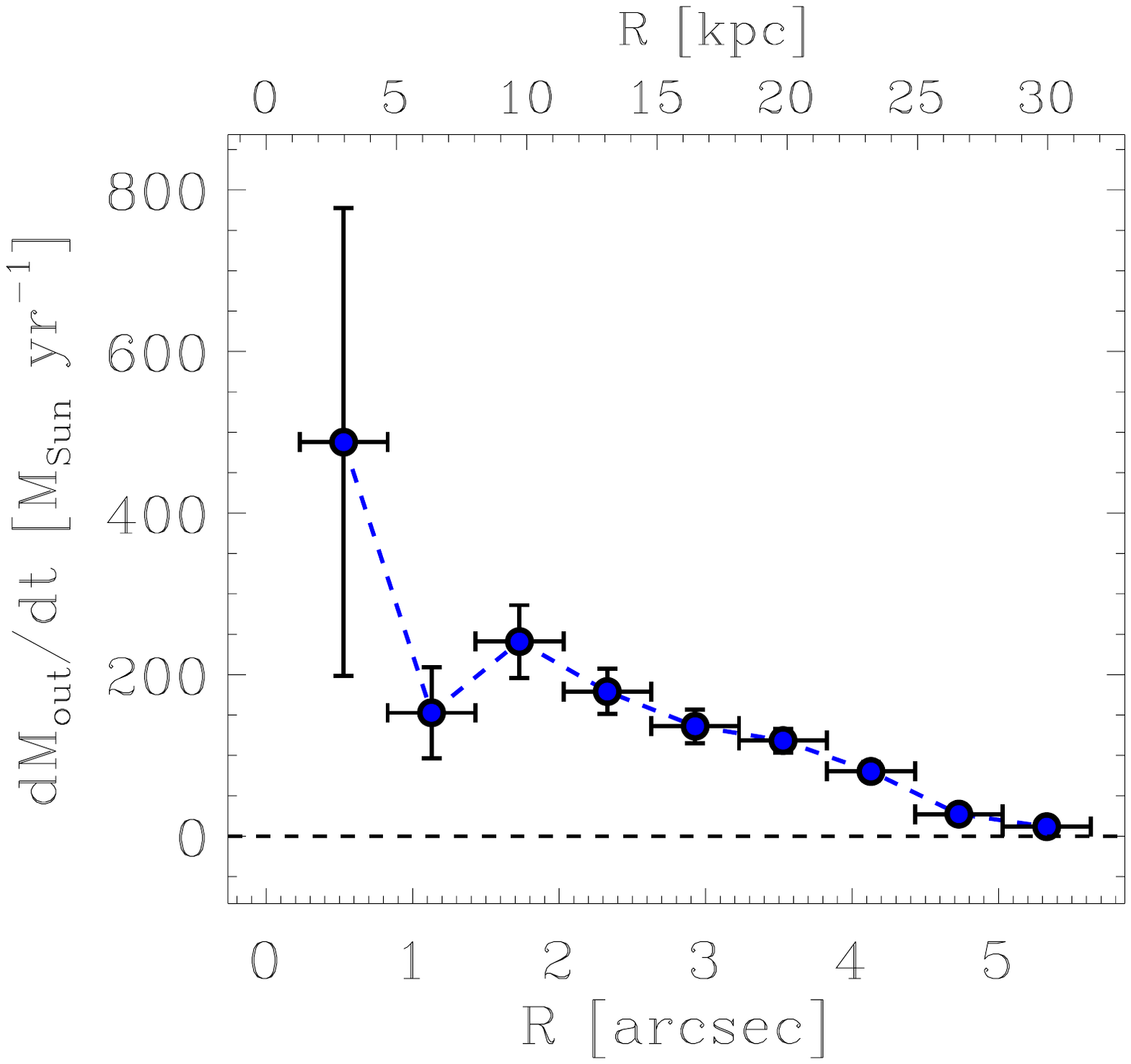}\quad
    \includegraphics[width=.58\columnwidth,trim=3.5cm 5cm 3cm 6cm,angle=90]{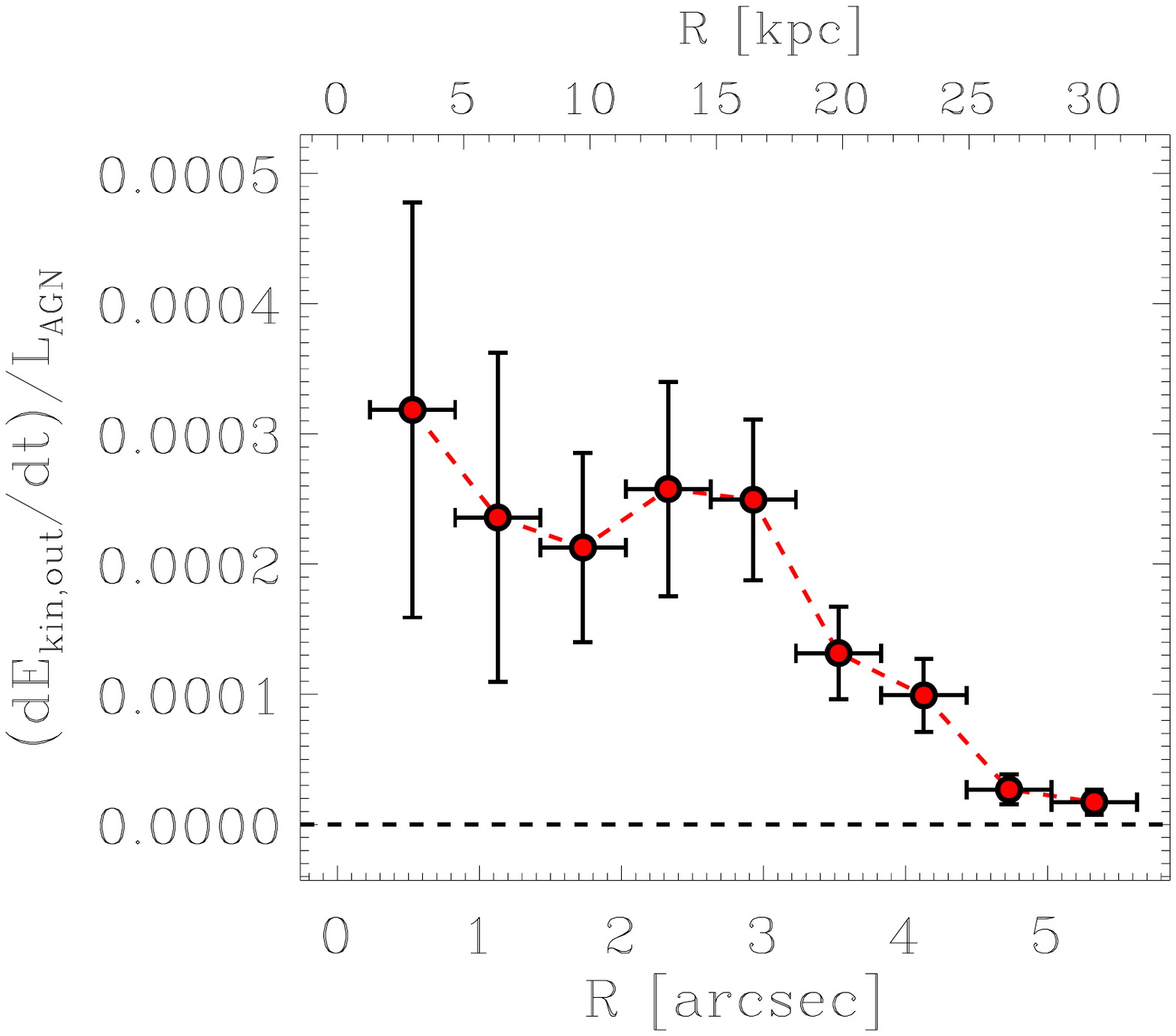}\quad
    \includegraphics[width=.58\columnwidth,trim=3.5cm 5cm 3cm 6.5cm,angle=90]{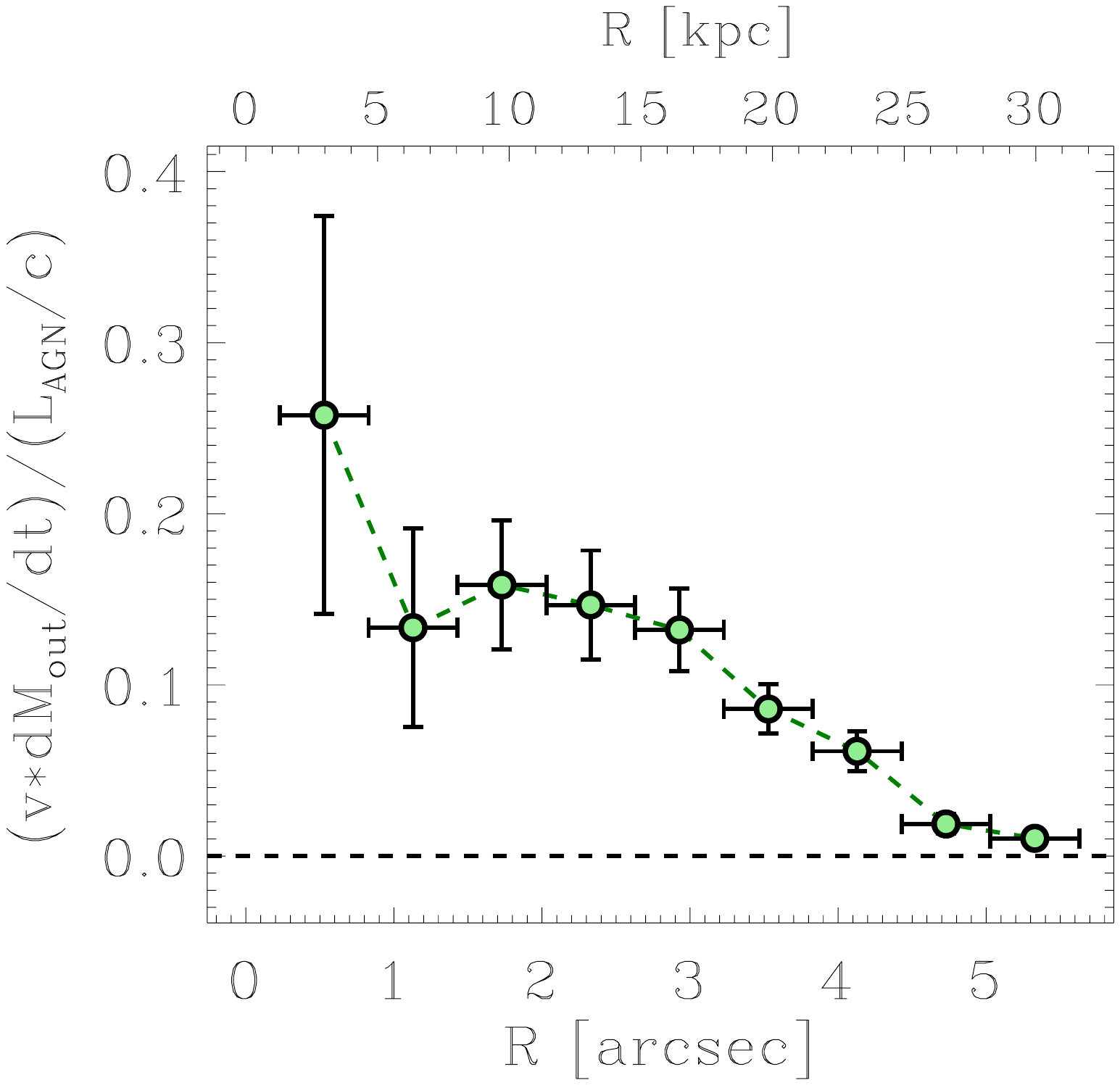}\\
     \caption{Outflow mass-loss rate $\dot{M}_{\rm out}$ ({\it left}), kinetic power ratio $(1/2\dot{M}_{\rm out}v^2)/L_{\rm AGN}$ ({\it centre}) 
     and momentum rate ratio $(\dot{M}_{\rm out}v)/(L_{\rm AGN}/c)$ ({\it right}) as a function of the projected
     distance from the QSO position. These plots have been obtained by dividing the outflowing clumps (see Section~\ref{sec:outflow}) in bins
     of $\Delta R = 0.6$~arcsec, corresponding to $\simeq3$~kpc.
     The plots show, in correspondence of the average $R$ of each bin, the total $\dot{M}_{\rm out}$,
     $(1/2\dot{M}_{\rm out}v^2)/L_{\rm AGN}$ and $(\dot{M}_{\rm out}v)/(L_{\rm AGN}/c)$ obtained by adding up the contribution of all the outflowing
     clumps within the bin. For an explanation of the error bars, see caption of Figure~\ref{fig:outflow_dyntime}.}
   \label{fig:outflow_radial}
\end{figure*}

\subsubsection{Integrated outflow mass-loss rate, kinetic power and momentum rate}\label{sec:outflow_integrated}

In order to calculate the outflow mass-loss rate, we need to estimate the mass of gas carried by the outflow.
To do this, we first measure the \cii flux associated with each outflowing clump, as explained in Appendix~\ref{sec:outflow_channel_maps}.
The relationship between \cii luminosity and gas mass depends on the gas density, temperature and
abundance of C$^+$ ions ($X_{\rm C^+}$). 
We convert the \cii luminosity of the outflowing clumps into gas mass by using the formula 
provided by \cite{Hailey-Dunsheath+10}, obtained under the hypothesis 
of optically thin \cii emission. To this regard, we note that the assumption of optically thin \cii has been challenged
by observations of high-z dusty star forming galaxies (\citealt{Neri+14}; Gullberg et al. submitted).
However, optically thick conditions for \cii have been proposed to occur in extremely compact and dense starburst systems. 
It is unlikely that the diffuse and very extended \cii outflowing component detected by us can be optically thick, 
even because the large velocity gradients in the outflow make self absorption even more unlikely.
As in \cite{Maiolino+12}, we assume that
$X_{\rm C^+} = 1.4 \times 10^{-4}$, ${\rm T}=200$~K, and $n\gg n_{\rm crit}$ ($n_{\rm crit} \sim 3 \times 10^3$ cm$^{-3}$), typical of 
photo-dissociated regions (PDRs), and that 70\% of the \cii flux arises from such medium
(while the remaining 30\% arises from other ISM phases).

The integrated outflow rate that we obtain by adding up the mass-loss rate 
contribution (i.e. ${\dot{M}_{\rm out} = M_{\rm out}/\tau_{\rm dyn}}$) 
from all of the outflowing clumps is $1400 \pm 300~M_{\odot}~{\rm yr}^{-1}$. 
The error on ${\dot{M}_{\rm out}}$ (see Appendix~\ref{sec:outflow_channel_maps} for the details), does not
take into account the uncertainty on the conversion from \cii luminosity to atomic gas mass.
We stress that our measurement should be considered as a 
{\it lower limit} on the total outflow mass-loss rate, in first place because of the conservative assumptions we made for estimating 
the atomic gas mass from the \cii luminosity, and in second place because it only accounts for the atomic gas phase of 
the outflow, while a significant fraction of the
mass in outflow may be in the molecular phase, similarly to local ULIRGs (e.g. \cite{Cicone+14} and refs therein).

The integrated outflow mass-loss rate measured using this ``resolved'' approach is over a factor of 2 lower than
the previous estimate by \cite{Maiolino+12}, which was based on lower resolution and lower 
sensitivity data and relied on a simplistic model for the outflow geometry. Conversely, our new method for estimating the outflow
mass-loss rate, described
in Section~\ref{sec:outflow_dyn} and in Appendix~\ref{sec:outflow_channel_maps}, 
does not assume any model for the outflow geometry. Furthermore, the outflow size estimated by \cite{Maiolino+12} by fitting the $uv$ 
data with a simple Gaussian model is of only $R \simeq 1.5$~arcsec ($\simeq 8$~kpc), i.e. half the
average $\langle R \rangle \simeq 3$~arcsec that we measure by applying our ``resolved'' approach to the new dataset. 
It is therefore not surprising that our revised estimate of the outflow mass-loss rate in SDSS~J1148+5251 is lower than the value that was
obtained by \cite{Maiolino+12}, although our estimate should be considered a lower limit as explained above.

We estimate the integrated kinetic power and momentum rate of the outflow 
by summing up the contribution from all outflowing clumps. Specifically,
the sum of ${1/2\dot{M}_{\rm out}v^2}$ and ${\dot{M}_{\rm out}v}$, evaluated separately
for each blob in outflow, gives, respectively, an integrated outflow kinetic power of $(3.3 \pm 0.5)\times 10^{44}$~erg s$^{-1}$ and
an integrated outflow momentum rate of $(7.2 \pm 1.0)\times 10^{36}$~g~cm~s$^{-2}$. 
It is common practice in studies of AGN-driven outflows to compare these quantities with the AGN bolometric luminosity ($L_{\rm AGN}$) and 
with the total radiative momentum output from the AGN ($L_{\rm AGN}/c$). In this case, by using 
the intrinsic AGN luminosity of $L_{\rm AGN} = 5.63 \times 10^{13}~L_{\odot}$ 
inferred by \cite{Schneider+14} for SDSS~J1148+5251, the ratio between our estimate of 
the outflow kinetic power and $L_{\rm AGN}$ is $(1.6 \pm 0.2)\times 10^{-3}$, and the ratio between the 
momentum rate and $L_{\rm AGN}/c$ is  $1.00 \pm 0.14$. These values and their implications will be
discussed in Section~\ref{sec:discussion_outflow}.
We also discuss in Section~\ref{sec:discussion_outflow} the possible contribution of star formation
to driving the observed outflow.

\subsubsection{Outflow properties as a function of $\tau_{\rm dyn}$ and $R$}\label{sec:outflow_profiles}

In Section~\ref{sec:outflow_integrated} we have derived the integrated outflow mass-loss rate, kinetic power and
momentum rate. However, one may argue that given the quite broad distribution in dynamical time-scales within the outflow
evidenced in Fig.~\ref{fig:t_dyn}, the clumps were not ejected all at the same time as a consequence of 
a single, explosive event, but, instead, they may have been expelled from the galaxy at different epochs.
It is therefore interesting to investigate the contribution to the mass-loss rate, kinetic power and momentum rate
of the outflow from clumps with different ``ages'', i.e.
with different $\tau_{\rm dyn}$.
To this end, we present in Figure~\ref{fig:outflow_dyntime} the mass-loss rate, 
the kinetic power ratio $(1/2\dot{M}_{\rm out}v^2)/L_{\rm AGN}$ and 
momentum rate ratio $(\dot{M}_{\rm out}v)/(L_{\rm AGN}/c)$ of the outflow
as a function of the dynamical time-scale of the outflowing clumps.
Figure~\ref{fig:outflow_dyntime} suggests that
the ejection of gas has not occurred at a constant rate in SDSS~J1148+5251. In particular,
the most extreme outflow event seems to have happened more recently, within the past $\sim$ 5~Myr, as indicated
by the first prominent peak of outflow mass-loss rate, kinetic power and momentum flux in 
Fig.~\ref{fig:outflow_dyntime}. Interestingly, Fig.~\ref{fig:outflow_dyntime}
shows the presence of a second, broader peak of these outflow properties between $\tau_{\rm dyn}\in (8-20)$~Myr.
Taken at face value, these findings would imply that, although the outflow may have been in place for 100~Myr, 
the bulk of the mass, energy and momentum have been released throughout two major outflow events happened more recently, within 
the past $\sim 20$~Myr (Figure~\ref{fig:outflow_dyntime}).

A similar non-uniform distribution of the outflow mass-loss rate, kinetic power and momentum rate is evidenced also as a function of
the (projected) distance from the QSO, as shown by Fig.~\ref{fig:outflow_radial}, in which we report the radial profiles
of the outflow properties. Figure~\ref{fig:outflow_radial} indicates
that, overall, ${\dot{M}_{\rm out}}$, $(1/2\dot{M}_{\rm out}v^2)/L_{\rm AGN}$ and $(\dot{M}_{\rm out}v)/(L_{\rm AGN}/c)$ decrease with the distance
from the QSO. The main peak can be identified at projected radial distance of $\sim$0.5~arcsec from 
the position of the central QSO, corresponding to $\sim$3~kpc. There is also marginal evidence for a second, broad peak
between $r\in (1.5-3)$~arcsec (i.e. $r\in(8-17)$~kpc).
Similarly to what suggested for Fig.~\ref{fig:outflow_dyntime}, the two peaks in Fig.~\ref{fig:outflow_radial} 
may be associated to major outflow events
during the recent life of the QSO-galaxy. However, the significance of the second peak at $r\in(8-17)$~kpc is low,
so this interpretation should be taken with caution.
Figure~\ref{fig:outflow_radial} can be readily compared with the predictions of recent theoretical studies of AGN feedback
\citep{Costa+14b,Thompson+14}. This discussion is presented in Section~\ref{sec:discussion_outflow}.

We stress that observations are naturally limited for large $R$ (and hence for high $\tau_{\rm dyn}$): 
indeed, outflow clumps at large distances from the galaxy centre are more difficult to probe due to the sensitivity of our maps.
The lowest outflow velocities (which can also give high values of $\tau_{\rm dyn}$) are also difficult to probe due to 
contamination by quiescent gas, i.e. the narrow component of the \cii line. 
In addition, ``old'' outflowing clumps, associated with longer dynamical time-scales and typically at larger distances, are more likely
to have dissolved or evaporated. Therefore, the decreasing outflow rate with $\tau_{\rm dyn}$ and radius 
may be in part the result of sensitivity issues in the observations and/or of 
the effect of outflowing clouds dissolving as they travel to larger
distances, or being decelerated by the interaction with the intergalactic and circumgalactic media.
For similar reasons, it is difficult to assess with the current data whether the outflow in SDSS~J1148+5251
manages to escape the viral radius of the host galaxy and to enrich the IGM.

\begin{figure*}[tb]
    \includegraphics[angle=270,width=.95\columnwidth]{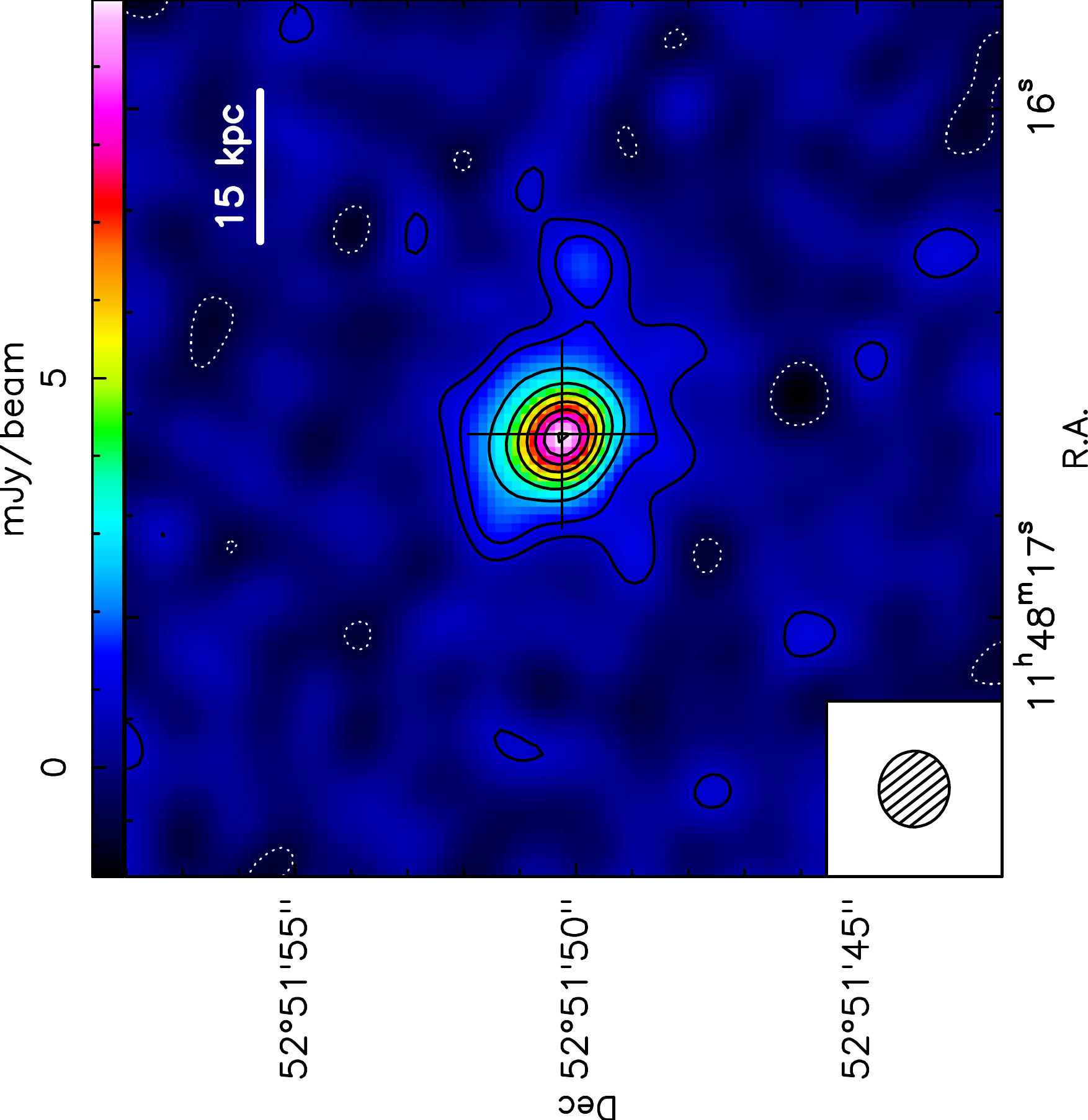}\quad
    \includegraphics[angle=270,width=.95\columnwidth,clip=true,trim=-1.5cm 0cm -0.5cm 0cm]{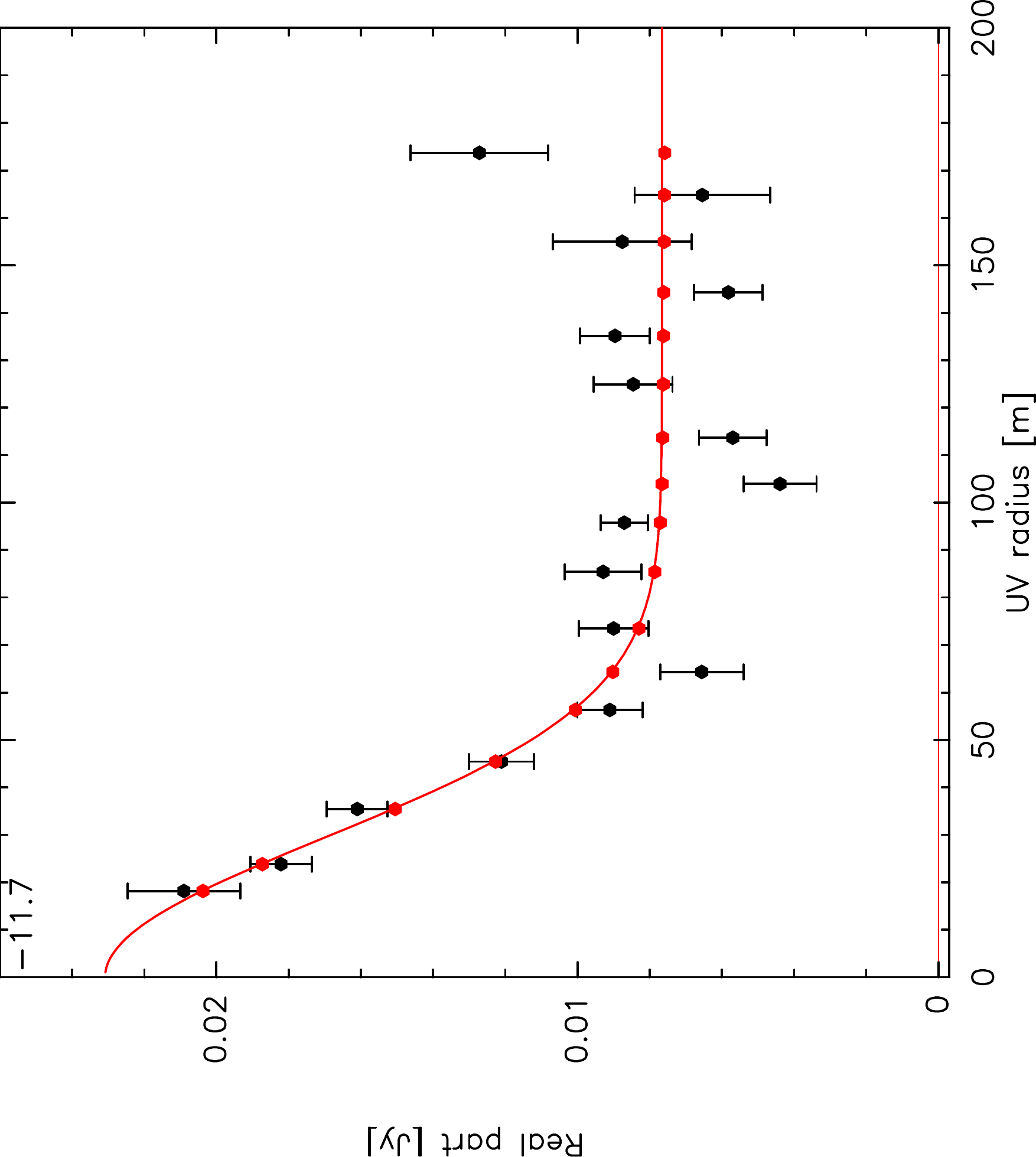}\\
     \caption{IRAM PdBI continuum-subtracted map ({\it left}) and $uv$ plane analysis ({\it right}) 
     of the core of the \cii 158$\mu$m emission line of SDSS~J1148+5251,
     integrated within $v\in(-200, 200)$~\kms. 
     {\it Left:}
     Negative and positive contours correspond to -3$\sigma$, 3$\sigma$, 5$\sigma$ 
     to 40$\sigma$ in steps of 5$\sigma$ (1$\sigma$ rms noise is 0.10 Jy beam$^{-1}$ km s$^{-1}$).
     The cross indicates the pointing and phase centre, corresponding to the optical position of the quasar.
     {\it Right:} Real part of visibilities as a function of the $uv$ distance ($uv$ plot), binned in baseline
     steps of 10m. The red curve indicates the best fit to the $uv$ plot, which corresponds to the sum of two components: 
     a point (unresolved) source, responsible for the flattening of the
     visibilities at large $uv$ radii ($uv$ radii $>50$~m), and an extended (resolved) source fitted with a circular Gaussian model, which is 
     responsible for the sharp decline of the visibilities at small $uv$ radii.}
   \label{fig:core}
\end{figure*}

\begin{figure}[tb]
    \includegraphics[angle=270,width=.9\columnwidth]{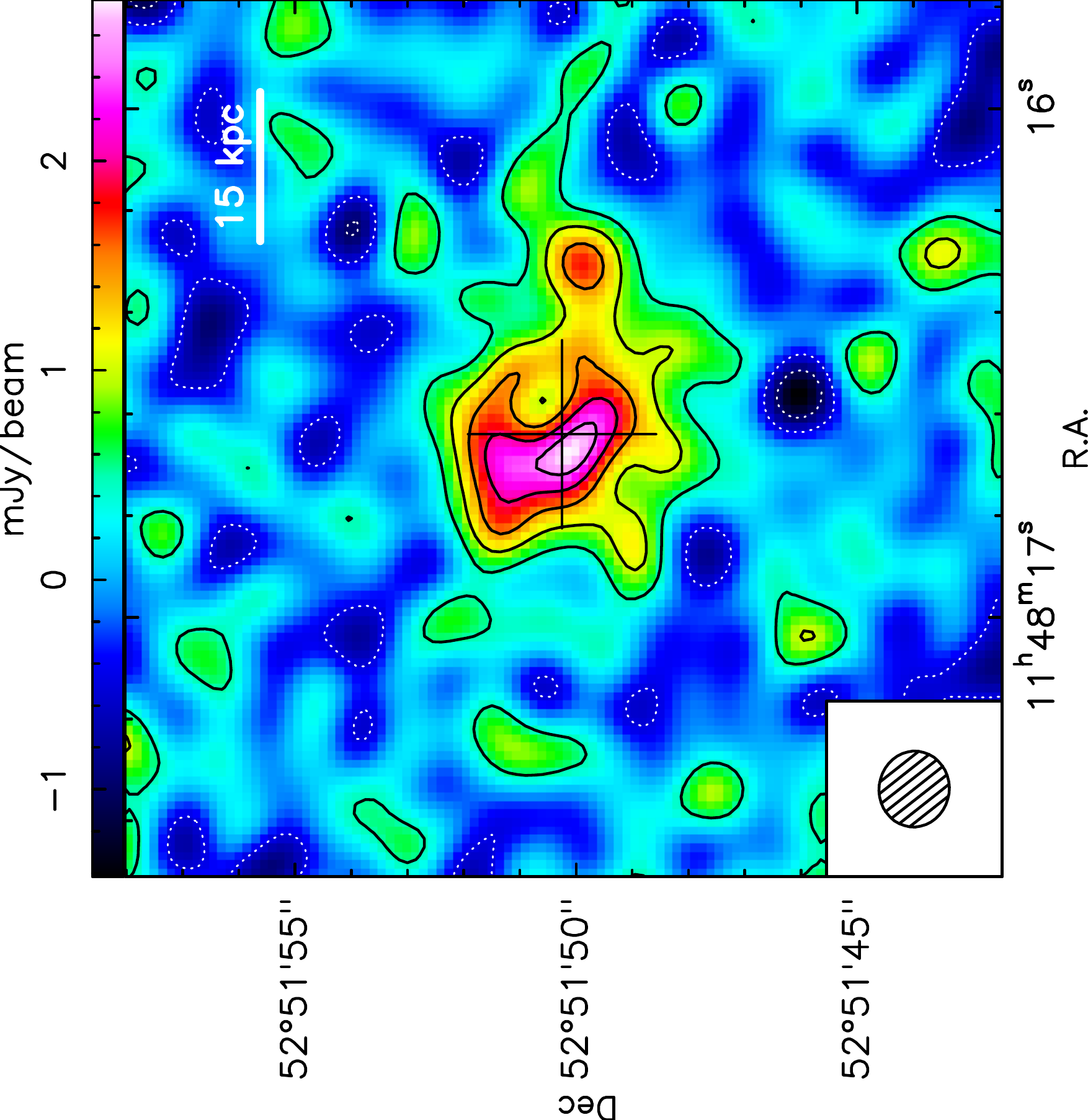}\quad
    \includegraphics[angle=270,width=.9\columnwidth]{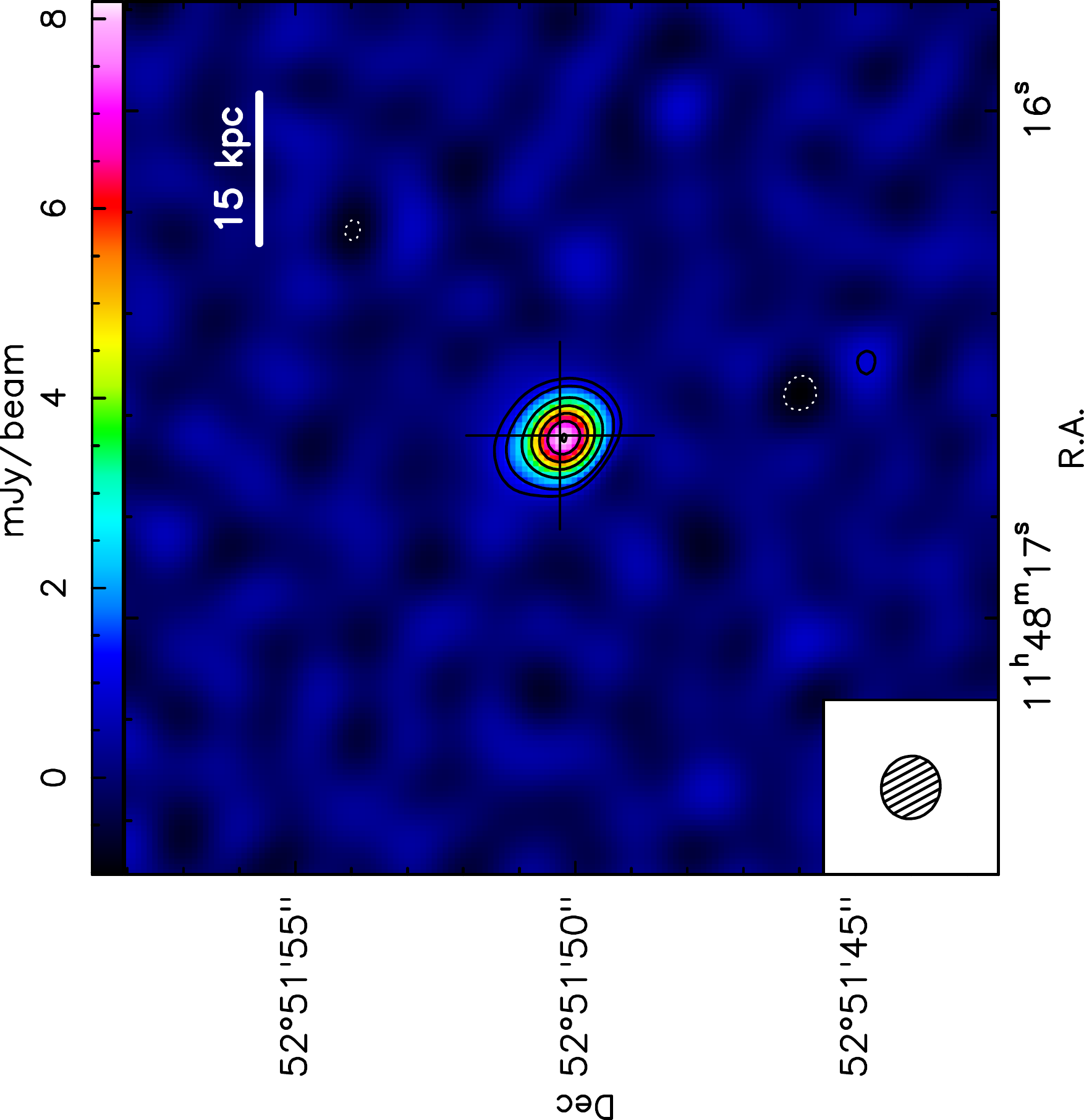}\\
     \caption{Maps of the extended ({\it top}) and compact ({\it bottom}) components that contribute
     		to the core \cii 158$\mu$m emission of SDSS~J1148+5251 
		within $v\in(-200, 200)$~\kms.		
		 The map of the extended component has been obtained by subtracting from 
		 the $uv$ data the point source model fit. 
		 The map of the compact component has been produced by applying a 
		 $uv$ tapering that cuts the visibilities with $uv$ radius $<80$~m; this operation increases 
		 the spatial resolution of the data (the resulting beam size is 1.1\arcsec$\times$1.1\arcsec).
		 Negative and positive contours are in steps of 2$\sigma$ for the extended component
		 in the top panel (1$\sigma$ rms noise of 0.10 Jy beam$^{-1}$ km s$^{-1}$). For the compact
		 component in the bottom panel, contours correspond to -3$\sigma$, 3$\sigma$, 5$\sigma$ 
		 to 30$\sigma$ in steps of 5$\sigma$
		 (1$\sigma$ rms noise of 0.11 Jy beam$^{-1}$ km s$^{-1}$).
		 In both panels, the cross
		indicates the pointing and phase centre, corresponding to the optical position of the quasar.}
   \label{fig:core_components}
\end{figure}

\begin{figure*}[tb]
   \includegraphics[angle=270,width=.95\columnwidth]{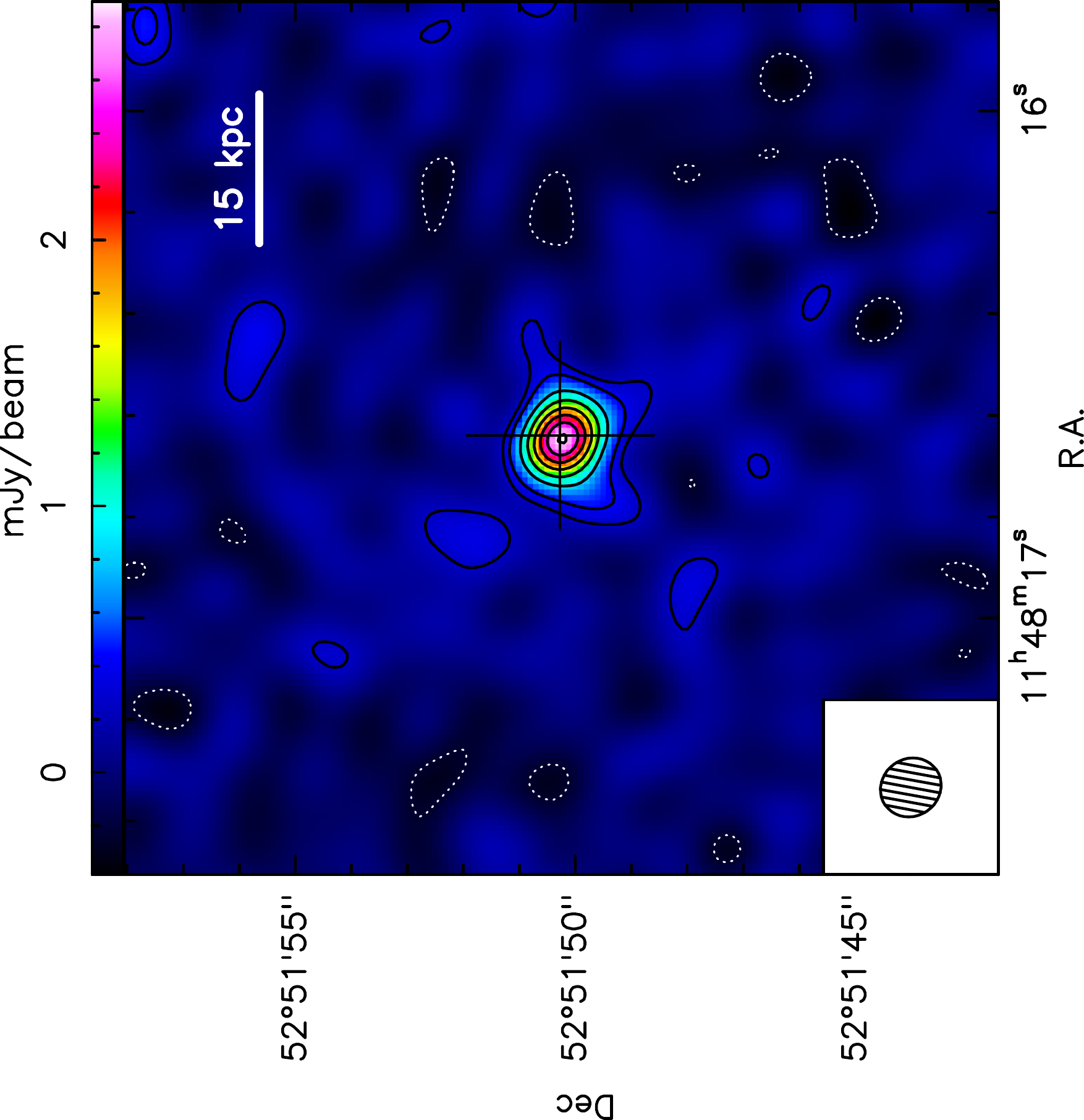}\quad
    \includegraphics[angle=270,width=.95\columnwidth,clip=true,trim=-1.5cm 0cm -0.5cm 0cm]{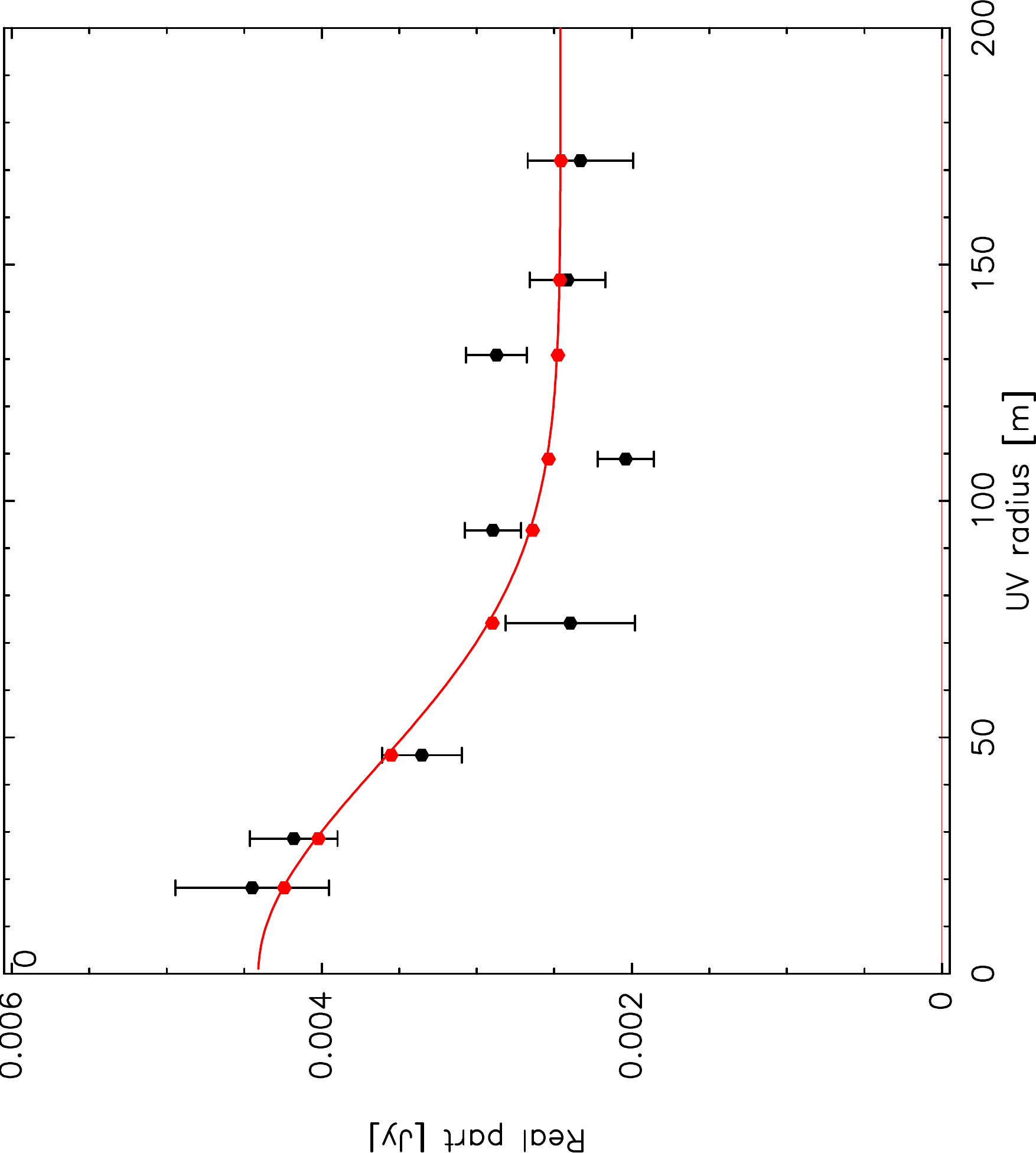}\\
     \caption{IRAM PdBI map ({\it left}) and $uv$ plane analysis ({\it right}) 
     of the FIR continuum of SDSS~J1148+5251 at average rest-frame wavelength $\lambda = 155 \mu$m
     (average observed frequency of 261~GHz).
     {\it Left:} 
     Negative and positive contours correspond to -3$\sigma$, 3$\sigma$, 5$\sigma$ to 35$\sigma$ in steps of 5$\sigma$ 
     (1$\sigma$ rms noise is 0.082 mJy beam$^{-1}$). 
     The synthesized beam is shown at the bottom left corner (1.1\arcsec$\times$1.0\arcsec) and
     the cross indicates the pointing and phase centre, corresponding to the optical position of the quasar.
     {\it Right:} Real part of visibilities as a function of the $uv$ distance ($uv$ plot), binned in baseline
     steps of 20m. The red curve indicates the best fit to the $uv$ plot, which corresponds to the sum of two components: 
     a point (unresolved) source, responsible for the flattening of the
     visibilities at large $uv$ radii ($uv$ radii $>100$~m), and an extended (resolved) source fitted with a circular Gaussian model, which is 
     responsible for the decreasing trend at smaller $uv$ radii.}
   \label{fig:fir_cont}
\end{figure*}

\begin{figure}[tb]
	\centering
    \includegraphics[angle=270,width=.9\columnwidth]{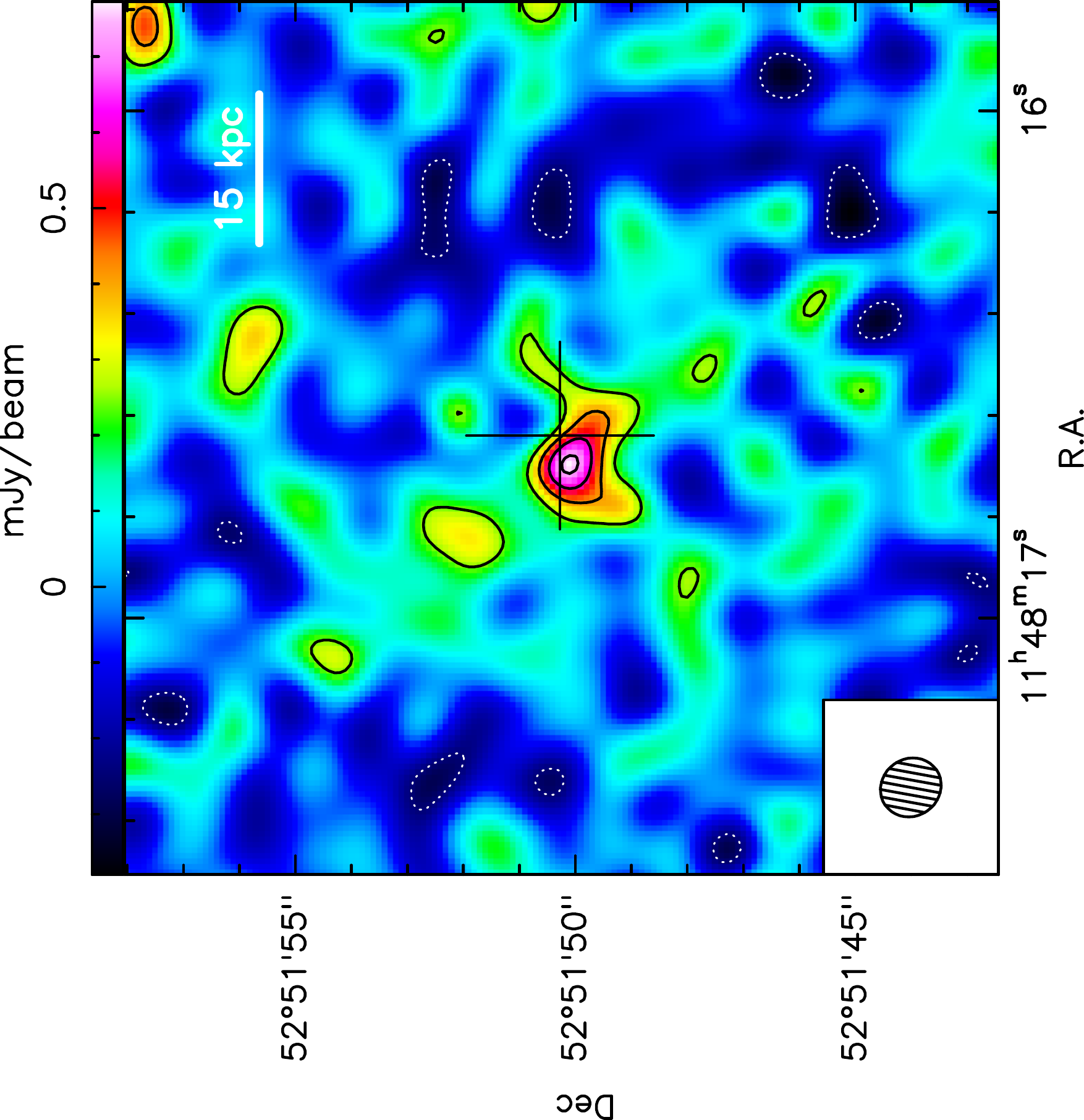}\quad
    \includegraphics[angle=270,width=.9\columnwidth]{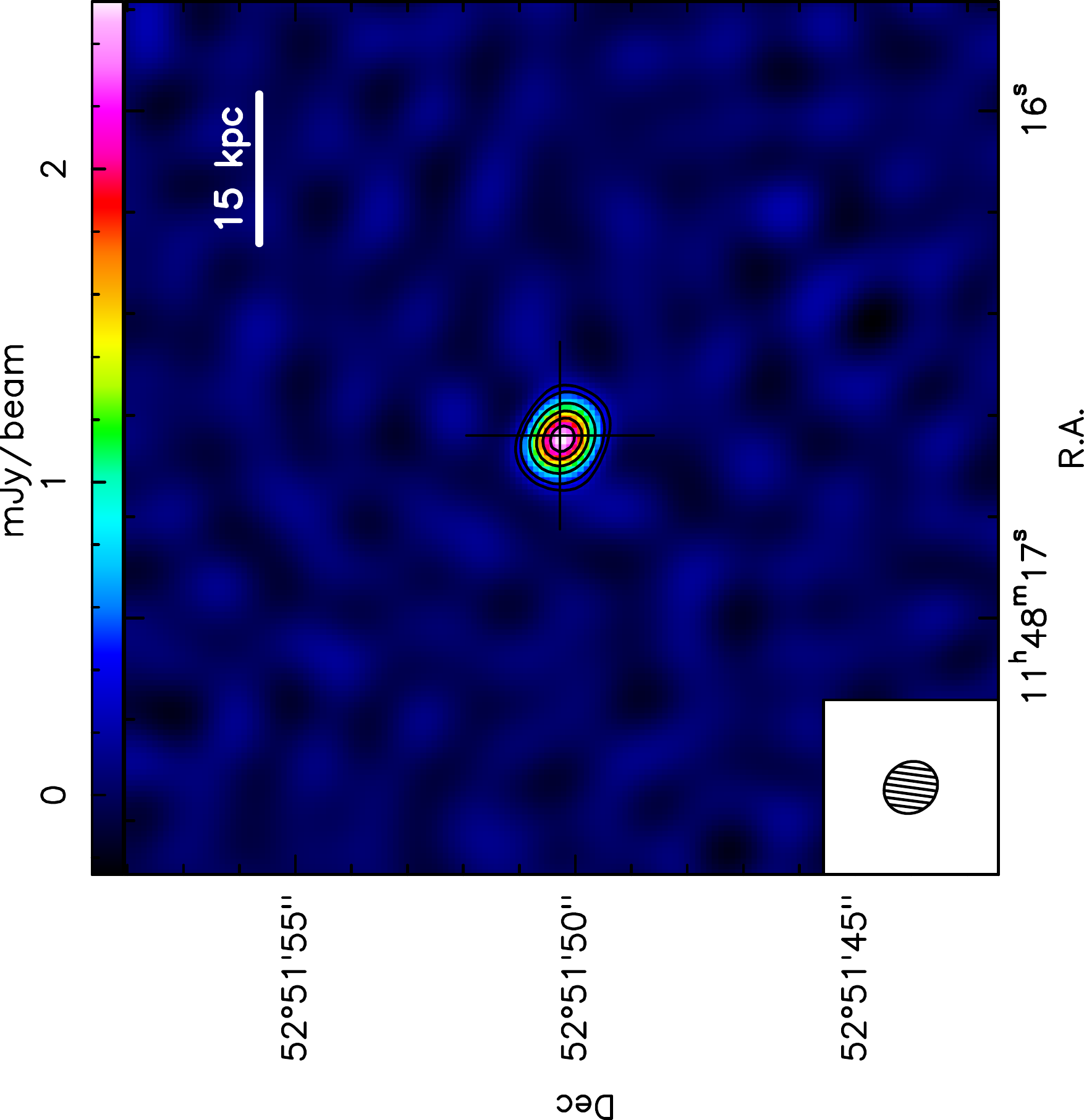}\\
     \caption{Maps of the extended ({\it top}) and compact ({\it bottom}) components that contribute
     to the FIR continuum emission of SDSS~J1148+5251 at average rest-frame wavelength $\lambda= 155\mu$m.
     The map of the extended component has been obtained by subtracting, from the $uv$ data, the point source
     model of the compact component resulting from the $uv$ fit reported in Fig.~\ref{fig:fir_cont}. The map of the compact
     component has been obtained by applying a $uv$ tapering that cuts visibilities with $uv$ radius $<100$~m, therefore slightly
     increasing the resolution of the data (the resulting beam size is 0.99\arcsec$\times$0.89\arcsec).
     In the map of the extended component, contours correspond to
     -3$\sigma$, 3$\sigma$, 5$\sigma$ to 9$\sigma$ in steps of 2$\sigma$
     (1$\sigma$ rms noise is 0.082 mJy beam$^{-1}$).
     For the compact component, contours correspond to 
     -3$\sigma$, 3$\sigma$, 5$\sigma$ to 25$\sigma$ in steps of 5$\sigma$ 
     (1$\sigma$ rms noise is 0.088 mJy beam$^{-1}$).
     In both panels, the cross indicates the pointing and phase centre, corresponding to 
     the optical position of the quasar.
     }
   \label{fig:fir_cont_components}
\end{figure}

\subsection{Quiescent (narrow) \cii emission}\label{sec:core}

The map of the narrow component of the \cii158$\mu$m emission, 
integrated within $v\in(-200, 200)$~\kms, is presented in Figure~\ref{fig:core}.
The source is resolved at high significance by our PdBI observations, and it
appears surprisingly extended, on scales $\gtrsim$ 20 kpc. This could already be inferred 
from the spectra extracted from different apertures,
which showed that the flux of the narrow component is a factor of $\sim$2 
larger in the 4 arcsec aperture spectrum with respect to the 2 arcsec aperture spectrum 
(Figure~\ref{fig:spectra} and Table~\ref{table:gauss_fit}). 

The $uv$ plot of the narrow \cii158$\mu$m emission, in the right panel
of Fig.~\ref{fig:core}, provides important information about the structure of the source.
In particular, the {\it uv} plot clearly shows the presence of two components: a compact
component (not resolved), which is responsible for the flattening of the $uv$ 
visibility amplitude at large ($>$50m) $uv$ radii, and an extended component 
(resolved at high significance), which is at the origin of the decreasing trend of the
$uv$ visibility amplitude with radius at small $uv$ radii \footnote{We note that the detection
of this spatially extended component has been confirmed also by splitting the dataset
into independent parts.}. The red curve represents our best fit model, which 
is obtained by fitting the compact component with a point source and the resolved component with
a circular Gaussian distribution. The resulting FWHM size of the circular Gaussian model is 
$3.1\pm0.3$~arcsec, corresponding to $17.4\pm1.7$~kpc. The velocity-integrated fluxes derived from the $uv$ modelling 
are (6.2 $\pm$ 0.6) Jy km s$^{-1}$ and (3.08 $\pm$ 0.12) Jy km s$^{-1}$ for the resolved/extended and point
sources, respectively. These fluxes can be translated into \cii luminosities of 
${L_{\rm [CII], ext} =  (6.7 \pm 0.7) \times 10^{9}~L_{\odot}}$ and 
${L_{\rm [CII], point} =  (3.33 \pm 0.13) \times 10^{9}~L_{\odot}}$. 
Strikingly, the $uv$ fit indicates that 
about 70\% of the total \cii emission at velocities $v\in(-200, 200)$~\kms
is in the extended component. 
We note that, however, the quality of the data does not allow us to distinguish between different models for the
extended source. In particular, an exponential intensity distribution would also provide a good fit to the resolved
component. 

The extended and compact components contributing to the narrow \cii emission within $v\in(-200, 200)$~\kms
can be disentangled in the $uv$ plane and, therefore, imaged separately, 
by using the model fit shown in Fig.~\ref{fig:core}. In the upper panel of 
Fig.~\ref{fig:core_components} we show the map of the extended 
component, obtained by subtracting from the $uv$ data the point source model fit.
The map shows an irregular ring-like and filamentary structure, with the most prominent filament extending up to
$r\sim28$~kpc (5 arcsec) in the western direction with respect to the QSO position. 
The centroid of the extended source is 
slightly shifted to the south-east with respect
to the optical QSO position (indicated by the cross).
The map of the compact component is shown in the bottom panel of Fig.~\ref{fig:core_components},
and it has been obtained by applying a $uv$ tapering that cuts the visibilities
with $uv$ radius $<$ 80m. This bright, compact source, unresolved in our data and peaking at the QSO position, 
had already been detected and imaged
by \cite{Walter+09} and it is likely tracing a kpc-scale hyper-starburst in the centre of SDSS~J1148+5251. 

We find no evidence for a regular rotation pattern in our IRAM PdBI \cii maps (Appendix~\ref{sec:moments}). 
A spatial shift between the blue- and the red-shifted sides of the \cii line
was found in previous high resolution interferometric \cii observations by \cite{Walter+09}, which
we now know that were resolving out a large fraction of extended \cii flux ($\sim$70\% within $v\in(-200, 200)$~\kms 
as shown by our $uv$ fit in Fig.~\ref{fig:core}). While the compact source, which is not resolved by our observations,
is probably rotationally-broadened as suggested by \cite{Walter+09}, the hypothesis of rotation does not seem to hold for the dominant 
extended component, as shown
in Fig.~\ref{fig:moments}, although some irregular velocity pattern is observed, possibly associated
with some minor mergers. This implies that it is difficult to estimate the 
dynamical mass of this system on large scales.

The spatially extended and kinematically quiescent (i.e. ``narrow'') \cii source discovered in 
SDSS~J1148+5251 (top panel of Fig.~\ref{fig:core_components}) is reminiscent of the giant ($r\gtrsim50$~kpc) low
surface brightness and kinematically unperturbed ($\sigma_v\lesssim$~300 \kms) haloes found in 
high redshift radio galaxies (HzRGs) by \cite{Villar-Martin+03}. These quiescent nebulae around HzRGs are detected in several
emission lines (Ly$\alpha$, \ion{C}{IV}, \ion{He}{II}, \ion{N}{V}) and are thought to trace
giant halo gas reservoirs (rather than companion galaxies) photoionised primarily by the radio-loud quasar continuum. Although 
SDSS~J1148+5251 is not radio-loud \citep{Carilli+04}, and although previous Hubble Space Telescope observations by \cite{Decarli+12}
did not find evidence for extended Ly$\alpha$ emission (possibly due to strong dust extinction on large scales, as supported by our 
FIR continuum observations described in Section~\ref{sec:FIR_cont}), our new IRAM PdBI observations of an extended and quiescent \cii halo
surrounding this high-z quasar show some similarities (e.g. the spatial extent and the velocity dispersion) with the gas haloes 
surrounding HzRGs. In the case of SDSS~J1148+5251, however, we also detect associated 
FIR continuum emission up to $r\simeq8-10$~kpc (Sec.~\ref{sec:FIR_cont}), suggesting that extended star formation, rather
than the QSO, is the main excitation source of the \cii halo, at least up to scales of $r\simeq10$~kpc.
We further discuss the origin of the extended \cii source and its connection with star formation in Sections~\ref{sec:cii_fir_lum_ratio}
and \ref{sec:cii_origin}.

\subsection{Analysis of the FIR continuum extent}\label{sec:FIR_cont}

We investigate the presence of extended dust emission in SDSS~J1148+5251 by using ancillary high sensitivity
IRAM PdBI observations of the FIR continuum in this source obtained as a part of a parallel project presented by 
\cite{Gallerani+14} (see also Section~\ref{sec:observations}). 
These deep ancillary continuum observations at 262~GHz and 259.4~GHz are particularly suited for studying the extent of the FIR
emission in SDSS~J1148+5251, as, once merged, they reach a continuum sensitivity of 0.082~mJy~beam$^{-1}$ and 
have a slightly higher spatial resolution (synthesised beam of 1.1\arcsec$\times$1.0\arcsec) than our main dataset at 256~GHz.

The left panel of Figure~\ref{fig:fir_cont} shows the map of the FIR continuum emission of SDSS~J1148+5251
at an average rest-frame wavelength of $\lambda = 155\mu$m (average observed frequency of 261~GHz), 
produced by merging the two datasets at 262~GHz and 259.4~GHz as explained in Section~\ref{sec:observations}
\footnote{A zoomed-out version of this map, showing the marginal detection of an additional continuum
source in the field, is shown in Fig.~\ref{fig:appendix_newsource} (Appendix~\ref{sec:additional_source}). In 
Fig.~\ref{fig:appendix_newsource} we also show a map of the 256~GHz continuum, obtained by collapsing the line-free 
channels of the main dataset and used for the purpose of continuum-subtraction as described in Section~\ref{sec:cont_256}.}. 
The corresponding $uv$ visibility diagram is shown in the right panel of Fig.~\ref{fig:fir_cont}. 
Similarly to the narrow \cii158$\mu$m emission (Figure~\ref{fig:core}), also the dust thermal emission 
can be disentangled in the $uv$ data into a compact (unresolved) and an extended (resolved) component.
Accordingly, we fit the $uv$ plot of the 155$\mu$m FIR continuum with a combination of a point source model 
and a circular Gaussian model, and show the best fit as a red curve in Fig~\ref{fig:fir_cont}.
We stress that, although the circular Gaussian model provides a good fit to the data, it
constitutes a simplistic approximation, and better quality data may reveal a more complex structure.
The $uv$ fit indicates that $45\pm9$~\% of the FIR emission at this wavelength is in the extended
source, and it provides a FWHM size for the circular Gaussian component of $2.0\pm0.5$~arcsec, corresponding to $11\pm3$~kpc. 
Taken at face value, this result suggests that the narrow \cii (FWHM = 17.4 $\pm$ 1.7 kpc) extends {\it beyond} the region of intensive and
high surface density star formation in the host galaxy of SDSS~J1148+5251, traced by the FIR continuum emission (FWHM =  $11\pm3$~kpc)
(further discussion in Section~\ref{sec:discussion}).

We use the $uv$ fit results to produce separate maps of the extended and compact 
FIR sources, which are presented in Fig.~\ref{fig:fir_cont_components}. Strikingly, the resolved component 
exhibits a tail extending by $r\sim10$~kpc in the western direction, i.e. the same direction as the $r\sim28$~kpc 
filament observed in the map of the extended \cii (Figure~\ref{fig:core_components}). This suggests 
that the western \cii filament embeds star formation heating dust on $r\sim$10~kpc scales.

In order to estimate the FIR luminosity associated with the extended and compact dust continuum sources,
we apportion the total FIR luminosity of SDSS~J1148+5251, 
$L_{\rm FIR (42.5-122.5 \mu m)}$=$(1.3 \pm 0.2) \times 10^{13}~L_{\odot}$ (Rosa Valiante,
private communication \footnote{This value of $L_{\rm FIR (42.5-122.5 \mu m)}$ has been estimated by using the
same method as in \cite{Valiante+11}, by restricting the integration range to the
wavelength range 42.5-122.5 $\mu$m.}), 
according to the flux ratio estimated from our $uv$ modelling of the FIR emission at
155 $\mu$m (Figure~\ref{fig:fir_cont}). This results into FIR $(42.5-122.5 \mu m)$ luminosities of 
 ${L_{\rm FIR, ext} =  (6.0 \pm 1.5) \times 10^{12}~L_{\odot}}$ and 
${L_{\rm FIR, point} =  (7.4 \pm 1.6) \times 10^{12}~L_{\odot}}$ for the extended and
compact component, respectively.

\section{Discussion}\label{sec:discussion}

\subsection{The driving mechanism of the outflow}\label{sec:discussion_outflow}

In Section~\ref{sec:outflow} we have studied with unprecedented detail the properties of the 
massive and extended \cii outflow of SDSS~J1148+5251. We have shown that the 
\cii-emitting gas in this source reaches extremely high projected outflow 
velocities of $v\sim1400$ \kms, with an {\it average}
outflow velocity of $v\simeq800$ \kms (Figure~\ref{fig:outflow_vel}).
The observed very high velocities strongly favour the AGN as the dominant driving 
mechanism of the outflow in SDSS~J1148+5251.

However, SDSS~J1148+5251, besides hosting a powerful unobscured quasar 
($L_{\rm AGN} = 5.63 \times 10^{13}~L_{\odot}$, \cite{Schneider+14}), 
is also a powerful starbursting galaxy \citep{Walter+09}. In particular,
our new deep FIR continuum observations show the presence of strong and spatially extended
warm dust continuum emission, indicative of intense star formation occurring up to scales
of at least $r\sim8-10$~kpc (see also discussion in Section~\ref{sec:cii_origin}).
By splitting the total infrared (TIR, 8-1000~$\mu$m) luminosity of 
$L_{\rm TIR (8-1000 \mu m)}$=$(2.2 \pm 0.3) \times 10^{13}~L_{\odot}$ \citep{Valiante+11} 
according to the flux ratio between extended and compact
component estimated from our {\it uv} analysis of the FIR continuum emission (Section~\ref{sec:FIR_cont}),
we obtain for the extended component ${L_{\rm TIR, ext} =  (10 \pm 2) \times 10^{12}~L_{\odot}}$.
By using the relation in \cite{Kennicutt98}, a Chabrier
IMF and a negligible AGN contribution (such assumption is valid only for the extended component,
as the AGN contribution to dust heating may be significant in the unresolved component, as shown
by \cite{Schneider+14}), the TIR luminosity can be converted
into a star formation rate of ${\rm SFR_{\rm ext} = (1000 \pm 200)~M_{\odot}~yr^{-1}}$. 
Supernova events associated with such an high star formation rate may also be 
an important source of feedback in SDSS~J1148+5251 (see also
\citealt{Costa+14a, Costa+14c}).
Following \cite{Veilleux+05}, the kinetic power injected by supernovae associated with 
${\rm SFR_{\rm ext}}$ is ${\rm P_{kin, SF}=(7.0 \pm 1.7) \times 10^{44}~erg~s^{-1}}$.
A comparison between ${\rm P_{kin, SF}}$ and our (lower limit) estimate of the integrated outflow kinetic power
of ${\rm P_{kin, outflow}=(3.3 \pm 0.5) \times 10^{44}~erg~s^{-1}}$ (Section~\ref{sec:outflow_integrated})
would give a lower limit on the coupling efficiency (by assuming a purely star formation-driven outflow) of 47$\pm$13\%.
Such lower-limit efficiency is already significantly higher than the value of 1-10\% expected for star formation-driven winds
(e.g. \citealt{Cicone+14}).

 In the following, based on the velocity and coupling efficiency arguments, we
will assume that, although the intense star formation on large
scales may also play a role, the AGN gives the dominant contribution to driving the outflow. 
This is also in agreement with the hierarchical semi-analytic models by \cite{Valiante+12}, 
who predicted quasar feedback to be the dominant mechanism in
SDSS~J1148+5251.
We will now discuss the properties of the major \cii outflow
in SDSS~J1148+5251 in the light of current models of AGN feedback.

The models that have been proposed so far to explain AGN-driven kpc-scale outflows
can be branched into two main categories: the so-called ``blast - wave'' models, in which large scale
outflows are generated by hydrodynamical coupling of the AGN with the galaxy ISM through fast winds
originating from the accretion disk (e.g. \citealt{King03,King05,King10,Lapi+05,Menci+06,Menci+08,
Zubovas+12,Zubovas+14,Faucher-Giguere+12,Costa+14b} and references therein), and the models invoking radiation 
pressure on dusty clouds
as the main mechanism at the origin of galactic-scale outflows (e.g. \citealt{Hopkins+Elvis10,Fabian12,Roth+12,Thompson+14} and refs therein). 
In the recent years, both classes of models have been tested against observations of AGN-driven outflows and,
in particular, of massive kpc-scale outflows of cold and dense molecular gas in nearby galaxies 
\citep{Feruglio+10,Feruglio+13a,Feruglio+13b,Alatalo+11,Sturm+11,Aalto+12a,Veilleux+13,Spoon+13,
Combes+13,Cicone+12,Cicone+14,Garcia-Burillo+14}.
A recent study by \cite{Cicone+14}
showed that the ``blast-wave'' scenario and,
specifically, the ``energy-conserving'' nature of the observed large-scale molecular outflows is favoured
when considering a consistent sample of (U)LIRGs and powerful QSOs in the local Universe. 
However, there are only a few studies
investigating the feedback mechanisms in high-z sources and, in particular, the detail
of information provided by our new \cii observations of SDSS~J1148+5251 is unprecedented.
In this section we discuss our results, presented in Section~\ref{sec:outflow}, in the light of the proposed theoretical models for
AGN-driven outflows.

In this context, theoretical works have shown that there are three main properties of outflows, namely
the mass-loss rate, the momentum rate (or momentum flux) and the kinetic power, that can be 
used as ``key'' indicators for the dominant feedback mechanism in action. In Section~\ref{sec:outflow} and Appendix~\ref{sec:outflow_channel_maps}
we have used our new \cii observations to infer these properties (both integrated and as a function of 
dynamical time-scale and radius)
for the very extended outflow in SDSS~J1148+5251. Bearing in mind that these are lower limits 
(because of the caveats discussed in Section~\ref{sec:outflow} and Appendix~\ref{sec:outflow_channel_maps}), 
the {\it integrated} values that we have obtained, 
do not help much in discriminating between different models. In particular, the modest integrated
momentum rate of $(1.00 \pm 0.14)~L_{\rm AGN}/c$ and kinetic power of $(1.6 \pm 0.2)\times 10^{-3}~L_{\rm AGN}$
would be in principle consistent with any of the proposed models. 
However, the unprecedented spatial extent of the outflow in SDSS~J1148+5251 has allowed us, for the first time, 
to go beyond the integrated values by investigating the dependence of the outflow properties as a function of 
the dynamical time-scale of the outflow and of the
distance from the QSO (Figs~\ref{fig:outflow_dyntime} and \ref{fig:outflow_radial}).
 
The plots in Fig.~\ref{fig:outflow_radial}, in particular, allow a direct comparison with the predictions of two recent studies, 
\cite{Costa+14b}, who incorporate the sub-grid physics representing the ``blast wave'' feedback prescriptions 
into cosmological simulations, and \cite{Thompson+14}, who further develop the physics of dusty radiation-pressure
driven shells and discuss it in the context of the most recent observational results.
Our results appear qualitatively inconsistent with the results obtained by \cite{Costa+14b} from cosmological simulations of 
energy-driven outflows generated by inner accretion disk winds, 
which instead show that, on the same physical
scales of $\lesssim 30$~(proper)~kpc, the
momentum flux and kinetic power of the outflow overall increase with radial distance from the AGN (Figure 15 of \cite{Costa+14b}). Moreover, 
quantitatively, the outflow momentum rate and kinetic power inferred for SDSS~J1148+5251 are about one order
of magnitude lower than the integrated values reported by \cite{Costa+14b} for the same simulation. 
We note however that the previously mentioned results by \cite{Costa+14b} were obtained using a central black hole with mass
of $\sim10^8 M_{\odot}$, i.e. significantly lower than the BH mass of
$\sim 3\times 10^9 M_{\odot}$ estimated for SDSS~J1148+5251 \citep{Willott+03}. 
Simulations may produce significantly different results for more massive halos and BHs. 
For example, \cite{Costa+14b} show that, in presence of a more massive BH ($M_{\rm BH}\sim 10^9 M_{\odot}$),
even in the ``momentum-driven'' limit, i.e. the limit in which the thermal energy input from the inner wind 
is not conserved, efficient large scale AGN-driven outflows may develop, comparable to those developed 
under ``energy-conserving'' conditions.
In this case, by definition, the momentum flux would be of the order of $L_{\rm AGN}/c$, consistently to what
we infer for SDSS~J1148+5251. 

\cite{Costa+14b} also show that, when an AGN-driven outflow develops, regardless of its momentum- or energy-conserving
nature, its morphology is highly anisotropic and it strongly depends on the cosmological environment. In particular,
the outflow develops preferably along ``lower density'' paths, free from filamentary cosmic inflows or from
regions characterised by large amounts of dense gas in the halo or in the host galaxy. Interestingly, by comparing the maps in 
Fig.~\ref{fig:wings_map} and the top panel of Fig.~\ref{fig:core_components}, we note that the peak of the outflow component
does not coincide with the peak of the narrow extended component. On the contrary,
the narrow extended component (top panel of Fig.~\ref{fig:core_components}) peaks to the south-east
with respect to the QSO position, where the outflow appears suppressed.
If assuming that the emission within $v\in (-200, 200)$~\kms is not significantly contaminated
by the outflow, this result would support the scenario in which the outflow develops preferably in 
directions avoiding high density regions in the galaxy or in the halo.

We now consider the second class of models of AGN-driven outflows, i.e. models of outflows generated
by coupling of the AGN radiation to dusty interstellar medium at galactic ($\sim$kpc) scales.
More specifically, we refer to the work by \cite{Thompson+14}, who further developed
the physics on which these models are based, demonstrating that dusty radiation-pressure driven shells
can reach high velocities ($v_{\rm out}\gg v_{\rm esc}$). 
The predictions of \cite{Thompson+14} are qualitatively consistent with our observations of 
the outflow in SDSS~J1148+5251, as shown by a comparison between our Fig.~\ref{fig:outflow_radial} and
Fig. 3 of their paper. In particular, the decreasing trend with $R$ and the modest (although lower limit) value of the momentum flux
inferred for SDSS~J1148+5251 could be explained by dusty radiation-pressure driven shells without
resorting to extreme ISM conditions such as $\tau_{\rm IR}\gg 1$. Our resolved FIR continuum observations
(Figure~\ref{fig:fir_cont_components}) show that large amounts of dust are indeed present in this source. However,
the FIR emission, indicating the presence of warm dust, extends only up to {\it maximum} radial distances of
$r\simeq10$~kpc from the nucleus, while the outflow extends up to $r\simeq30$~kpc.

We finally consider the possibility that a previous episode of radio-jets may have driven the observed outflow.
\cite{Carilli+04} imaged the radio continuum of SDSS~J1148+5251 at 1.4~GHz with the VLA, and 
they found that this source is radio-quiet, with a radio continuum 
emission consistent with being primarily powered by star formation. Therefore, if radio jets are responsible
for the observed large-scale \cii outflow, they have now disappeared.
However, the complex outflow morphology (Figure~\ref{fig:wings_map}) and the rather flat trend of velocity as a function of radius
(Figure~\ref{fig:outflow_vel}) suggest that a significant contribution from radio jets is unlikely, as radio jets 
tend to drive very collimated outflows with large velocity gradients across the kpc-scale jets (e.g. \citealt{Nesvadba+08}).

\subsection{The \cii-to-FIR luminosity ratio}\label{sec:cii_fir_lum_ratio}

 \begin{figure}[tb]
    \includegraphics[width=.82\columnwidth,trim=4cm 9cm 4cm 5.5cm,angle=90]{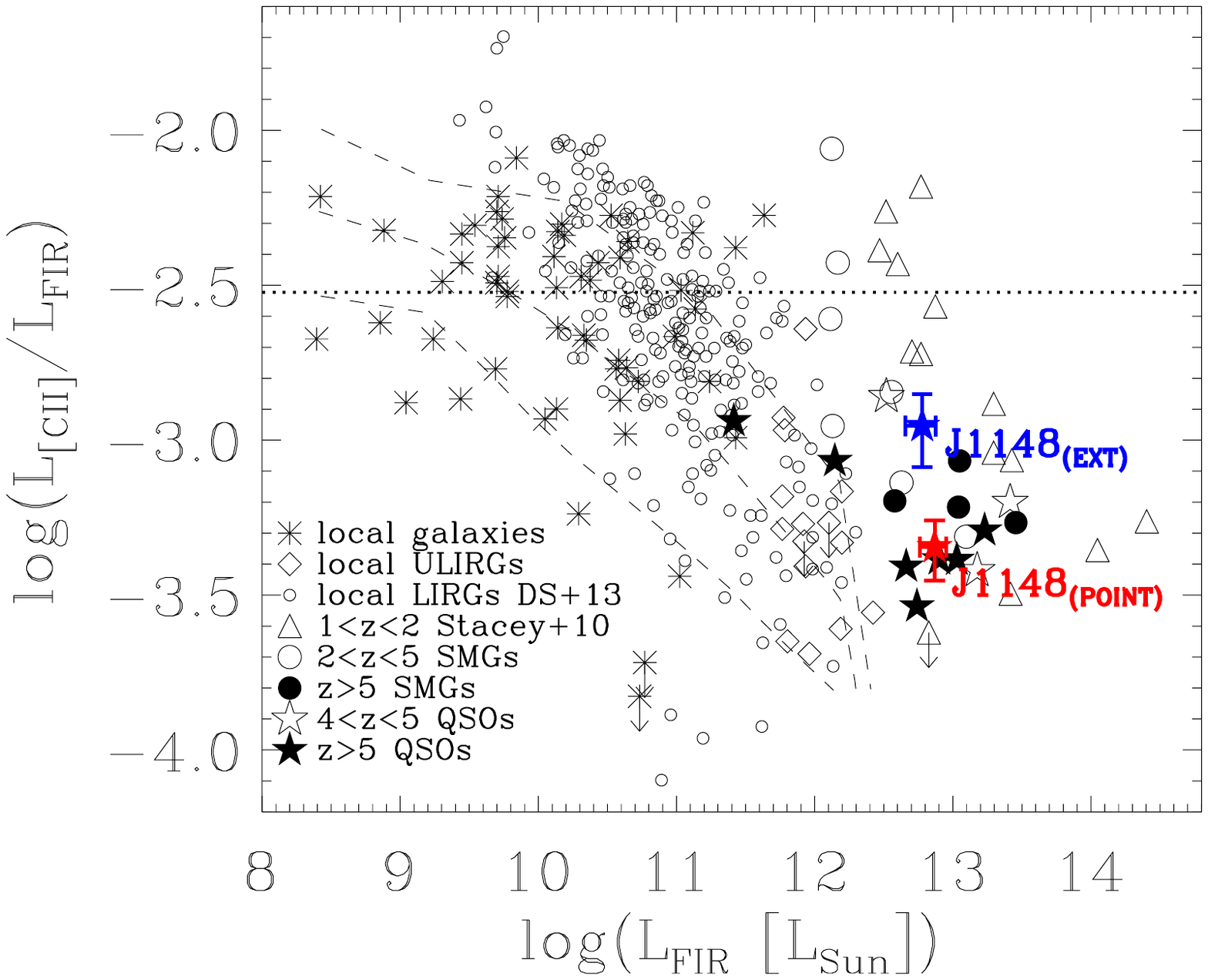}\\
     \caption{\cii/FIR luminosity ratio as a function of FIR luminosity, with our 
    new estimate of $L_{\rm [CII]}/L_{\rm FIR}$  in SDSS~J1148+5251, obtained separately for the extended and compact 
     components, as explained in Section~\ref{sec:cii_fir_lum_ratio}. For comparison, the plot also includes
     previous ISO observations of local star-forming galaxies (asterisks, \citealt{Malhotra+01}) and of local ULIRGs	 
     (diamonds, \citealt{Luhman+03}) along with their median and scatter (dashed lines), recent Herschel-PACS observations of local LIRGs 
    (small open circles, \citealt{Diaz-Santos+13}), the CSO-ZEUS survey of star forming and active galaxies at redshift 
     $1<z<2$ (triangles, \citealt{Stacey+10}), observations of 
     sub-millimeter galaxies (SMGs) at $2<z<5$ (open circles, \citealt{Ivison+10,DeBreuck+14,Valtchanov+11, 
     Cox+11,Swinbank+12,Carniani+13}), sub-millimeter galaxies (SMGs) at $z>5$ (filled circles, \citealt{Riechers+13, Riechers+14,Neri+14,Rawle+14}),
     QSOs at $2<z<5$ (open stars, \citealt{Maiolino+09,Wagg+10,Carniani+13}), and $z>5$ QSOs (filled stars, \citealt{Venemans+12,Wang+13,Willott+13}).
     The dotted horizontal line indicates the Milky Way value  (${L_{\rm [CII]}/L_{\rm FIR}\sim 3 \times 10^{-3}}$, \citealt{Carilli+Walter13}).  
     All measurements refer to the FIR luminosity as integrated over 42.5-122.5 $\mu$m. Where necessary, we have converted 
    $L_{\rm FIR (40-500 \mu m)}$ and $L_{\rm TIR(8-1000 \mu m)}$ into $L_{\rm FIR (42.5-122.5 \mu m)}$ by using the conversion
     factors provided by \citet{Carilli+Walter13}, i.e. $L_{\rm FIR (40-500 \mu m)} \simeq 1.2~L_{\rm FIR (42.5-122.5 \mu m)}$ and
     $L_{\rm TIR (8-1000 \mu m)} \simeq 1.3~L_{\rm FIR (40-500 \mu m)}$.
      }
   \label{fig:cii_fir_compilation}
\end{figure}

In Section~\ref{sec:core} and \ref{sec:FIR_cont} we have shown a method, based on the $uv$ visibility analysis, 
for decomposing both the narrow \cii emission 
and the FIR continuum emission of SDSS~J1148+5251 into an extended (resolved) and a compact (unresolved) component.
Following this decomposition, 
we estimate the \cii-to-FIR luminosity ratios separately for these two components.
 We obtain ${\rm log}(L_{\rm [CII]}/L_{\rm FIR})_{\rm EXT}$ = -2.95$^{+0.10}_{-0.13}$ and
${\rm log}(L_{\rm [CII]}/L_{\rm FIR})_{\rm POINT}$ = -3.34$^{+0.09}_{-0.11}$. These estimates 
are compared in Fig.~\ref{fig:cii_fir_compilation}
with literature $L_{\rm [CII]}/L_{\rm FIR}$ measurements obtained for local and distant galaxies.

We stress that the $L_{\rm [CII]}/L_{\rm FIR}$ values reported in Fig.~\ref{fig:cii_fir_compilation} refer to 
the compact and extended components of the {\it narrow} \cii emission in SDSS~J1148+5251,
integrated within (-200, 200)~\kms. By selecting only the emission within this narrow velocity range, we intend to 	
avoid contamination by the outflow, whose contribution to the \cii flux may be significant already at 
velocities $|v|>$200~\kms, as shown in Appendix~\ref{sec:outflow_channel_maps}. However, the reader should 
be aware that, by selecting a wider velocity range for the \cii emission, the ratio $L_{\rm [CII]}/L_{\rm FIR}$ would
naturally increase, especially for the extended component. In particular, by selecting the \cii emission within (-300, 400)~\kms, we obtain 
${\rm log}(L_{\rm [CII]}/L_{\rm FIR})_{\rm EXT}$ = -2.79$^{+0.10}_{-0.14}$ and 
${\rm log}(L_{\rm [CII]}/L_{\rm FIR})_{\rm POINT}$ = -3.26$^{+0.09}_{-0.11}$, corresponding
to an increase of $L_{\rm [CII]}/L_{\rm FIR}$ by a 
factor of (1.5 $\pm$ 0.5) and by a factor of (1.2 $\pm$ 0.4) for the extended and compact components, respectively.

The \cii-to-FIR luminosity ratio is related to the contribution of the \cii line emission to the cooling of the (neutral) ISM 
in comparison to the cooling via dust grains as traced by the FIR emission. 
In local galaxies, early ISO observations and, more recently, Herschel observations, have
established a clear trend for decreasing \cii-to-FIR ratio with increasing FIR luminosity  
\citep{Luhman+03, Brauher+08, Diaz-Santos+13}. The decreasing trend is even more evident
for $L_{\rm [CII]}/L_{\rm FIR}$ as a function of  $L_{\rm FIR}/M_{\rm H_2}$ \citep{Gracia-Carpio+11}. In particular, in
local LIRGs and ULIRGs, $L_{\rm [CII]}/L_{\rm FIR}$ drops by about one order of magnitude with respect to normal disk galaxies, and this
deficiency is also extended to other FIR fine structure lines such as the \oi 63$\mu$m and \nii 122$\mu$m emission lines.

Lower \cii-to-FIR ratios are most likely associated with intense star formation activity occurring in 
compact and warm starbursts \citep{Diaz-Santos+13}. More specifically, in compact starbursts, the main factors that can act to
reduce the \cii emission with respect to the FIR emission are: (i) high gas volume densities ($n\gg n_{\rm cr}^{\rm [CII]}$),
for which the level populations of the C$^+$ atoms ratio as their statistical weights, and thus the \cii intensity
becomes almost insensitive to higher gas volume densities and higher FUV fluxes \citep{Wolfire+89}; (ii) high ratios of FUV intensity field to
the total hydrogen gas volume density, which would increase the positive charge of dust grains, thereby reducing
the efficiency of gas heating via photo-electrons released by dust grains and so the relative efficiency of cooling via \cii emission with respect to
dust cooling \citep{Luhman+03, Diaz-Santos+13}; (iii) a high ionisation parameter $U$ (which is essentially linked to (ii)) 
that reduces the opacity
of hydrogen atoms (because a high fraction of HI is ionised) with respect to the dust opacity, 
hence increasing the fraction of UV photons
absorbed by dust. Additionally, the incident UV radiation heats the dust to higher temperatures \citep{Abel+09, Gracia-Carpio+11}.
Lower \cii-to-FIR ratios may also be associated with an increased AGN contribution to the dust heating (and so to
 $L_{\rm FIR}$) for sources with higher FIR luminosities. According to \cite{Sargsyan+14}, the \cii luminosity is little
 affected by the presence of an AGN and it traces mostly star formation, while the FIR luminosity can have a significant AGN contribution
 in addition to the starburst component, resulting therefore
 in lower $L_{\rm [CII]}/L_{\rm FIR}$ for AGN-dominated sources.
 
However, a relation between the \cii luminosity and the 
FIR luminosity has not been clearly established for high redshift infrared luminous galaxies, which
display a large scatter in the $L_{\rm [CII]}/L_{\rm FIR}$ vs $L_{\rm FIR}$ diagram \citep{Gracia-Carpio+11}, 
in some cases with values that are completely consistent with local 
disk galaxies, as also shown by Fig.~\ref{fig:cii_fir_compilation}. 
We find that, in the case of SDSS~J1148+5251, the \cii-to-FIR ratio in the compact source is lower 
than in the extended component, by a factor of $\sim$2.5 (Figure~\ref{fig:cii_fir_compilation}). 
We note that a similar result has been obtained by \cite{Diaz-Santos+14} for the nuclear and extended 
components of local LIRGs observed with Herschel-PACS. 
This finding is consistent with a picture in which the nuclear \cii and FIR emissions in 
SDSS~J1148+5251 are excited by a compact ``hyper-starburst'' \citep{Walter+09}, similar to the
nuclear regions of local ULIRGs, while
the extended components trace more diffuse, ``disk-like'' star formation.

\begin{figure}[tb]
	\centering
        \includegraphics[width=.82\columnwidth,trim=3.8cm 3.5cm 3.cm 5cm,angle=90]{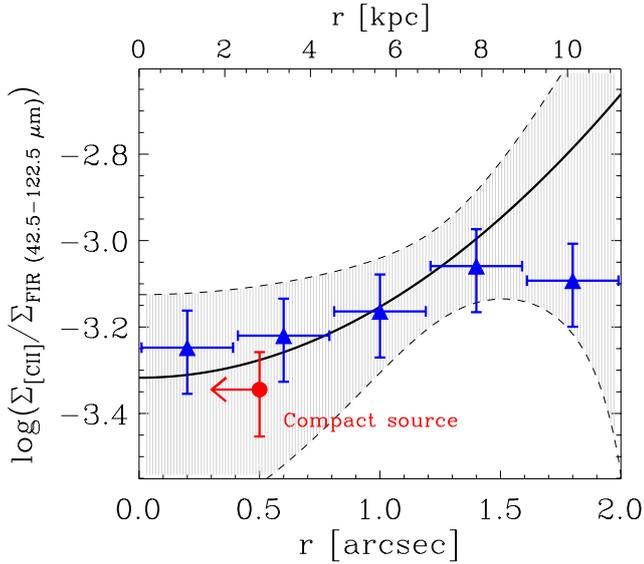}\\
     \caption{\cii/FIR luminosity surface density ratio as a function of the projected distance from the QSO optical position. 
     		The black curve represents our model of $\Sigma_{\rm [CII]}/\Sigma_{\rm FIR}(r)$ 
		for the extended source, obtained from the analysis of the $uv$ data as explained in
     		Section~\ref{sec:cii_fir_lum_ratio}. The grey shaded region indicates the 
		statistical uncertainty on the $\Sigma_{\rm [CII]}/\Sigma_{\rm FIR}$ predicted by the model
     		based on the $uv$ analysis. The blue triangles show the values of $L_{\rm [CII]}/L_{\rm FIR}$
		calculated at different radii by integrating the extended \cii and FIR emission over concentric annuli. 
		For comparison, the \cii/FIR luminosity ratio 
     		of the compact (unresolved) component is also shown, plotted at a representative 
     		upper limit radius of ${\rm FWHM_{beam}/2}$, where ${\rm FWHM_{beam}}$ 
		refers to the beam-size of the FIR continuum observations (Figure~\ref{fig:fir_cont}).
     } 
   \label{fig:cii_fir_ratio}
\end{figure}

In Figure~\ref{fig:cii_fir_ratio} we show the \cii-to-FIR surface density ratio as a function
of the projected distance from the QSO. 
The black curve is obtained by modelling in the $uv$ data the \cii and FIR extended
components with circular Gaussian distributions (Sections~\ref{sec:core} and \ref{sec:FIR_cont}), and it corresponds to:
\begin{equation}
\frac{\Sigma_{\rm [CII]}}{\Sigma_{\rm FIR}}(r)=\frac{L_{\rm [CII], ext}}{L_{\rm FIR, ext}}\Bigg(\frac{\sigma_{\rm FIR}^2}{\sigma_{\rm [CII]}^2}\Bigg)\exp\Bigg[-r^2\Bigg(\frac{1}{2\sigma_{\rm [CII]}^2}-\frac{1}{2\sigma_{\rm FIR}^2}\Bigg)\Bigg],
\end{equation}
where $L_{\rm [CII], ext}$ =$(6.7\pm 0.7)\times 10^{9} L_{\odot}$,   $L_{\rm FIR, ext}$ = $(6.0 \pm 1.5)\times 10^{12} L_{\odot}$, 
$\sigma_{\rm [CII]}$ = $(1.32 \pm 0.13)$ arcsec and 
$\sigma_{\rm FIR}$ = $(0.9 \pm 0.2)$ arcsec, as derived from the $uv$ analysis.
We remind that this model of $\Sigma_{\rm [CII]}/\Sigma_{\rm FIR}(r)$ 
is based on the narrow \cii emission integrated within (-200, 200)~\kms, because in this velocity range
the \cii flux contamination by the outflow is low ($\lesssim$20\%, according to the 
spectral decomposition shown in Fig.~\ref{fig:spectra}). 
Considering a larger velocity range for the \cii emission
would have as a main effect an upward shift of the relationship in Fig.~\ref{fig:cii_fir_ratio}. In particular,
if taking into account the \cii emission within (-300, 400)~\kms, $\Sigma_{\rm [CII]}/\Sigma_{\rm FIR}$
increases, at every radius, by a factor of $\sim 1.4$.

The blue triangles in Fig.~\ref{fig:cii_fir_ratio} show the $\Sigma_{\rm [CII]}/\Sigma_{\rm FIR}(r)$ ratio obtained by integrating the 
\cii extended emission in the top panel of Fig.~\ref{fig:core_components} over concentric annuli with
$\Delta r=0.4$~arcsec. These values follow quite well the black curve, at least up to $r\simeq 8$~kpc. 
We also show in Fig.~\ref{fig:cii_fir_ratio} the $L_{\rm [CII]}/L_{\rm FIR}$ ratio obtained for the compact source,
plotted in correspondence of an upper limit-radius of 0.5~arcsec (2.8~kpc), corresponding to half beam size.
Interestingly, the \cii-to-FIR ratio measured for the compact source is consistent within the one-sigma uncertainty with the
value obtained for the extended source at a comparable $r$, suggesting that the two components
may be powered by a similar mechanism (i.e. intense and compact star formation), at least up to $\sim$5 kpc distance from the nucleus.
At larger $r$, i.e. $5<r {\rm[kpc]} <10$, $\Sigma_{\rm [CII]}/\Sigma_{\rm FIR}(r)$ increases to values more compatible
with the disks of local galaxies (although still lower than the values measured in the Milky Way and in the 
bulk of local galaxies, see Figure~\ref{fig:cii_fir_compilation}), hinting
at a less extreme star formation activity occurring on these scales. 

The discovery of extended star formation (at least up to scales of $r\sim10$ kpc) in SDSS~J1148+5251 challenges
the scenario presented by previous observations of high redshift quasar host galaxies, which suggested that in this
class of sources both the star formation and the cold gas reservoirs are confined within less than a few kiloparsecs from the nucleus 
\citep{Carilli+Walter13}. Quite interestingly, the possibility of a diffuse and extended mode of star formation in high-z
massive galaxies was predicted by \cite{Hatch+08}, who discovered a large (60 kpc) halo of diffuse ultraviolet light, powered
by a young stellar population, surrounding the Spiderweb Galaxy, a radio galaxy at $z=2$. 
Understanding the origin of the \cii emission
at $r\gtrsim10$ kpc and its connection with star formation is however more complex, 
and possible explanations are suggested in Section~\ref{sec:cii_origin}.

\subsection{Origin of the \cii emission at $r\gtrsim10$~kpc}\label{sec:cii_origin}

Figure~\ref{fig:cii_fir_ratio} shows that our model uncertainty for the \cii-to-FIR surface density ratio increases significantly
for $r\gtrsim10$~kpc, and, in particular, no constraints can be put on $\Sigma_{\rm [CII]}/\Sigma_{\rm FIR}$ at such large radii. 
This is due to the fact that, 
above $r\sim 8-10$ kpc, the FIR continuum
drops below our sensitivity and only \cii is detected, as it can be clearly seen by comparing the upper panels of Figures~\ref{fig:core_components} 
and \ref{fig:fir_cont_components}. 
Previous studies have suggested that \cii emission is a good tracer of star formation in galaxies regardless
of their classification as starbursts, AGN hosts or composite galaxies (e.g. \citealt{Sargsyan+14, DeLooze+14}).
However, the abrupt drop of FIR surface density that we observe in SDSS~J1148+5251 at $r\gtrsim10$~kpc may indicate that
the \cii flux in this source and on such large scales does not correlate with star formation (and so with the FIR emission)
in the same way as in the nuclear regions. In other words, extended \cii emission at $r\gtrsim10$~kpc may trace a {\it different
regime of star formation} with respect to the central regions of the galaxy ($r<5$~kpc) and also with respect to the
regions at $5<r{\rm[kpc]}<10$. 

Furthermore, at $r\gtrsim10$~kpc, a significant fraction of \cii may
trace a different ISM phase than dense PDRs. For example, \cii may arise from 
diffuse HI (mostly cold neutral medium, CNM), diffuse molecular clouds and, in minor
measure, diffuse warm ionised medium (WIM). This behaviour of \cii has been observed in
low metallicity systems such as Haro 11 and in the outskirts of the
Milky Way and other local galaxies \citep{Kramer+13,Parkin+13,Pineda+13,Langer+14}. 
In particular, \cite{Kramer+13} have shown that the CNM contribution
to \cii in the outer regions of M33 increases by $\sim$25\% with respect to the inner 
region (i.e. $r<2$~kpc). This effect is accompanied by an increase of $L_{\rm [CII]}/L_{\rm FIR}$ by almost a factor of 4 
for $r>4.5$~kpc, similarly to what we observe in SDSS~J1148+5251, although in M33 the transition occurs at much smaller scales.
Moreover, \cite{Pineda+13} and \cite{Langer+14} have shown that, in our Galaxy,
a significant fraction of \cii stems from CNM ($\sim$20\%) and from ``CO-dark'' molecular gas 
($\sim$30\%), which dominates the gas mass of diffuse and ``warm'' molecular clouds (i.e. a
transition phase between diffuse HI clouds and dense CO-bright molecular clouds), and these 
fractions increase with increasing galactocentric distance.
In addition, \cite{Vallini+13} have shown through radiative transfer simulations of the ISM
in Himiko \citep{Ouchi+13} that \cii emission may arise from the CNM residing at the periphery ($r\geq5$~kpc)
of the galaxy.

However, we note that the \cii surface density that we measure in the extended component of 
SDSS~J1148+5251 is still three orders of magnitude higher than what observed in the outskirts
of local galaxies. This implies that the contribution of different ISM phases to the \cii emission discussed
above is not sufficient to justify the extended \cii flux - with no significant associated FIR emission - observed 
at $r\gtrsim10$~kpc in SDSS~J1148+5251.
Therefore, although a comparison with the local Universe may help 
understanding what we observe in SDSS~J1148+5251, we should keep in mind that
the galactic environment of a luminous QSO at $z>6$ can be very different
even from the most extreme galaxies that we observe locally, and even more so from
the Milky Way or other local disks such as M33 and M51. 

In order to clarify the 
origin of the extended \cii emission at $r>10$~kpc, high sensitivity and multi-wavelengths observations
are needed. In particular, while our IRAM PdBI observations prove, for the first time, that cold gas (both in outflow and not)
extends up to $r\sim$30~kpc in the host galaxy of SDSS~J1148+5251, corresponding to about
$1/3$ of the virial radius estimated for its massive halo (\citealt{Valiante+11, Valiante+14}), it is still not clear 
what is the primary mechanism exciting \cii up to such large distances from the nucleus.
In summary, whether associated with dense PDRs or with a more diffuse ISM phase, 
the extended - and relatively high surface density - 
\cii emission discovered in SDSS~J1148+5251 requires an excitation source.
This may be provided by in situ star formation, by the nuclear starburst, by the QSO and/or also by strong shocks
(e.g. see \citealt{Appleton+13}).

\section{Summary and conclusions}

We have presented follow-up C and D configuration IRAM PdBI observations 
of the \cii 158$\mu$m fine structure line and
FIR continuum in the host galaxy of the luminous quasar SDSS~J1148+5251 at redshift z=6.4189. 
The total \cii velocity-integrated flux of $34\pm9$~Jy \kms, estimated by using 
a spectrum extracted from a circular aperture with a diameter of 8 arcsec, is
significantly larger than any previous measurement of the same source. This clearly
indicates the presence, in the QSO host galaxy, of additional baryonic mass (in cold gas)
that was missed by previous observations with a lower sensitivity to large scale structures.
The \cii spectrum shows broad emission extending from -1400~\kms to 1200~\kms 
from the systemic velocity, confirming the detection of a powerful outflow by \cite{Maiolino+12}.
The total continuum-subtracted map shows a very extended \cii source, with filamentary
structures extending up to the extreme distance of $r\sim30$~kpc from the position of the QSO.

We have further investigated the properties of the {\it outflow}, and we have found that:
\begin{itemize}
\item The outflow is resolved at high significance by our observations,
revealing an extremely complex morphology, butterfly-like, with filaments extending 
up to $r\simeq30$~kpc from the nucleus. 
\item Although the median outflow dynamical time-scale is $\tau_{\rm dyn}^{\rm median} = 25$~Myr, the
distribution of $\tau_{\rm dyn}$ within the outflow is quite broad, ranging from $\sim$4~Myr to $\sim$100~Myr, indicating that the
outflow has been in place for at least 100~Myr, which is a non-negligible fraction of the age of the Universe at this redshift.
\item Our new estimates of the lower limits on the {\it integrated} mass-loss rate, momentum rate and kinetic power of the outflow
are of $1400\pm 300~M_{\odot}~{\rm yr}^{-1}$, $(1.00 \pm 0.14)~L_{\rm AGN}/c$ and $(1.6 \pm 0.2)\times 10^{-3}~L_{\rm AGN}$,
respectively.
\item We have investigated, for the first time in an external galaxy, the outflow properties as a function of the dynamical
time-scale and of the projected distance from the QSO. The results suggest that the ejection of gas in this source has not
occurred at a constant rate, but most likely through multiple outflow events. In particular, we have found that
the bulk of the mass, energy and momentum in this source have been released within the past $\sim20$~Myr.
\end{itemize}

We have discussed possible driving sources for the observed \cii outflow, finding that 
quasar feedback is likely the dominant mechanism driving the outflow in SDSS~J1148+5251 (although the contribution
from extended star formation may be significant). We have discussed the outflow properties in the light of theoretical models for AGN-driven outflows.
We have noted that the {\it integrated} outflow momentum rate of $(1.00 \pm 0.14)~L_{\rm AGN}/c$ and kinetic power 
of $(1.6 \pm 0.2)\times 10^{-3}~L_{\rm AGN}$ are 
quite modest compared to local ULIRGs /QSOs and in principle 
consistent with any of the proposed models of AGN-driven outflows.
By considering the 
mass-loss outflow rate, momentum rate and kinetic power of the outflow as a function of the
distance from the QSO, we have found that 
observations are qualitatively consistent with radiation-pressure driven dusty shells.
However, we have also found that the morphology and distribution of the outflow is consistent with recent
cosmological simulations involving energy-driven outflows. Yet, a more quantitative comparison with the latter
studies requires simulations matching the properties of this specific object (halo mass, SMBH mass).

Quite surprisingly, we have discovered that not only the outflow, but also the {\it quiescent gas} traced by \cii
at systemic velocity is extremely extended in this source. 
More specifically,
we have found that the \cii emission within $v\in(-200, 200)$ \kms traces two components: 
\begin{description}
\item[(i)]  A compact source,
not resolved at our spatial resolution of $\sim$1.2~arcsec, which corresponds to the 1.5~kpc size 
nuclear ``hyper-starburs'' detected by previous observations;
\item[(ii)] An extended source with 
a FWHM size of 3.1$\pm$0.3~arcsec, corresponding
to 17.4$\pm$1.7~kpc. The map of this component reveals a 
complex structure, with a prominent filament extending up to $r\simeq28$~kpc to the west (with respect to the QSO position).
Strikingly, this extended component contributes to almost $\sim$70\% of the 
\cii flux within $v\in(-200, 200)$ \kms.
\end{description}

By using ancillary deep FIR continuum observations at $\lambda_{\rm rest} = 155~\mu$m, we have found that 
also the FIR emission from dust can be disentangled 
into a compact and a more extended component contributing to 45\% of the total FIR emission at this wavelength.
The extended FIR component has a FWHM size of ($11\pm3$) kpc, with a tail extending by $r\sim10$~kpc to the west, 
and it has therefore a smaller extent than the extended \cii source.

We have estimated the \cii-to-FIR surface density ratio as a function of the distance from the QSO, and we have
found that, within $r\lesssim 5$~kpc, $\Sigma_{\rm [CII]}/\Sigma_{\rm FIR}$ is low ($\simeq6\times10^{-4}$)
and not significantly different between the compact and the extended components, suggesting that both components may be powered by an intense and 
compact starburst. At intermediate radii, i.e. $5<r[{\rm kpc}]<10$, $\Sigma_{\rm [CII]}/\Sigma_{\rm FIR}$ 
increases up to $\simeq$1.0$\times$10$^{-3}$, supporting the hypothesis of less extreme star formation activity on these scales. 
At larger radii, i.e. $r\gtrsim10$~kpc, the FIR continuum drops below our sensitivity and only \cii is detected. This indicates
that at such large distances from the galaxy centre, the \cii emission does not correlate with FIR continuum and
with star formation in the same way as in the central regions. Moreover, similarly to the outskirts of local
galaxies and to low metallicity systems, a large fraction of \cii at $r\gtrsim10$~kpc 
may arise from different phases of the ISM than dense PDRs, such as diffuse atomic or molecular gas clouds. 
However, how \cii emission can be excited up to such large distances from the QSO
and from the intense central starburst remains obscure, and additional observations with the future facilities such as NOEMA
are needed to answer 
this question.

\begin{acknowledgements}
We thank the referee, Carlos De Breuck, for providing constructive comments that have significantly
helped to improve the paper. We also thank Rosa Valiante and Christian Leipski for useful discussion.
CC acknowledges support by the Isaac Newton Studentship and by STFC funding.
This work is based on observations carried out with the IRAM Plateau de Bure Interferometer.
IRAM is supported by INSU/CNRS (France), MPG (Germany) and IGN (Spain).
In this paper we use the IDL SED fitting code developed by Caitlin M. Casey \citep{Casey12}.
\end{acknowledgements}

\bibliography{ref}
\bibliographystyle{aa}

\appendix 
\section{Additional FIR continuum source}\label{sec:additional_source}

\begin{figure}[tbc]
	\centering
    \includegraphics[angle=90,clip=true,width=.48\textwidth]{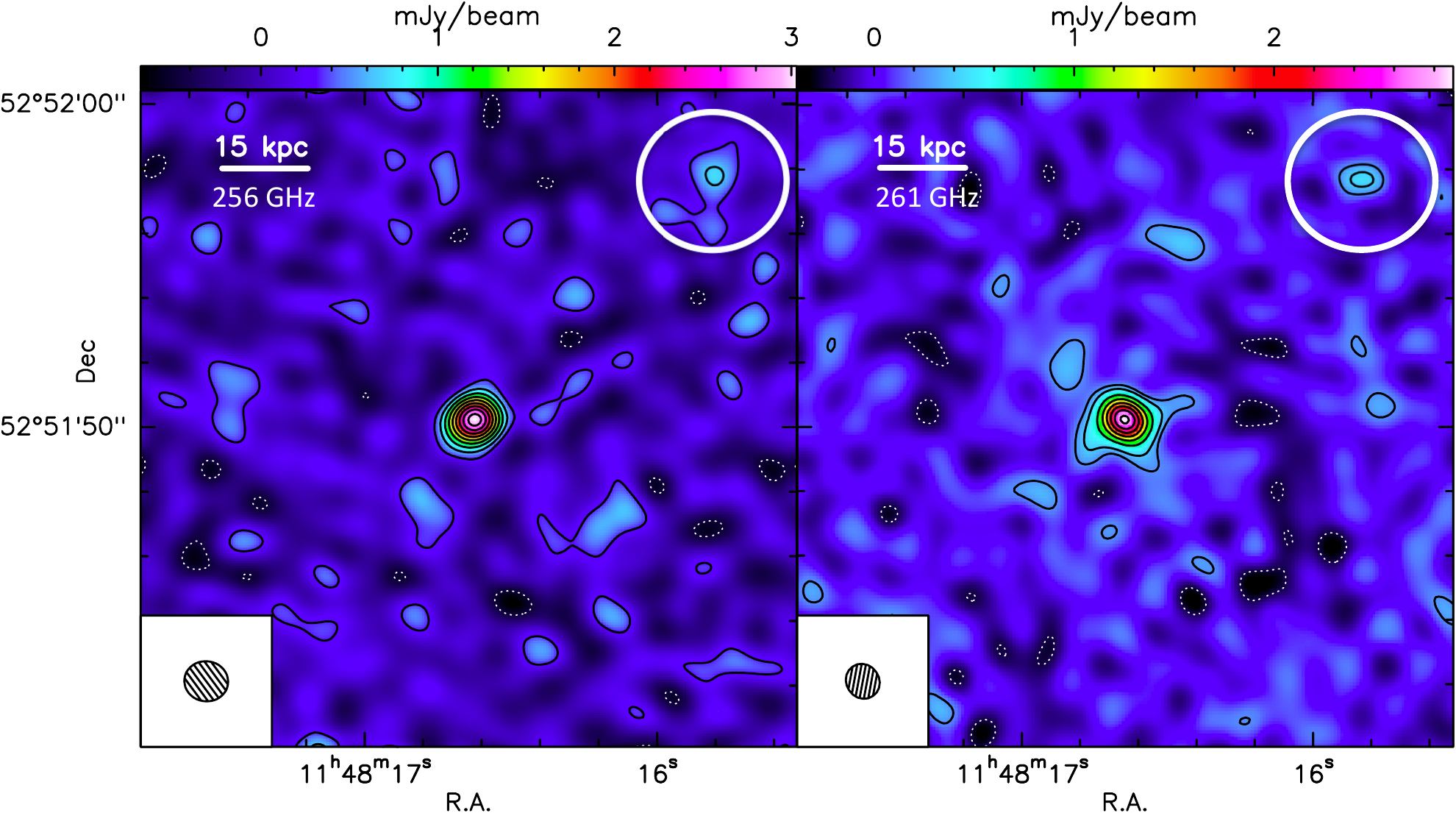}\\
       \caption{IRAM PdBI maps of the FIR continuum showing the marginal detection of an additional source
       in the field, north-west of SDSS~J1148+5251, at a projected distance of 10.5~arcsec, originally 
       discovered by Herschel. {\it Left:} The map on the left panel
       has been obtained by collapsing the line-free channels of the dataset at 256 GHz as explained in Section~\ref{sec:cont_256}.
       The synthesized beam is 1.3\arcsec$\times$1.2\arcsec. Negative and positive contours correspond to $-3\sigma$, $2\sigma$,
       $4\sigma$ to $18\sigma$ in steps of $2\sigma$ ($1\sigma$ rms noise is 0.159~mJy~beam$^{-1}$). {\it Right:} The map on the right panel
       has been obtained by merging the two ancillary datasets at 262 GHz and 259.4 GHz (average observed frequency of 260.7 GHz). 
       This is the same map as Figure~\ref{fig:fir_cont}, which is employed to study the spatial extent of the FIR continuum.
       The synthesized beam is 1.1\arcsec$\times$1.0\arcsec. Negative and positive contours correspond to $-3\sigma$, $3\sigma$, $5\sigma$
       to $40\sigma$ in steps of $5\sigma$ ($1\sigma$ rms noise is 0.082~mJy~beam$^{-1}$).
        }
   \label{fig:appendix_newsource}
\end{figure}

\begin{figure}[tbc]
\centering
    \includegraphics[width=.7\columnwidth,trim=4cm 5cm 4cm 5.5cm,angle=90]{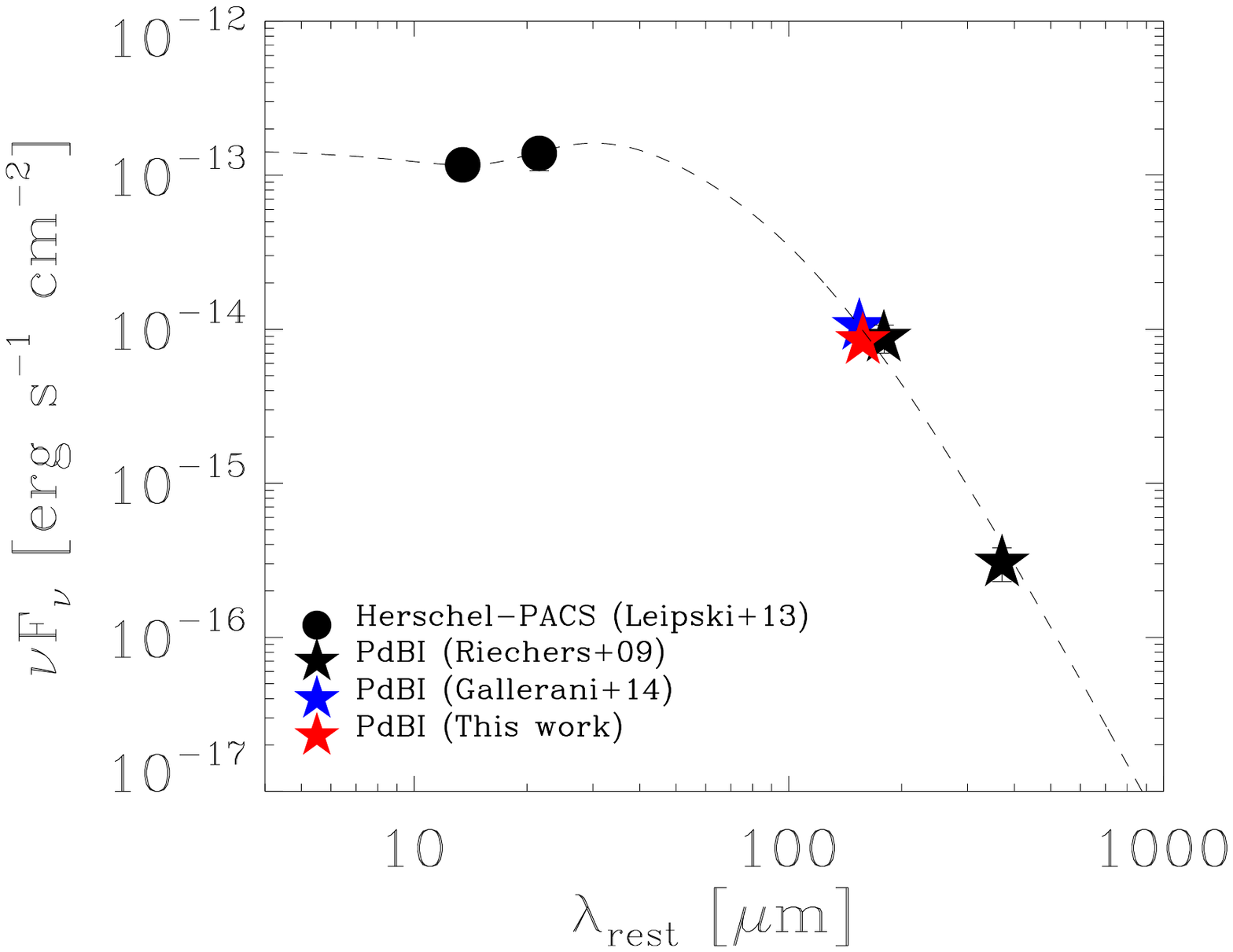}\\
       \caption{Far-infrared Spectral Energy Distribution (SED) of SDSS~J1148+5251, where only the 
       		continuum measurements
       		corrected for the contaminating north-west source are shown, i.e. the Herschel-PACS observations 
		 by \cite{Leipski+13}, the PdBI observations by \cite{Riechers+09} and \cite{Gallerani+14}, and our new PdBI observations at
		 256~GHz (Section \ref{sec:cont_256}). Different symbols have been used for different datasets (see legend at the bottom-left corner of the plot).
		 The FIR SED has been fit with a single dust temperature modified blackbody with fixed emissivity 
		 $\beta=1.6$, combined with a
		 MIR power-law, following \cite{Casey12}. Our new PdBI data point at 256 GHz has been excluded from the fit, as explained in the text.
		}
   \label{fig:fir_sed}
\end{figure}

Herschel observations revealed the presence of an additional FIR continuum source,
north-west of SDSS J1148+5251, at a distance of $\sim$10\arcsec \citep{Leipski+10,Leipski+13,Leipski+14}. This
north-western source is detected at the $\sim$5$\sigma$ level in our PdBI observations. More specifically, Figure~\ref{fig:appendix_newsource} shows
that a FIR continuum source at a distance of 10.5\arcsec from SDSS J1148+5251 is independently detected in both our FIR continuum 
maps at 256~GHz (4.5$\sigma$) and
at 261~GHz (5.6$\sigma$). 

In Section~\ref{sec:cont_256} we have evidenced a 2$\sigma$ inconsistency between our new FIR continuum flux density estimate
at 256~GHz of (3.3 $\pm$0.7)~mJy and the flux density of (4.8 $\pm$ 0.6)~mJy expected at the same observed frequency
from the MAMBO-II bolometric observations by \cite{Bertoldi+03}. Such small discrepancy can be fully
explained by the presence of the north-western source contaminating the MAMBO-II flux measurement.
We further test this hypothesis by performing a fitting of the FIR SED of SDSS J1148+5251, in which we select
only the continuum measurements available in the literature that are not contaminated by the 
north-western source or that have been corrected for this effect (Figure~\ref{fig:fir_sed}). In particular, we select the
100~$\mu$m and 160~$\mu$m Herschel-PACS observations by \cite{Leipski+13} (corrected for the contaminating source), 
the 1.1mm PdBI observations by \cite{Gallerani+14} (observed frequency
of 262~GHz), and the 1.3mm and 2.8mm PdBI observations by 
\cite{Riechers+09} (observed frequencies of 225~GHz and 109~GHz, respectively).
We exclude from the SED fitting our PdBI data point at 256~GHz (shown as a red star in Figure~\ref{fig:fir_sed}),
as the purpose of the fit is to show the consistency of our continuum measurement with the 
previous observations, once corrected for the contaminating north-western source.
The FIR SED in Figure~\ref{fig:fir_sed} is fit with a single temperature modified blackbody combined with a MIR power-law, following
\cite{Casey12}. By fixing the emissivity $\beta=1.6$, we obtain a dust temperature of 64~K.
We note that our PdBI continuum 
observation at 256~GHz sits perfectly on the the SED model in Fig.~\ref{fig:fir_sed}.

We then use the FIR SED fit model to estimate the expected MAMBO-II flux density,
by taking into account the spectral response of the bolometer (transmission curve). 
We obtain an expected MAMBO-II flux density of 3.8 mJy, i.e. significantly lower than the value
reported by \cite{Bertoldi+03}, which would in turn predict a flux density at 256~GHz of
(3.6 $\pm$ 0.4) mJy, consistent with our PdBI measurement.

\section{Spectral features in the extended \cii source}\label{sec:position_spectra}

\begin{table}
\centering
 \begin{minipage}{\columnwidth}
  \caption{Results of the Gaussian fits to the spectra extracted from different positions on the source shown in Fig.~\ref{fig:position_spectra}}
  \label{table:pos_fit}
\begin{tabular}{@{}lcccc@{}}
\hline
 Pos$^{\dag}$     & $v$     & $\sigma_v$ &  $S_{\rm peak}$ & $I_{\nu}$ \\
		          & [\kms] 	& [\kms] 	     &  [mJy] 		  & [Jy \kms] \\
\hline
A 			&  -700 $\pm$ 300 & 800 $\pm$ 300  &   0.8 $\pm$ 0.2 &  1.7 $\pm$ 0.8 \\
			& 1100 $\pm$ 140   & 230 $\pm$ 150   & 0.8 $\pm$ 0.4 &  0.5 $\pm$ 0.4 \\
B			& -170 $\pm$ 40 & 110 $\pm$ 50 & 2.7 $\pm$ 1.0 & 0.7 $\pm$ 0.4 \\
			&  -300 $\pm$ 400 & 800 $\pm$ 400 & 0.9 $\pm$ 0.5 & 1.7 $\pm$ 1.3 \\		
C 			& -1120 $\pm$ 110 & 210 $\pm$ 120 & 1.5 $\pm$ 0.7 & 0.8 $\pm$ 0.6 \\
			& 700 $\pm$ 500 & 700 $\pm$ 400 & 1.0 $\pm$ 0.5 & 1.6 $\pm$ 1.2 \\
			& 130 $\pm$ 100 & 140 $\pm$ 120 & 1.4 $\pm$ 1.0 & 0.5 $\pm$ 0.6 \\
D 			& -900 $\pm$ 400 & 600 $\pm$ 400 & 1.1 $\pm$ 0.5 & 1.5 $\pm$ 1.3 \\
			& 800 $\pm$ 200  & 500 $\pm$ 200 & 1.6 $\pm$ 0.5 & 1.9 $\pm$ 1.1 \\
E		        & -1430 $\pm$ 90 & 130 $\pm$ 90 & 1.1 $\pm$ 0.6 & 0.4 $\pm$ 0.3 \\
			& 100 $\pm$ 200 &  500 $\pm$ 200 & 1.5 $\pm$ 0.3 & 2.0 $\pm$ 0.9 \\
			& 1200 $\pm$ 200 & 300 $\pm$ 200 &  0.7 $\pm$ 0.6 &  0.5 $\pm$ 0.5 \\
F			& 110 $\pm$ 80 & 220 $\pm$ 80 & 1.9 $\pm$ 0.5 & 1.0 $\pm$ 0.5 \\
			& 1100 $\pm$ 90 & 100 $\pm$ 90 & 1.0 $\pm$ 0.7 & 0.3 $\pm$ 0.3 \\
			& -800 $\pm$ 500 & 400 $\pm$ 600 & 0.4 $\pm$ 0.4 & 0.3 $\pm$ 0.7 \\	
 \hline
\end{tabular}
\end{minipage}
\begin{flushleft}
\small
\textbf{Notes:} 
$^\dag$ Each row
corresponds to a Gaussian component. We note that in
all cases the spectral fit requires multiple Gaussian components. 
\end{flushleft}
\end{table}

In Table~\ref{table:pos_fit} we report the results of the spectral fitting to the bright features A-F
of the extended \cii 158$\mu$m source, shown in Fig.~\ref{fig:position_spectra}.
It is immediately evident from Table~\ref{table:pos_fit} that the
features A-F are characterised, on average, by high velocity dispersions, which
can be as high as $\sigma_v\sim800$~\kms, suggesting that they are mostly associated with the 
outflow discovered by \cite{Maiolino+12}. This is confirmed by the analysis of the maps of the broad wings
(Section~\ref{sec:outflow}). However, we note that, in correspondence of some positions
(e.g. B, C, F), there is also a significant
contribution from ``narrow'' emission at the systemic velocity, hinting at  
the presence of a very extended ``quiescent'' (i.e. non outflowing) \cii component, which is 
investigated in Section~\ref{sec:core}.

\section{Outflow channel maps}\label{sec:outflow_channel_maps}

\begin{figure*}[tb]
	\centering
    \includegraphics[angle=90,clip=true,width=.99\textwidth]{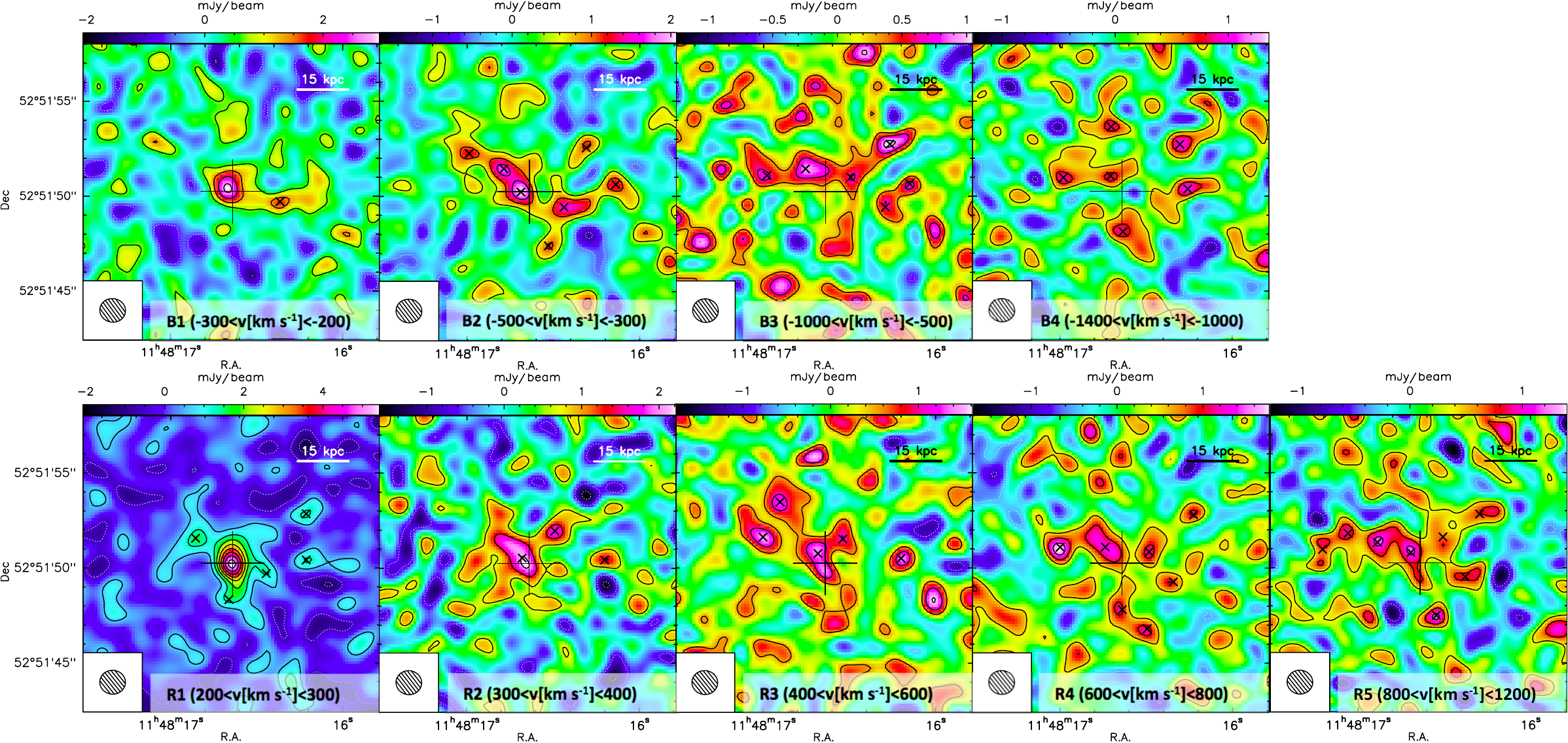}\\
       \caption{IRAM PdBI continuum-subtracted maps of the blue- ({\it top panels}) and red- ({\it bottom panels}) shifted 
       		\cii158$\mu$m emission of SDSS J1148+5251, obtained at different velocities.
		The corresponding velocity integration ranges 
		are indicated on each map. Negative and positive contours are
		in steps of 1.5$\sigma$. 
		In all panels the big central cross
		indicates the pointing and phase centre, corresponding to the optical position of the quasar.
		The small crosses mark the position of the 48 blobs
		that we identified as belonging to the outflow. For each of these clumps in outflow we estimate the dynamical time-scale 
		${\tau_{\rm dyn} = R/v}$, where $R$ is the distance from the QSO position, 
		and $v$ is the average velocity of each map.}
   \label{fig:outflow_channel_maps}
\end{figure*}

In this section we explain the method we employ to estimate the outflow mass-loss rate using the resolved
information provided by our IRAM PdBI observations.
We produce channel maps of the \cii 158$\mu$m blue- and red-shifted emission at velocities $v\leq -200$ \kms and 
$v\geq 200$ \kms, respectively. The maps are shown in Fig.~\ref{fig:outflow_channel_maps}. The channel widths vary
with velocities, because we aim to obtain an approximately constant signal-to-noise for the extended emission in each map. For this reason,
at lower velocities, i.e. closer to the \cii emission peak, we use narrow channels with $\Delta v$ = 100 \kms, while
at higher velocities, tracing the fainter high-velocity component of the outflow, we integrate over wider channels of $\Delta v\sim$ 200-500 \kms.
The centroid positions of the 48 clumps that we ascribe to the outflow are indicated in Fig.~\ref{fig:outflow_channel_maps}.
These are selected to be spatially included within the region corresponding to the total \cii source shown in Fig.~\ref{fig:total_map} and
to have in the maps a signal-to-noise ratio ${\rm SNR}\geq 3$.
We note that we have conservatively excluded from the outflow computation the central components of maps B1 ($\langle v \rangle = -250$~\kms) and 
R1 ($\langle v \rangle = 250$~\kms), because, due to low velocities probed by these two maps, it is difficult to assess to which extent
the central extended emission belongs to the outflow. For similar reasons we have not taken into account the
central core emission within $\pm 200$~\kms, although there may be a significative outflow contribution of $\sim$20\% even at these low
velocities, as suggested by the Gaussian fits to the line profile (Figure~\ref{fig:spectra}). However, 
we have included in the outflow the central blobs
visible in maps B2 and R2 (mean velocities of $\pm 350$~\kms), because they are offset with respect to the
QSO position and appear to be co-spatial with the higher velocity emission in maps B3 and R3, which is undoubtedly tracing the outflow.

For each blob (belonging to the outflow) we calculate from the maps 
its projected distance from the central QSO and, accordingly, obtain a measure of its dynamical time-scale. The dynamical time-scale 
is defined as ${\tau_{\rm dyn} = R/v}$, where $R$ is the distance travelled by the clump (assumed equal to the projected distance 
between the centroid of each clump and the optical position of the QSO) and 
$v$ is the velocity of the gas. 
The error on $R$ is estimated differently for resolved and unresolved clumps; in particular, for unresolved clumps,
$\sigma_R$ is the FWHM of the synthesised beam divided by the signal-to-noise ratio. 
For resolved clumps, instead, we set $\sigma_R = (r_{\rm max}^2-({\rm FWHM}_{\rm beam}/2)^2)^{1/2}$, where $r_{\rm max}$ is the 
maximum (projected) radius of the blob.  
We conservatively
adopt for $v$ the average (projected) velocity of each map. We note that
the uncertainty on $v$, defined as $\sigma_v$=$\Delta v$/2, is obviously higher in the higher velocity maps,
where we have integrated the emission over larger velocity channels ($\Delta v$).
The distribution of ${\tau_{\rm dyn}}$ that we measure within the outflow is shown in Fig.~\ref{fig:t_dyn}.

To estimate the mass of (atomic) gas in outflow, we first measure the \cii flux associated with each outflowing clump, by using
apertures centred on the positions of the crosses (indicated in Fig.~\ref{fig:outflow_channel_maps}). 
We then convert the \cii fluxes (integrated
over their corresponding velocity ranges $\Delta v$) into \cii 158$\mu$m luminosities, and
the \cii luminosities into (lower limits on the) atomic gas mass, following \cite{Hailey-Dunsheath+10} as explained in Section~\ref{sec:outflow}.
By adding up the mass-loss rate contribution, i.e. ${\dot{M}_{\rm out} = M_{\rm out}/\tau_{\rm dyn}}$,
from all the 48 clumps in outflow, we obtain a total integrated outflow rate of $1400 \pm 300~M_{\odot}~{\rm yr}^{-1}$. 
The error is simply $(\sum_{i=1}^{i=48} \sigma_{i,\dot{M}_{\rm out}}^2)^{1/2}$, where $\sigma_{\dot{M}_{\rm out}}$ is
the uncertainty on $\dot{M}_{\rm out}$ for a given clump in outflow, obtained by propagating
errors on $\tau_{\rm dyn}$ and on the velocity-integrated flux associated with the clump. We note, however, that 
the uncertainty on the conversion from \cii luminosity to atomic gas mass is not taken into account in
our error estimate.

\section{Moments maps}\label{sec:moments}

\begin{figure}[tb]
	\centering
    \includegraphics[angle=270,width=.8\columnwidth,clip=true]{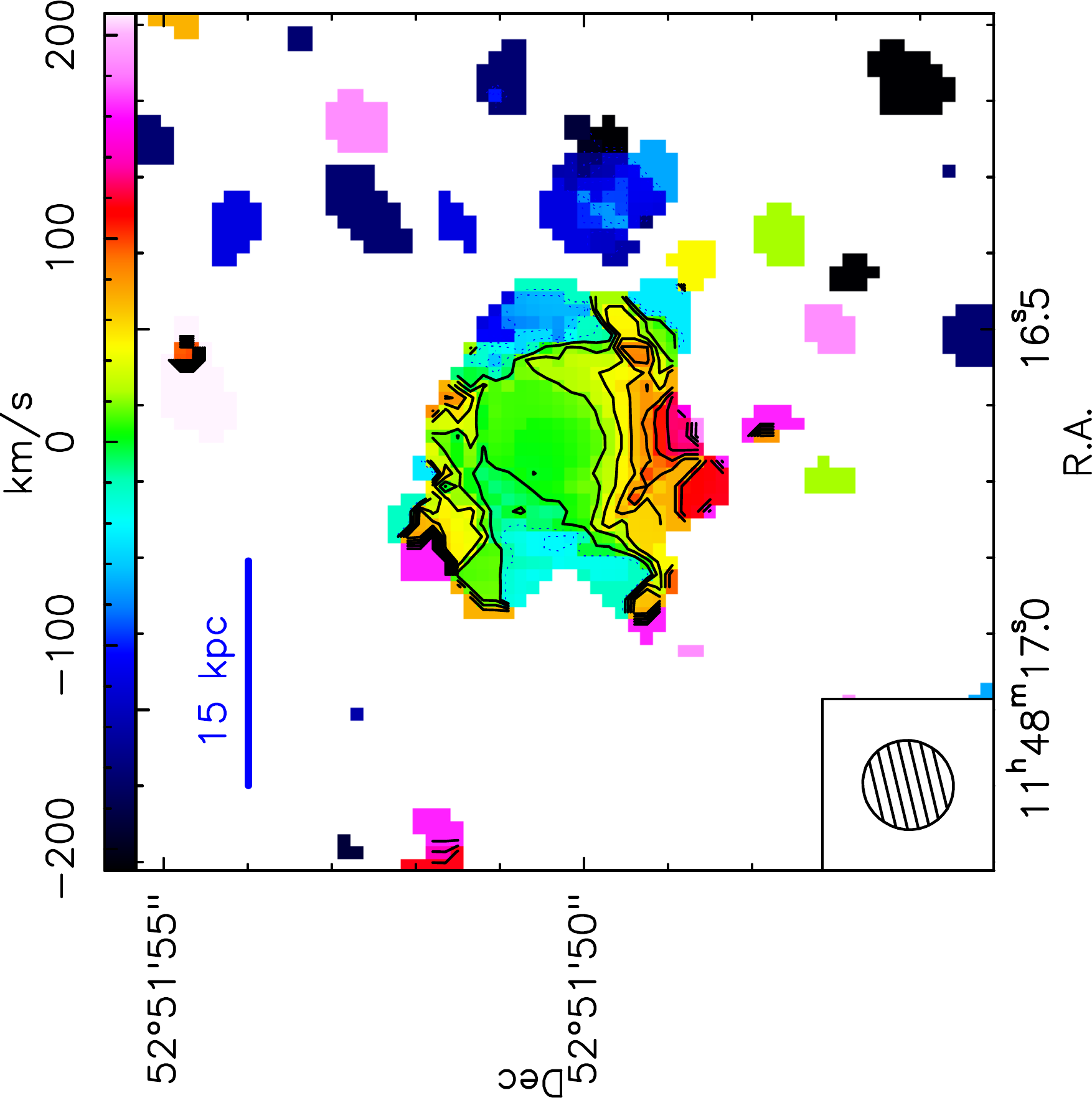}\quad
    \includegraphics[width=.8\columnwidth,clip=true]{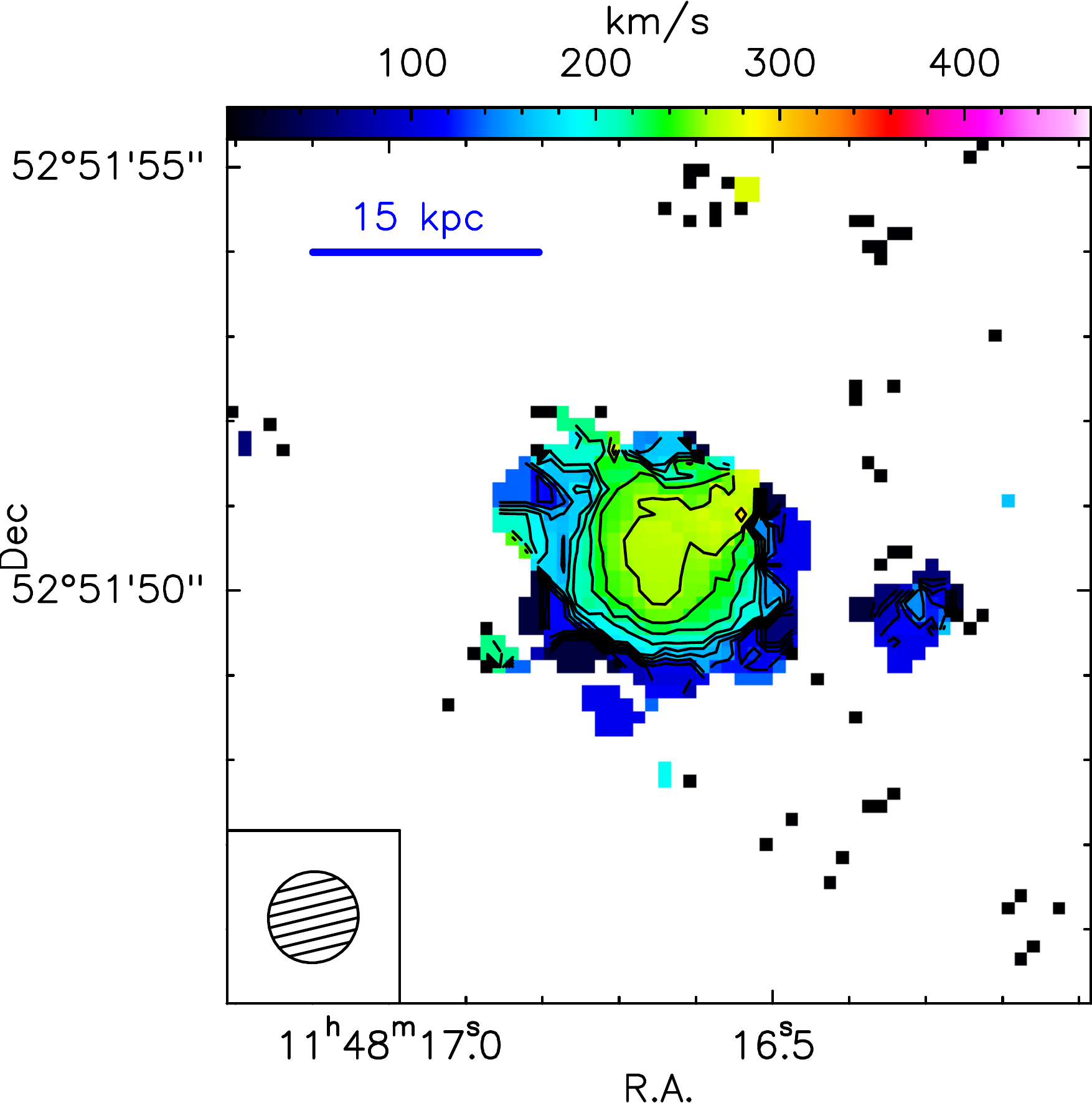}\\
     \caption{First ({\it top}) and second ({\it bottom}) moment maps 
    obtained within $v\in$(-200, 200)~\kms by applying a flux threshold of 2.8~mJy. Velocity contours in both maps are in steps of 20 \kms.
     }
   \label{fig:moments}
\end{figure}

We show in Fig.~\ref{fig:moments} the first and second moment maps of the \cii emission within 
$v\in$(-200, 200)~\kms, obtained by applying a flux threshold of 2.8 mJy. The moments maps suggest that 
the bulk of \cii does not trace gas in a regularly rotating disk. Our IRAM PdBI observations
are dominated by the extended \cii component which, in this velocity range, contributes to $\sim$70\% of the
total flux (Section~\ref{sec:core}). It is however possible that the compact source, unresolved by our observations, 
is rotationally-supported, as suggested by \cite{Walter+09}.
Moreover, the first moment map shows 
a north-south velocity gradient south of the QSO position, with velocities ranging from 20 \kms to 100 \kms, whose
origin is not clear.
In conclusion, no constrains can be put on the total dynamical mass of the system from our
\cii observations.

\end{document}